\documentclass[final,3p,12pt]{elsarticle}
\makeatletter
\patchcmd{\ps@pprintTitle}
  {Preprint submitted to}
  {Preprint accepted to}
  {}{}
\makeatother

\usepackage{epsfig}
\usepackage{amssymb}
\usepackage{amsmath}
\usepackage{amsthm}

\usepackage{lineno}

\usepackage[nonumberlist, style=long, toc, acronym, symbols]{glossaries}

\usepackage{orcidlink}

\usepackage{amsmath,amssymb,amsfonts}
\usepackage{graphicx}
\usepackage{textcomp}
\usepackage{xcolor}
\usepackage{bbm,bm}

\usepackage{svg}
\usepackage[ruled]{algorithm}
\usepackage[algo2e]{algorithm2e} 
\usepackage{algorithmic}
\usepackage{ragged2e}
\usepackage[nonumberlist, style=long, toc, acronym, symbols]{glossaries}
\usepackage{orcidlink}
\usepackage{multirow}
\usepackage{colortbl}
\usepackage{tabulary}
\usepackage{etoolbox}
\usepackage{subcaption}  
\usepackage{tcolorbox}
\usepackage{empheq}
\usepackage[normalem]{ulem}
\usepackage{tabularx} 
\usepackage{booktabs} 

\usepackage{setspace}
\usepackage{color}
\usepackage{tabularray}
\definecolor{JapaneseLaurel}{rgb}{0,0.501,0}
\usepackage{graphicx,natbib,amssymb}

\newcommand{\lam}[1]{\textcolor{black}{{#1}}}

\journal{Computer Communications}

\begin{document}
\newacronym{cdf}{CDF}{Cumulative Distribution Function}
\newacronym{sdwan}{SD-WAN}{Software Defined - Wide Area Network}
\newacronym{sdn}{SDN}{Software Defined Networking}
\newacronym{wan}{WAN}{Wide Area Network}
\newacronym{mpls}{MPLS}{Multi-Protocol Label Switching}
\newacronym{ecmp}{ECMP}{Equal-Cost Multiple Path}
\newacronym{ucmp}{UCMP}{Unequal-Cost Multiple Path}
\newacronym{qos}{QoS}{Quality-of-Service}
\newacronym{vpn}{VPN}{Virtual Private Networks}
\newacronym{nfv}{NFV}{Network Function Virtualisation}
\newacronym{ar}{AR}{Access Router}
\newacronym{ip}{IP}{Internet Protocol}
\newacronym{lb}{LB}{Load Balancer}
\newacronym{spr}{SPR}{Smart Policy Routing}
\newacronym{tcp}{TCP}{Transmission Control Protocol}
\newacronym{udp}{UDP}{User Datagram Protocol}
\newacronym{mlu}{MLU}{Maximum Link Utilization}
\newacronym{lu}{LU}{Link Utilization}
\newacronym{cpe}{CPE}{Customer Premises Edge}
\newacronym{sla}{SLA}{Service Level Agreement}
\newacronym{api}{API}{Application Programming Interface}
\newacronym{nlp}{NLP}{Non Linear Programming}
\newacronym{ns3}{NS3}{Network Simulator 3}
\newacronym{sls}{SLS}{System Level Simulation}
\newacronym{cpu}{CPU}{Central Processing Unit}
\newacronym{gpu}{GPU}{Graphics Processing Unit}
\newacronym{cuda}{CUDA}{Compute Unified Device Architecture}
\newacronym{ml}{ML}{Machine Learning}
\newacronym{drl}{DRL}{Deep Reinforcement Learning}
\newacronym{dl}{DL}{Deep Learning}
\newacronym{rl}{RL}{Reinforcement Learning}
\newacronym{cpo}{CPO}{Constrained Policy Optimization}
\newacronym{rcpo}{RCPO}{Reward Constrained Policy Optimization}
\newacronym{cbf}{CBF}{Control Barrier Function}
\newacronym{mdp}{MDP}{Markov Decision Process}

\newacronym{ddpg}{DDPG}{Deep Deterministic Policy Gradient}
\newacronym{td3}{TD3}{Twin-Delayed Deep Deterministic Policy Gradient}
\newacronym{trpo}{TRPO}{Trust Region Policy Optimization}
\newacronym{ppo}{PPO}{Proximal Policy Optimization}
\newacronym{scip}{SCIP}{Solving Constraint Integer Programs}
\newglossaryentry{As}{%
name=\ensuremath{\mathcal{A}^s},
description={set of safe actions}
}

\newglossaryentry{Ap}{%
name=\ensuremath{\mathcal{A}^p},
description={set of predicted actions}
}

\newglossaryentry{acbf}{%
name=\ensuremath{a^{CBF}},
description={safe action}
}

\newglossaryentry{arl}{%
name=\ensuremath{a_\theta},
description={predicted action}
}

\newglossaryentry{pk}{%
name=\ensuremath{\mathbb{P}_k},
description={set of available paths in tunnel k \in K}
}

\newglossaryentry{de}{%
name=\ensuremath{d_e},
description={delay on each edge/link e \in p \forall p\in \mathbb{P}_k }
}

\newglossaryentry{dpe}{%
name=\ensuremath{d_{e,prop}},
description={propagation delay on each edge/link e \in p \forall p\in \mathbb{P}_k }
}

\newglossaryentry{dpk}{%
name=\ensuremath{d_p^k},
description={delay on each path p of tunnel k }
}

\newglossaryentry{dk}{%
name=\ensuremath{d^k},
description={delay of tunnel k }
}

\newglossaryentry{Dtk}{%
name=\ensuremath{T^k},
description={Source traffic of tunnel k }
}

\newglossaryentry{Dtkh}{%
name=\ensuremath{\widehat{T_t^k}},
description={Admitted traffic of tunnel k }
}

\newglossaryentry{xpk}{%
name=\ensuremath{{x_p^k}},
description={split ratio on path p of tunnel k }
}

\newglossaryentry{lpk}{%
name=\ensuremath{{l_p^k}},
description={load on path p of tunnel k }
}

\newglossaryentry{bpk}{%
name=\ensuremath{b_p^k},
description={boolean variable indicating flow k has traffic routed through path p }
}

\newglossaryentry{mlue}{%
name=\ensuremath{\mu_e},
description={\acrfull{mlu} of link e}
}

\newglossaryentry{lue}{%
name=\ensuremath{\nu_e},
description={\acrfull{lu} of link e}
}

\newglossaryentry{ce}{%
name=\ensuremath{c_e},
description={capacity of link/edge e }
}

\newglossaryentry{le}{%
name=\ensuremath{l_e},
description={traffic load of link/edge e }
}

\newglossaryentry{ncbf}{%
name=\ensuremath{N_{cbf}},
description={CBF solutions per iteration }
}

\newglossaryentry{mcbf}{%
name=\ensuremath{M_{cbf}},
description={maximum iteration to find optimal CBF solution}
}

\newglossaryentry{xv}{%
name=\ensuremath{\bm{x}},
description={weight for link i}
}

\newglossaryentry{N}{%
name=\ensuremath{\mathcal{N}},
description={set of local links attached in WAN network}
}

\newglossaryentry{exp}{%
name=\ensuremath{\mathcal{D}},
description={experience array}
}

\newglossaryentry{DeltaT}{%
name=\ensuremath{\Delta T},
description={measurement period}
}

\newglossaryentry{ttl}{%
name=\ensuremath{du_{mf}},
description={Time To Live of each licro-flow}
}

\newglossaryentry{dmf}{%
name=\ensuremath{D_{mf}},
description={Data rate of each micro-flow}
}

\begin{frontmatter}



\title{Safe Load Balancing in Software-Defined-Networking\tnoteref{paper}}

\tnotetext[paper]{Extended version of our paper presented at IEEE/IFIP NOMS (South Korea, 2024).}
\author[hw]{Lam Dinh\corref{cor}\orcidlink{0000-0003-3043-3627}} 
\cortext[cor]{Corresponding author.}
\ead{lam.dinh@huawei.com; nlam.dinh@gmail.com}

\author[hw]{Pham Tran Anh Quang\orcidlink{0000-0002-5961-8134}}
\ead{phamt.quang@huawei.com}

\author[hw]{J\'er\'emie Leguay\orcidlink{0000-0002-4670-6389}}
\ead{jeremie.leguay@huawei.com}

\affiliation[hw]{organization={Huawei Technologies France},
            addressline={18 Quai du Point du Jour},
            city={Boulogne-Billancourt},
            postcode={92100},
            country={France}}



\begin{abstract}
High performance, reliability and safety are crucial properties of any Software-Defined-Networking (SDN) system. Although the use of \acrfull{drl} algorithms has been widely studied to improve performance,  their practical applications are still limited as they fail to ensure safe operations in exploration and decision-making. To fill this gap, we explore the design of a \acrfull{cbf} on top of \acrfull{drl} algorithms for load-balancing. We show that our \acrshort{drl}-\acrshort{cbf} approach is capable of meeting safety requirements during training and testing while achieving near-optimal performance in testing. We provide results using two simulators: a flow-based simulator, which is used for proof-of-concept and benchmarking, and a packet-based simulator that implements real protocols and scheduling. Thanks to the flow-based simulator, we compared the performance against the optimal policy, solving a \acrfull{nlp} problem with the \acrshort{scip} solver.  
Furthermore, we showed that pre-trained models in the flow-based simulator, which is faster, can be transferred to the packet simulator, which is slower but more accurate, with some fine-tuning. Overall, the results suggest that near-optimal \acrfull{qos} performance in terms of end-to-end delay can be achieved while safety requirements related to link capacity constraints are guaranteed. In the packet-based simulator, we also show that our \acrshort{drl}-\acrshort{cbf} algorithms outperform non-RL baseline algorithms. When the models are fine-tuned over a few episodes, we achieved smoother \acrshort{qos} and safety in training, and similar performance in testing compared to the case where models have been trained from scratch.
\end{abstract}




\end{frontmatter}


\newpage
\section{Introduction}

Leveraging \acrfull{sdn} to control network overlay, many enterprises are adopting \acrfull{sdwan}~\citep{yang2019software}  to trade-off between cost-effectiveness and \acrfull{qos} satisfaction. 
This paradigm allows businesses to interconnect multiple sites (enterprise branches, headquarter, data centers) without the need to deploy their own physical infrastructure, making it cost-effective. A centralized controller maintains a set of policies deployed at access routers that send traffic
to their peers over several transport networks (e.g., private lines, broadband internet, 5G).
Typically, access routers are responsible for enforcing traffic engineering, for instance load-balancing,
and queuing policies to meet \acrfull{sla} requirements in terms of
end-to-end \acrshort{qos}, security, etc. At a slow pace, the controller maintains policies, while the access devices make real-time decisions for every flow. \acrshort{sdwan} provides a cost-effective alternative to private lines and drastically improves \acrshort{qos} compared with best-effort solutions such as \acrfull{vpn}. Beyond \acrshort{sdwan}, \acrshort{sdn} is also widely used to control traffic steering in carrier networks. Centralized load-balancing solutions have been proposed to control traffic splitting with \acrfull{ucmp}~\citep{medagliani2016global}.

\begin{figure}[ht]
   \centering
   \includegraphics[scale=0.6]{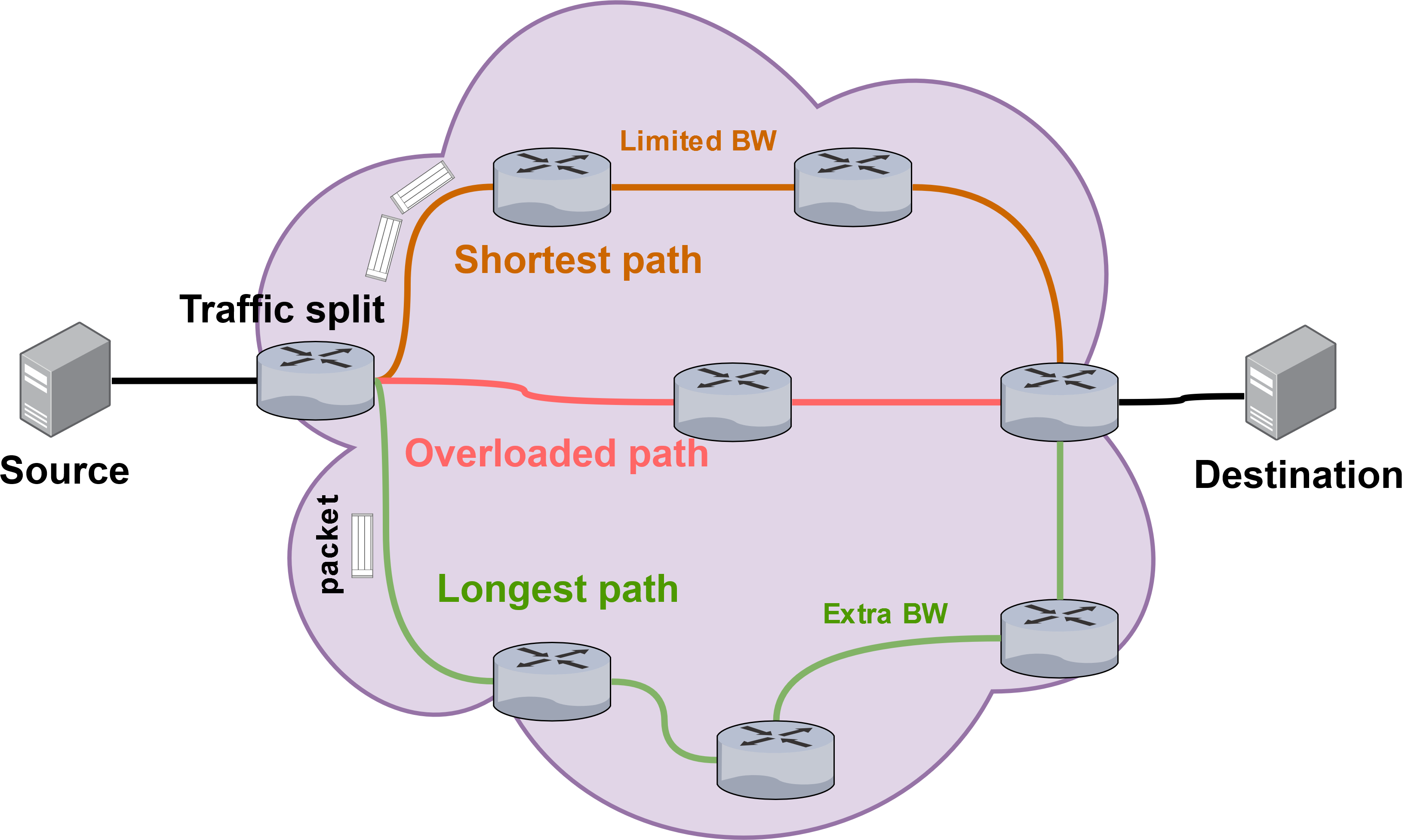}
   \caption{Traffic steering in a network including 3 paths.}
   \label{fig:traf_steer}
\end{figure}

\lam{Traffic steering or load-balancing plays an important role in guaranteeing network \acrshort{qos}, by properly splitting part of the traffic and sending it to a particular path. As illustrated in Figure \ref{fig:traf_steer}, traffic splitting is performed on the source access router, and packets are forwarded in only two paths in a network that has three available paths to avoid congestion. Given that the shortest path provides the best \acrshort{qos} yet is resource-limited, and the longest path offers extra bandwidth at the cost of lower \acrshort{qos}, an optimal traffic steering policy should appropriately split source traffic to better utilize the high-performance connection of the limited bandwidth path and the available resources of the extra path.} 

To optimize the QoS performance of \acrshort{sdn}-based load balancing, \acrfull{drl} algorithms have been shown to outperform existing approaches because of their ability to predict network performance and proactive management \citep{troiaDeepReinforcementLearning2021}. However, this method lacks a safety bound while performing trial-and-error to explore the environment or even sampling actions to make final decisions. This issue may cause instabilities and poor performance. Considering these issues, our work seeks to complement current \acrshort{drl}-based load-balancing solutions with an additional safety shield.
Based on the safe learning approach presented primarily in \citep{amesControlBarrierFunctions2019}, which is designed for critical robotic systems, we propose a safe load-balancing solution based on a \acrfull{cbf}. Our solution achieves 
near-optimal \acrfull{qos} performance in terms of average end-to-end delay, while safety concerns related to the violation of link capacity constraints are fully considered. The contributions of our work can be categorized into three main studies:
\begin{itemize}
    \item We first analyzed our solution in a simplified environment as a proof-of-concept:
    \begin{itemize}
        \item  We designed a dedicated \acrfull{cbf} based on local search to provide safety on top of gradient-based \acrfull{drl} algorithms (e.g., off/on policy learning).
        \item  We implemented the \acrshort{drl}-\acrshort{cbf} algorithms over GPU to drastically decrease computational time and make the solution usable in practice.
        \item We compared gradient-based \acrshort{drl} algorithms (e.g., off/on policy learning), with and without \acrshort{cbf} on top, and benchmarked them against the optimal policy, solving an \acrfull{nlp} problem with the \acrshort{scip}~\citep{bestuzhevaSCIPOptimizationSuite2021} solver.
    \end{itemize}
    \item[] Compared to our previous conference paper~\citep{dinhSafeLoadBalancing2024a} where we introduced the simplified environment (and only focused on it), we considered a general network scenario for the evaluation, instead of an SD-WAN overlay. 
    \item Then, we evaluated the \acrshort{drl}-\acrshort{cbf} algorithms in a packet-level simulator (\acrshort{ns3}~\citep{ns3}) for a more realistic  evaluation:
    \begin{itemize}
        \item  We designed an interface between \acrshort{ns3} and our learning algorithms, which are based on stable-baseline 3 \citep{stable-baselines3} and were already developed in~\citep{dinhSafeLoadBalancing2024a}. 
        \item We accelerated training using multiple CPU cores for the parallelization of \acrshort{ns3} simulations. 
        \item We compared the performance of \acrshort{drl} algorithms, with and without \acrshort{cbf}, to  existing non-RL baselines. The results show that our CBF function efficiently bounds the \acrshort{drl} algorithms in the safe region in both training and testing. In testing, our algorithms achieve better \acrshort{qos} performance than the baselines. 
    \end{itemize}
    \item Finally, we showed that pre-trained models in the simplified environment can be transferred to the \acrshort{ns3} environment. In this way, models can be trained faster and directly operate in a more complex environment close to their convergence, as they only require fine-tuning over a few episodes. The results obtained show that smoother \acrshort{qos} performance is achieved and safety is respected during training. At  convergence, they also achieve similar performance to models that have been trained from scratch.    
\end{itemize}

The remainder of this paper is organized as follows: Section~\ref{sec:lb_req} discusses the requirements for designing ideal load-balancing algorithms in \acrfull{sdn}.  Related work is discussed in Section~\ref{sec:related}. Section~\ref{sec:system} presents the system model and describes the load-balancing problem. Section~\ref{sec:CBF} describes the proposed solution. Section~\ref{sec:simulation} details our simulation campaign to validate our algorithms, and Section~\ref{sec:results} provides numerical results that demonstrate the performance of our algorithm. Subsequently, Section~\ref{sec:limitations} provides our perspectives for future works. Finally, Section~\ref{sec:conclusion} concludes the paper.

\section{Requirements for Load Balancing algorithms}
\label{sec:lb_req}
The use of the \acrshort{sdn} paradigm in \acrshort{sdwan} offers multiple benefits for the design of load-balancing algorithms. Thanks to its centralized view, the network controller can collect network statistics (e.g., path delay, packet loss, etc.) for global load-balancing. The following sections describe the intended properties when performing path assignments while optimizing \acrshort{qos}. 

\subsection{Desired properties}
\begin{itemize}
    \item \textit{\textbf{Proactive management}}: Proactivity is crucial for an \acrshort{sdn} controller to manage the underlying network. To enhance availability and reliability, the \acrshort{sdn} controller should anticipate future network issues (i.e., performance degradation, network faults, etc.) and attempt to prevent them from occurring. 

    \item \textbf{\textit{Flexibility}}: Network traffic is often bursty, and the network topology can be frequently changed. Any robust load-balancing algorithm should cope well with these changes while optimizing \acrshort{qos}.   
    \item \textbf{\textit{Network stability}}: In addition to \acrshort{qos} optimization, stable network operations are also of paramount importance. In load-balancing, congestion and \acrshort{qos} degradation are among the common causes of network instability \citep{pourgheblehComprehensiveSystematicReview2020}. Therefore, any algorithm designed should take these into account, given the dynamic of network changes.
\end{itemize}

\subsection{Why is DRL  used for Load Balancing?}
Prior to the era of \acrshort{ml} algorithms, analytical models have been used to estimate end-to-end \acrshort{qos} performance metrics. For instance, centralized and distributed path selection mechanisms have been proposed to maximize a utility function~\citep{magnouche2021distributed} or \acrshort{sla} satisfaction using a performance model~\citep{quang2022intent}. To optimize latency and other \acrshort{qos} parameters for a specific set of flows, closed-form performance models, using network calculus or queuing models, can be embedded into routing optimization algorithms. The authors in \citep{benAmeur2006} considered the Kleinrock function~\citep{kleinrock2007communication} to minimize latency. However, it can be difficult to derive and integrate them into routing optimization algorithms. Indeed, unknown scheduling parameters, behaviors within the \acrshort{wan} and transport-layer mechanisms are, in practice, too difficult to be captured by tractable performance models. The challenge of integrating accurate analytical models makes \acrshort{qos} routing and load-balancing a great opportunity for model-free solutions.

Instead of having a predefined performance model, \acrfull{rl} agents can interact with the environment and evaluate the outcomes of their actions through a reward function. 
\acrfull{drl}~\citep{xu2018experience} combines \acrfull{dl} and \acrfull{rl} principles (e.g., parameterize policies with neural networks). It has been first applied for routing to optimize network utility
under the umbrella of \textit{experience-driven networking}~\citep{xu2018experience}. Since then, several single agent and multi-agent \acrshort{drl} solutions have been proposed to tune queues and load-balancing policies to satisfy \acrshort{qos} requirements or minimize congestion~~\citep{mai2020multi, kim2021deep, houidi2022constrained, fawaz2023graph, troiaDeepReinforcementLearning2021, RILNET}.

\subsection{Why is DRL  not sufficient in practical systems? }
Although \acrshort{drl} has demonstrated tremendous potential to improve networking performance, trial-and-error, which is the basis of any RL algorithm, causes serious problems for practical networking systems. In particular, in the beginning phase of the stochastic learning process, unsafe actions are often sampled to explore the environment. 
Nevertheless, at the same time, these unsafe actions cannot be deployed as they degrade network quality in production systems.

Unsafe requirements in any network can be explicitly described, depending on the user's demands. Explicit requirements can be formulated using closed form functions, whereas implicit requirements are rarely expressed \citep{dalalsafeexploration2018}.
For instance, congestion due to overloaded links in the network can be seen as an explicit unsafe behavior because it is directly related to \acrfull{mlu}, which can be modeled. Then, the network safety issue can be translated into the assurance of the \acrshort{mlu} under a certain value. Guaranteed network delay can also be seen as another explicit constraint and any learning policy should maintain the end-to-end delay below a predefined threshold.

Taking safe load-balancing into account, any poor policy can accidentally put more traffic load on a path that is already congested because it might seek to globally minimize the average \acrshort{qos} performance. This harmful action causes more severe congestion and accidentally blocks any ongoing traffic on that path. Consequently, it degrades the overall network performance.  Therefore, considering safety during both the training phase, particularly for more practical on-policies, and testing is key for the wide adoption of \acrshort{rl} algorithms in network systems.

\section{Related Work}
\label{sec:related}

For years, great effort has been devoted to the study of improving traffic engineering in \acrshort{sdn}, using oblivious routing and/or load-balancing \citep{anticObliviousRoutingScheme2008}. \textit{Oblivious} traffic engineering with load-balancing does not require any prior network information and is based on the shortest route \citep{tsunodaLoadBalancedShortestPathBasedRouting2011}, edge-disjoint \citep{jainB4ExperienceGloballydeployed2013} or optimal path \citep{liuTrafficEngineeringForward2014}. However, they are usually prone to excessive congestion in bottleneck links and poor performance in practical systems \citep{kumarKulfiRobustTraffic2016a}. Semi-oblivious approaches, such as COPE \citep{wangCOPETrafficEngineering2006} and SMORE \citep{kumarSemiobliviousTrafficEngineering2018}, which take advantage of traffic history, have been proposed to optimize the worst-case performance, which is related to the minimization of \acrshort{mlu} in the network links. On the basis of the prediction across the set of traffic demand matrices, COPE achieves high efficiency and close-to-optimal performance at the cost of extremely high computing resources. SMORE \citep{kumarSemiobliviousTrafficEngineering2018} employed a centralized controller and Racke's algorithm \citep{rackeMinimizingCongestionGeneral2002b}
to minimize congestion by computing multiple paths and adapting the sending rate on those paths. As a result, SMORE is low-stretch and naturally balances the load. However, this solution can miserably fail to achieve good performance even with a slight change in the traffic matrices \citep{valadarskyLearningRoute2017}. 

Instead of using traffic-prediction-based solutions which are not always accurate \citep{perryDOTERethinkingPredictive2023a}, model-free solution based on \acrfull{drl} agent is a good-fit to make sequential-decision-making process and to tackle with non-linear behaviors of networking systems.
In particular, the effective use of \acrshort{drl} algorithms for \acrshort{sdn} controllers has been proven to improve overall network performance~\citep{mai2020multi, kim2021deep, houidi2022constrained, fawaz2023graph, troiaDeepReinforcementLearning2021, RILNET}. For instance, Troia et al. \citep{troiaDeepReinforcementLearning2021} have shown a target \acrshort{qos} can be achieved through the proper design of the reward function. Similar results have been achieved using multiple agents~\citep{houidi2022constrained}.
However, as mentioned before, most of the literature only focuses on 1) off-policies and 2) network performance without paying attention to safety.

Indeed, RL-based load-balancing systems may violate capacity constraints in both training and testing phases. When global minimization of end-to-end delay is achieved, unsafe actions that create congestion and violate capacity can be taken to eliminate portions of traffic and improve reward. To deal with such abnormal behaviors, the LearnQueue~\citep{bouacidaPracticalDynamicBuffer2019} reward has been introduced to minimize end-to-end delay while penalizing traffic rejections.
However, it requires proper weighting of the two objectives, which can be tedious in practice because it is highly dependent on the environment.
To address these limitations and avoid manual parameter tuning, the first work to systematically optimize \acrshort{qos} under safety constraints for load-balancing has been presented by Kamri et al.~\citep{kamriConstrainedPolicyOptimization2021b}.
This work applies the \acrfull{rcpo} algorithm~\citep{tesslerRewardConstrainedPolicy2018}, where the reward integrates traffic rejection as a constraint using  Lagrangian relaxation. During training, the algorithm determines the optimal Lagrangian multiplier. Following this work, Zhang et al.~\citep{zhangPathPlanningModel2022} investigated a path planning problem based on constrained policy iteration to improve the performance of multiple path selection. Huang et al.~\citep{huangProactiveLoadBalancing2022} also discussed a safe load-balancing strategy for ultra-dense networks. In this study, they proposed a proactive load-balancing algorithm on top of \acrfull{cpo}~\citep{achiamConstrainedPolicyOptimization2017a}, which has been proven to guarantee optimal policy under (safety) constraints. 

Although several papers have presented \acrshort{drl} algorithms for load-balancing ~\citep{troiaDeepReinforcementLearning2021,  bouacidaPracticalDynamicBuffer2019} with some constraints~\citep{kamriConstrainedPolicyOptimization2021b,huangProactiveLoadBalancing2022}, they all provide \textit{soft} guarantees during exploration and mostly ensure safety during exploitation. Furthermore, most of the literature only focuses on off-policies and their performance once training has converged, without paying attention to safety during both learning and testing. In addition, as  network environments often drift, frequent retraining of the full model to adapt to changes can be cumbersome. To mitigate these issues, we propose to employ a \acrfull{cbf}~\citep{amesControlBarrierFunctions2019}  on top of current \acrshort{drl} algorithms for (safe) load-balancing optimization with \textit{hard} guarantees. 

Our previous work \citep{dinhLoadBalancingSafe2024,dinhSafeLoadBalancing2024a} was the first to guarantee that training and testing performance are safely bounded in terms of capacity constraints while providing near-optimal \acrshort{qos}. However, the algorithms are evaluated on the basis of a simple simulator, which does not take full networking protocols into account.  Thus, the main goal of this work is to examine our algorithms in a realistic environment, which involves networking protocols and packet scheduling. Furthermore, we study the tangible relationship between the two simulators by fine-tuning the pre-trained models from the simplified environment in the \acrshort{ns3} environment.

\begin{table}[ht!]
\centering
\caption{Table of notations.}
\label{tab:notation}
\resizebox{260pt}{!}{%
\begin{tabular}{|l|l|} 
\hline
\multicolumn{1}{|c|}{\textbf{Math Notation}} & \multicolumn{1}{c|}{\textbf{Meaning}}  \\
\hline
    $K$     &  Set of OD Tunnels         \\ 
\hline
    \gls{pk}     &  Set of available paths in tunnel $k \in K$         \\
\hline
    $p$     &  Current path $p \in \gls{pk}$ of tunnel $k$         \\
\hline
    \gls{Dtk}     &  Traffic demand at tunnel $k$         \\
\hline
    \gls{de}     &  Delay on link $e$ of  path $p$       \\
\hline
    \gls{dpe}     &  Propagation delay on link $e$         \\
\hline
    \gls{ce}     &  Capacity of link $e$        \\
\hline
    \gls{le}     &  Traffic load of link $e$        \\
\hline
    \gls{dpk}     &  Delay on path $p$ of tunnel $k$         \\
\hline
    \gls{dk}     &  Tunnel delay of tunnel $k$         \\
\hline
    \gls{xpk}     & Split ratio applied on path $p$ of tunnel $k$         \\
\hline
    $\mu$     &  \acrlong{mlu}         \\
\hline
    $\sigma$     &  Reward parameter between delay and \acrshort{mlu}         \\
\hline
    $\theta$     &  Parameters of neural network for actor         \\
\hline
    $\mathcal{L}_b^A(\theta)$      &  Actor loss's objective  to be maximized         \\
\hline
    $\phi$     &  Parameters of neural network for critic         \\
\hline
    $\mathcal{L}_b^C(\phi)$     &  Critic loss's objective  to be minimized         \\
\hline
    $\gamma$     &  Discounted factor        \\
\hline
    $\delta_s$     &  Local search radius         \\
\hline
    \gls{ncbf}     &  Number of \acrshort{cbf} solutions         \\
\hline
    \gls{mcbf}     &  Number of maximum local search iterations          \\
\hline
    $\tau$     & Delay network update in \acrshort{ddpg}  algorithm         \\
\hline
    $\varepsilon$     &  Clipping objective ratio in \acrshort{ppo}  algorithm         \\
\hline
    $\delta_{KL}$     &  Target KL divergence in \acrshort{ppo}  algorithm          \\
\hline
    $B$     & Training  Batch          \\
    \hline
    \gls{DeltaT} [second]    & Time period per policy update           \\
\hline
\end{tabular}
}
\end{table}

\section{System architecture and problem formulation}
\label{sec:system}

Before we introduce the system architecture, we present a typical \acrshort{sdn} use case which involves intelligent traffic steering in \acrshort{sdwan} networks. Note that without loss of generality and for completeness, we consider a more general underlay scenario in the Sec.~\ref{sec:results} for the performance evaluation. Results for the \acrshort{sdwan} scenario in the flow-based simulator can be found in~\citep{dinhSafeLoadBalancing2024a}. The mathematical notations, that are used in this paper, are described in Table~\ref{tab:notation}.

\subsection{SD-WAN Use Case }

\begin{figure}[ht]
    \centering
    \includegraphics[scale=0.25]{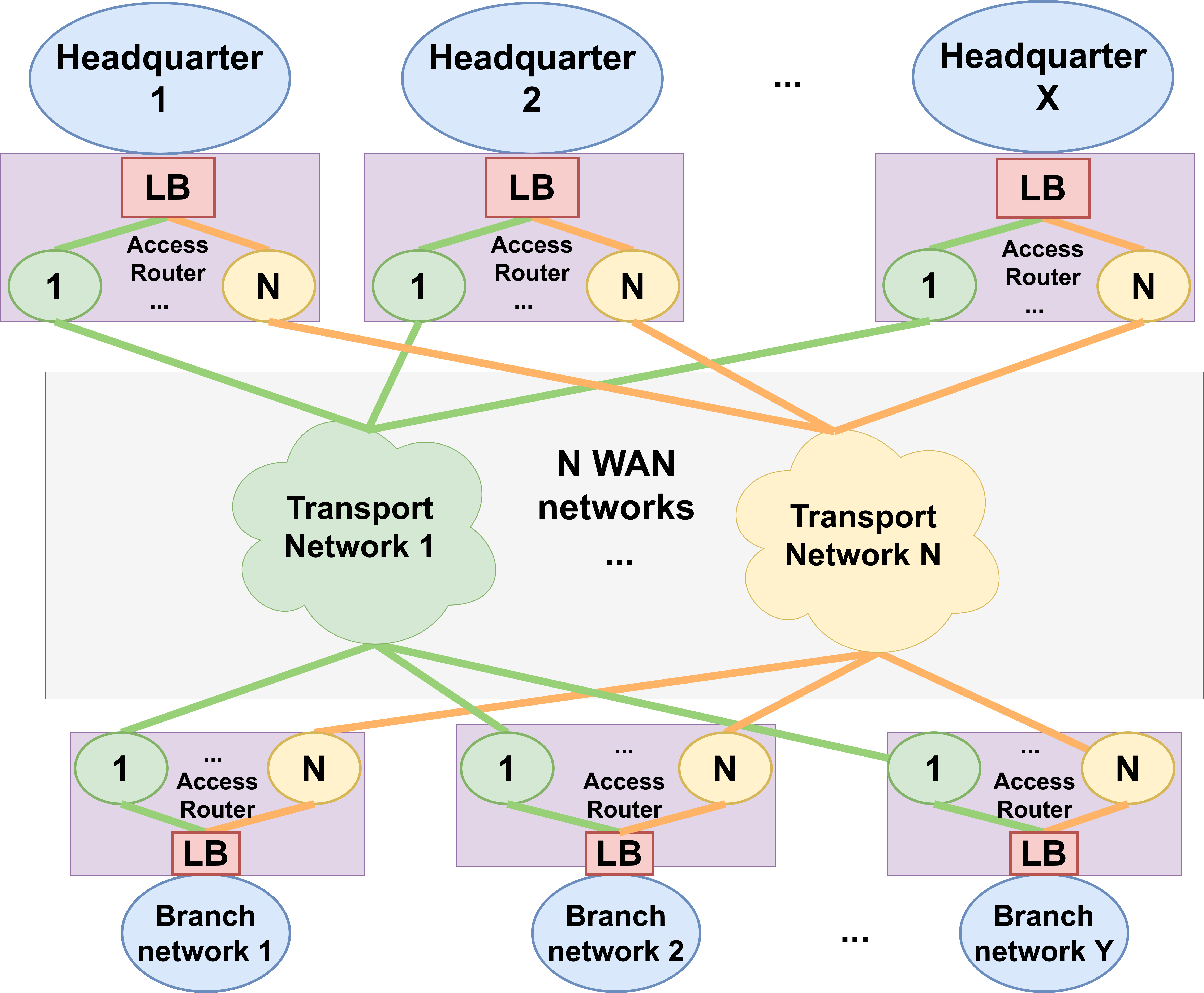}
    \caption{SD-WAN network with $X$ headquarters and $Y$ branches via $N$ transport networks.}
    \label{fig:scenario}
\end{figure}

Figure~\ref{fig:scenario} presents a typical \acrshort{sdwan} architecture in which the $X$ headquarters and $Y$ branches of an enterprise are interconnected via $N$ transport networks (e.g., Internet connection, \acrfull{mpls} private line, 5G .etc). We consider in this work the bidirectional communication between headquarter $x$ ($\forall x = \{1,...,X\}$) and branch $y$ ($\forall y = \{1,...,Y\}$). Traffic, which is generated stochastically in the application layer of each source, is then sent towards the destination, forming a flow (tunnel). Then, $2|X||Y|$ OD (Origin-Destination) flows, also called \textit{tunnels}, are considered between all headquarter and branch pairs in both directions via the \acrshort{wan} networks. Each tunnel has $N$ paths corresponding to the $N$ transport networks. A \acrfull{lb} agent at each \acrfull{ar} splits the traffic into $N$ paths according to the policy received by the centralized controller. 

\subsection{Intelligent Load Balancing Workflow}

\begin{figure}[ht]
    \centering
    \includegraphics[scale=0.50]{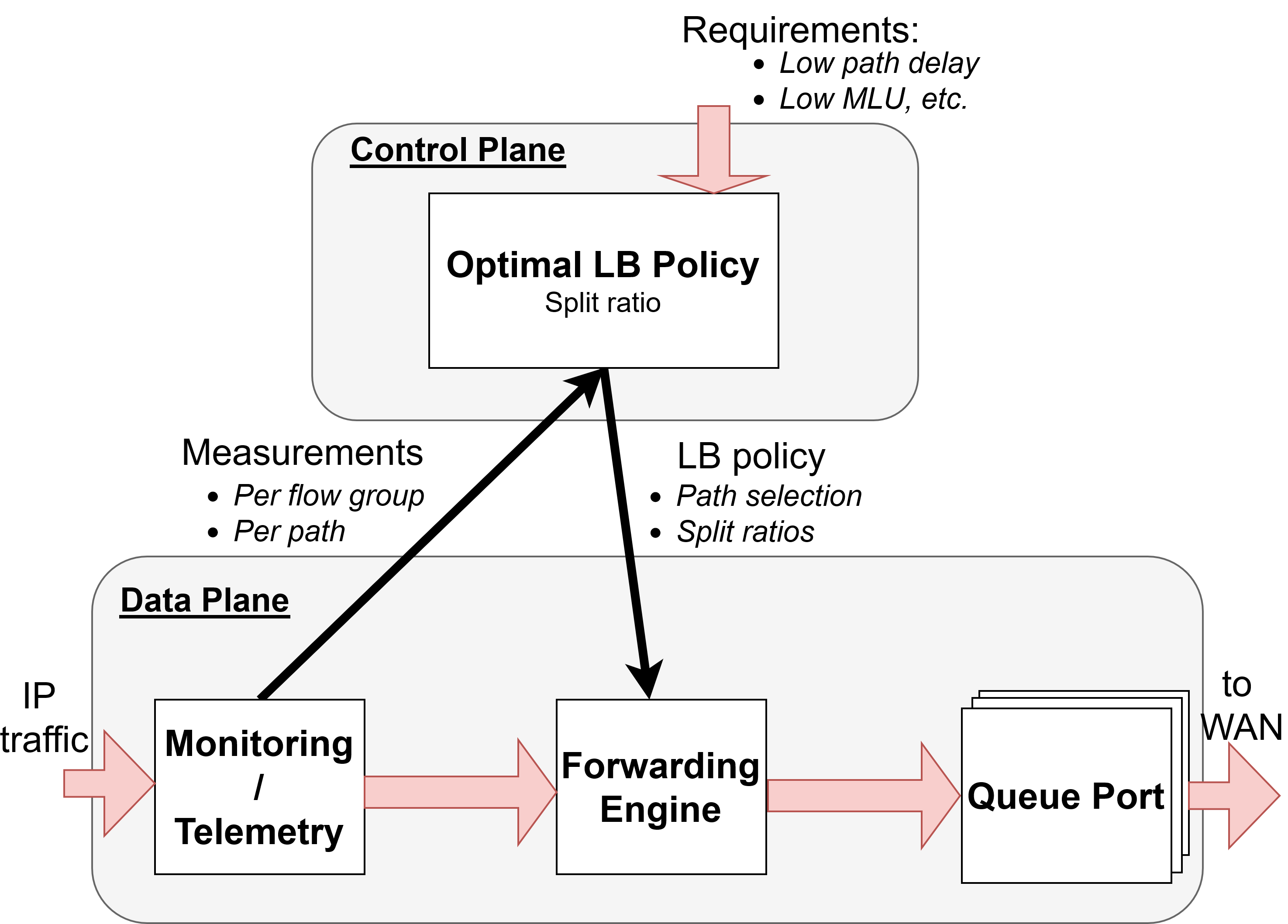}
    \caption{Load balancing workflow.}
    \label{fig:lb_arch}
\end{figure}
Figure \ref{fig:lb_arch} illustrates the load-balancing workflow, which comprises a control plane and a data plane in the general \acrshort{sdn} architecture.  In a fast-paced loop, \acrshort{ar}s perform real-time decisions on path selection following the evolution of \acrshort{ip} traffic and network states. In the data-plane, there are three main blocks:  \textit{(1)} Monitoring/Telemetry is responsible for measuring the statistics of each flow and path (e.g., delay, packet loss, jitter, etc.) and frequently reports them to the routing agent in the control plane. \textit{(2)} The forwarding engine  receives the \acrshort{lb} policy from the controller and recent network monitoring to perform path selection. In the \acrshort{ar} devices, each \acrshort{wan} path is connected to an unique networking port, and the \acrshort{ip} packets are then forwarded to the queue port \textit{(3)} corresponding to a transport network path. The \acrshort{lb} policy is calculated at the controller, where the control plane is hosted. The load-balancing agent collects the measurements from the \acrshort{ar}s and  decides a policy to split traffic over available paths to satisfy the predefined requirements. 
\lam{In practice, ARs are generally equipped with built-in traffic monitoring capabilities (e.g. iFiT \cite{ huaweiIFIT}, which allow them to monitor service traffic, QoS of tunnels (i.e. delay, packet loss, and jitter) and link utilization using NetFlow,  NetStream \cite{ciscoOverviewNetflow, huaweiOverviewNetstream}. On the other hand, the controller periodically collects statistics from AR routers.  To mitigate load on the control channel between the controller and routers, those statistics are collected at the granularity of minutes (e.g. $5$ minutes in Huawei controller \cite{huaweiNCEIP}). There are two time scales in our system: \textit{(i)} a slow control loop in the controller to collect data measurement and to update current policy (i.e., target split ratios), and \textit{(ii)} a fast control loop inside ARs to forward incoming packets / flows according to the current policy. Our work only focuses on the slow control loop.}

\subsection{Problem Formulation}
This overlay network can be modeled as a graph in which edge links correspond to \acrshort{wan} ports at \acrshort{ar} routers. Inside each transport network $n$ ($\forall n =1,2,...N$), the core links represent overlay links. As depicted in Figure \ref{fig:scenario}, ports $n$ ($\forall n =1,2,...N$) are used to connect a AR router to the corresponding transport network $TN_n$, and this link models the egress/ingress capacity provisioned at each transport network. In practice, \acrshort{wan} ports can receive traffic from multiple sites, potentially of higher capacity, and they might be congested under high load conditions. Let formally consider a graph $G=(V,E)$ where $V$ is the set of nodes and $E$ is the set of edges.
Each tunnel $k$ in a set of tunnels $K$ can use a set of \textit{candidate paths} denoted as  $\gls{pk}$ to load balance traffic. Each edge $e$ carries an instantaneous load $\gls{le}$ and has a capacity $\gls{ce}$. Let denote  $\gls{Dtk}$ 
 as the traffic demand 
of tunnel \textit{k}  at time \textit{t}. Each \acrshort{lb} agent applies at time $t$ a split ratio $\gls{xpk}$ for each tunnel $k$ over each path $p \in \gls{pk}$ 
($\gls{xpk} \in [0,1]$ and $\sum_{p \in \gls{pk}} \gls{xpk}=1\ \forall k \in K$).

The delay on each path $p$ for a tunnel $k$ is denoted $\gls{dpk}$ and  tunnel delay, denoted $\gls{dk}$ is computed as follows:
\begin{equation}
    \gls{dk} = \underset{p \in \gls{pk} }{\max} \quad \gls{dpk}
\end{equation}

In practice, $\gls{dpk}$ is continuously measured by access routers. 

The main objective is to derive an optimal load-balancing policy so that the \acrshort{sdn} delivers the best \acrshort{qos}. In the rest of the paper, we consider as the primary target the  minimization of the average tunnel delay under the constraint that link 
capacity constraints are not violated (safety constraint). Indeed, this safety measure prevents that 1) congestion is induced by misconfigured split ratios and that 2) the average tunnel delay is not artificially minimized by rejecting traffic (i.e., by intentionally creating some congestion). 
To prevent high link delays, or very heterogeneous ones even if the average delay is low, a common practice is to enforce a \acrfull{mlu} $\mu \in [0,1]$ over all links.
\acrshort{lb} agents at the headquarters and branches are configured by a network controller. 

In this case, load-balancing policies can be derived by solving the following optimization problem: 

\begin{align*}
    \underset{\gls{xpk}}{\min} ~~~~~~~&   \frac{\sum_{k\in K}\gls{dk} }{|K|}\tag{$\mathcal{P}$} \label{eq:P0}\\
     \text{s.t.}~~~~~& \sum_{k \in K}  \sum _{i=0, p \in \gls{pk}} ^{|p|} \gls{Dtk}.x^k_i \leq \mu . c_e ~~~~~ \forall e \in E \tag{$\mathcal{C}_0$} \label{eq:C0} \\
     & \sum _{i=0} ^{|p|} x^k_i =1  ~~~~~~~~~~~~~~~~~~~~~~~~\forall x^k_i \in [0,1]\tag{$\mathcal{C}_1$} \label{eq:C1} \\
     & \gls{dpk} \geq \sum_{e \in p, p\in \gls{pk}} \gls{de} (\gls{xpk})  ~~~~~~~~~~~~~~~~~~~~~~~~\tag{$\mathcal{C}_2$} \label{eq:C2} \\
\end{align*}


where problem 
\eqref{eq:P0} minimizes the average tunnel delay. Constraints \eqref{eq:C0} guarantee that the traffic over each edge $e$ in the network is maintained under the \acrshort{mlu} $\mu$.
Constraints \eqref{eq:C1} ensure that split ratios sum to $1$. 
Constraint \eqref{eq:C2} explains the relationship between path delay $\gls{dpk}$ (and thus tunnel $\gls{dk}$) and edge delay $\gls{de}$, which is represented as a function of split ratio $\gls{xpk}$ on each path $p$ of tunnel $k$. This constraint highlights the importance of using model-free algorithms over model-based assumption. In particular, the calculation of edge delay $\gls{de}$ can be mathematically derived using queuing theory (e.g., M/M/1 queuing model). Nevertheless, this model-based queuing for edge delay calculation might not be accurate and it failed to capture behaviors of more complex networking environment, for example priority queuing is not considered in the M/M/1 model. 
It opens the door for \acrshort{ml} algorithms, as they take into account any non-linear network behavior during the learning process (e.g., priority queuing at egress port of devices, transport protocol, etc. ), to derive an optimal policy. 

In our work, the M/M/1 model plays an instrumental role in the simplified environment for approximating edge delay and obtaining an optimal \acrshort{nlp} solution. The results obtained from this optimal model are then benchmarked to our learning-based models, which do not know the queuing model for decision-making. On the other hand, \acrshort{nlp} is not considered to be an optimal solution in the packet-based environment, i.e. NS3, where the underlying scheduling is more difficult to approximate.

\section{Learning-based and safe load-balancing}
\label{sec:CBF}

In this section, we propose a constrained policy optimization for load-balancing based on  \acrfull{drl} and a \acrfull{cbf} \citep{amesControlBarrierFunctions2019}. 

\subsection{Learning-based optimization}

Our optimization problem can be formulated as a \acrfull{mdp} which is defined by the tuple  $\left<\mathcal{S},\mathcal{A},\mathcal{R},\mathcal{T},\gamma  \right>$ where $\mathcal{S}$ represents the set of states, $\mathcal{A}$ is the set of available actions, $\mathcal{R}: \mathcal{S} \times \mathcal{A} \times \mathcal{S} \rightarrow \mathbb{R}$ is the reward function which gives the reward for the transition from one state to another given an action, $\mathcal{T}: \mathcal{S} \times \mathcal{A} \times \mathcal{S} \rightarrow [0,1]$ is the transition matrix, which gives the probabilities of transitioning from one state to another given an action 
and $\gamma \in [0,1]$ is the discount factor. A policy $\pi: S \rightarrow  A$ refers to the probability of taking an action $a \in \mathcal{A}$ under state $s \in \mathcal{S}$. The agent iteratively interacts with the (networking) environment to learn an optimal policy $\pi^*$ that chooses actions with the best payoff. To solve the \acrshort{mdp} associated with problem \eqref{eq:P0}, a centralized controller is employed in the network.  The controller can periodically collect at every time $t$ information from the different \acrshort{ar} routers: the traffic demand $\gls{Dtk}$, the delay $\gls{dk}$ of each tunnel $k\in K$ and the maximum link utilization $\mu$. In this context, we consider the observation space or state $s_t$ as the set of traffic demands $\gls{Dtk}$ for all tunnels $k \in K$. The action space is determined as the set of split ratios $\gls{xpk}$ for all paths $p \in \gls{pk}$ of each tunnel $k \in K$. \lam{In addition to traffic demands $D_t^k$, other metrics can be monitored and considered as the observation space for DRL algorithms, such as the current queue occupancy at each network node, the maximum utilization of links on each path, packet loss, etc. However, in this work, we decided to only stick to traffic demands $D_t^k$ for two reasons: \textit{(i)} keep the design as simple as possible, and \textit{(ii)} focus on very common statistics that are widely available in existing AR devices. Other features can be useful to better describe traffic or network conditions to potentially improve QoS, but their integration comes at the cost of an extra complexity as a larger/deeper model size is needed to learn more complex features, and it might deteriorate convergence times. }


By using \acrshort{rl} algorithms, we learn the optimal stochastic control policy $\pi(a|s)$ that maximizes the performance measure (i.e., $J(\pi)$) which is expressed as follows:
\begin{equation}
    J(\pi) = \mathbb{E}_{\tau \sim \pi} \left[ \sum_t ^\infty \gamma^t r_t(s_t,a_t) \right] 
\end{equation}
where $\tau \sim \pi$ denotes a trajectory during which actions are sampled according to the policy $\pi(a|s)$. 

Given that policy gradient  \acrshort{rl} methods have shown good performance in continuous control problems \citep{suttonReinforcementLearningIntroduction2018}, in the rest of this paper, we consider off-policy learning (e.g. \acrfull{ddpg}) and on-policy learning (e.g., \acrfull{ppo}) algorithms, both of which are based on the actor-critic architecture \citep{silverDeterministicPolicyGradient2014}. In the former technique, an exploratory policy (e.g., Gaussian, Ornstein-Urlenbeck, etc.) is generally combined with a deterministic policy to facilitate learning from a broader set of state-action-reward  combinations (i.e., training samples). These training data, which are stored in a replay buffer \citep{andrychowiczHindsightExperienceReplay2018}, can be re-sampled in a batch to update the target (deterministic) policy. Despite efficiently re-using the training data, policy updates might experience instability in this way. 
On the other hand, on-policy methods update the policy in the next iteration with respect to the data collected from the current iteration. It is achieved by solving the following optimization problem: 

\begin{align}
    & \pi_{i+1} = \underset{\pi}{argmax}  \sum_s \rho_{\pi_i} (s) \sum_a \pi(a|s) A^{\pi_i}(s,a) \\
    & \text{s.t.} ~~~~ KL\left[\pi_{i+1}(.|s_t),\pi_{i}(.|s_t) \right]  \leq \delta_{KL}
\end{align}
where $A^{\pi_i}(s,a)$ is the advantage of performing action $a$ (sampled by old policy $\pi_i$) at state $s$, $\rho_{\pi_i} (s)$ represents the frequency of visit of state $s$ under policy $\pi_i$  and $KL\left[\pi_{i+1}(.|s_t),\pi_{i}(.|s_t) \right]$ refers to Kullback-Leibler divergence between the new and old policies \citep{Kullback51klDivergence}, and this value is bounded by a target $\delta_{KL}$. Both on-policy and off-policy gradient algorithms do not natively consider safety, therefore our goal is to complement these model-free learning algorithms with a \acrfull{cbf}, which guarantees safety during exploration and exploitation. 

\begin{figure}[ht!]
\centering
\includegraphics[scale=0.25]{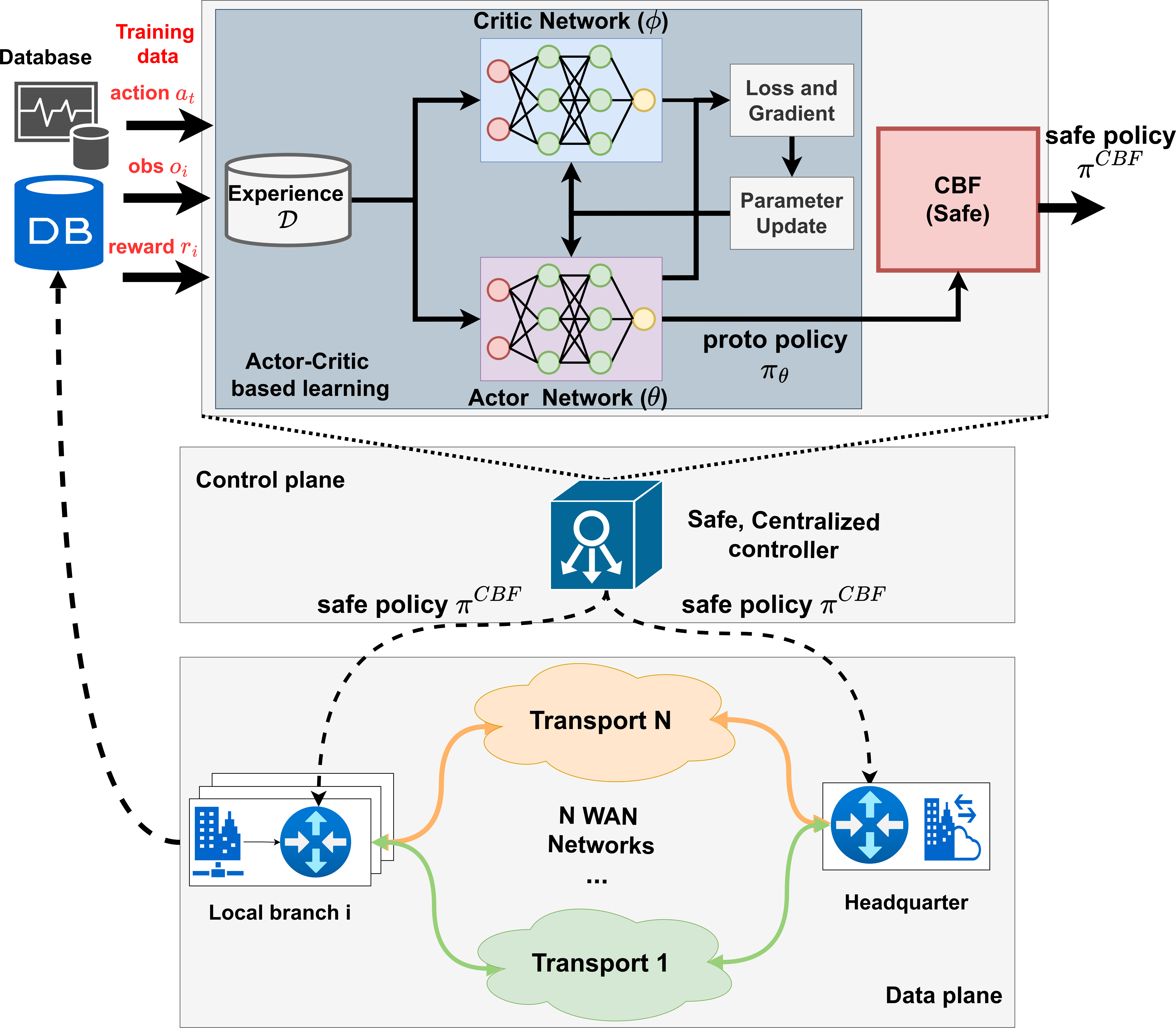}
\caption{Safety-based actor-critic learning architecture.} 
\label{fig:central_dm}
\end{figure}

Figure \ref{fig:central_dm} introduces the control plane architecture where each \acrshort{lb} agent is centrally configured by the controller. The controller can periodically query the network state (i.e., tunnel delays, OD traffic, \acrshort{mlu}) at each \acrshort{lb}, and its control system is composed of two main blocks that focus on policy optimization and safety, respectively. The former block, which is based on an actor-critic  architecture\citep{silverDeterministicPolicyGradient2014}, learns the (unconstrained) optimal policy.  
Consequently, the value function and the policy are approximated using two different neural networks, parameterized by $\phi$ and $\theta$, respectively. Each learning parameter is responsible for either maximizing the actor's objective function (i.e., $\mathcal{J}_b^A$) or minimizing the critic's loss function (i.e., $\mathcal{L}_b^C$). The specific expressions for $\mathcal{J}_b^A$ and $\mathcal{L}_b^C$ heavily depend on the on/off policy reinforcement learning algorithms being used. 

\begin{algorithm}[tp]
\small
\footnotesize 
\caption{\acrshort{drl}-based optimization algorithm}\label{alg:rlAlgo}

Initialize  parameterized \acrshort{drl} policy $\pi_{\theta_0}$ \\
Initialize value function parameter $\phi_0$ \\
Define episode length $L_{eps}$, batch size $B$, total episodes $N_{eps}$\\
Initialize experience array $\gls{exp}=\{\oslash\}$\\
\If{episode $k < N_{eps}$}{
\For {time step $t=1,...,L_{eps}$}{
Observe state $s_t$ \\
Sample action $a_t \sim \pi_{\theta_k}(.|s_t)$ \\
Perform local search algorithm \\
\begin{equation*}
{\colorbox{lightgray}{$\gls{acbf}_t = Local\_Search(s_t,a_t)$}}
\end{equation*}

Deploy action $\gls{acbf}_t$ and obtain reward $r_t$, next state $s_{t+1}$\\
Store experience tuple $<s_t,\gls{acbf}_t,s_{t+1},r_t>$ in $\gls{exp}$\\
}

\For {batch sampling $B$ experiences in \gls{exp}}{
Perform gradient \textbf{ascent} in the actor network\\
\begin{equation*}
    \theta_{k+1} \leftarrow \frac{1}{|B|}\underset{\theta_k}{\arg \max} \sum _{b=0}^{|B|} \mathcal{J}_b^A(\theta_k)
\end{equation*}
Perform gradient \textbf{descent} on critic network\\
\begin{equation*}
    \phi_{k+1} \leftarrow \frac{1}{|B|}\underset{\phi_k}{\arg \min} \sum _{b=0}^{|B|} \mathcal{L}_b^C(\phi_k)
\end{equation*}
}
}
\textbf{return} parameterized policy $\pi_\theta$, parameterized value function $\phi$ 
\end{algorithm}

In particular, actor objective (i.e., $\mathcal{J}_b^A$) and critic loss (i.e., $\mathcal{L}_b^C$) of \acrshort{ddpg} algorithm are respectively expressed as follows \citep{silverDeterministicPolicyGradient2014}: 
\begin{align}
\begin{split}
    &\mathcal{J}_b^A=Q_{\phi_k}(s_b,\pi_{\theta_k}(.|s_b)) \\
    &\mathcal{L}_b^C=Q_{\phi_k}(s_b,a_b)-(r_b+\gamma Q_{\phi_{k,target}}(s'_b,\pi_{\theta_k}(.|s'_b)))
\end{split}
\end{align}
where $Q_{\phi_k}(.)$ is the parameterized (i.e., $\phi$) action-value Q-function at episode $k$ and $\pi_{\theta_k}$ is the parameterized (i.e., $\theta$) policy.
On the other hand, the actor objective and critic loss formulation of the clipped \acrshort{ppo} algorithm \citep{schulmanProximalPolicyOptimization2017} are shown as belows:
\begin{align}
    \begin{split}
        &\mathcal{J}_b^A=\frac{1}{L_{eps}}\sum_{t=0}^{L_{eps}}\min \{\frac{\pi_{\theta_{k-1}}(a_{b,t}|s_{b,t})}{\pi_{\theta_k}(a_{b,t}|s_{b,t})}A^{\pi_{\theta_k}}(s_{b,t},a_{b,t}), \\
        & clip\left(\varepsilon,A^{\pi_{\theta_k}}(s_{b,t},a_{b,t})\right)\} \\
        & \mathcal{L}_b^C=\frac{1}{L_{eps}}\sum_{t=0}^{L_{eps}}(V_{\phi_k}(s_{b,t}) - r_{b,t})
    \end{split}
\end{align}
where $L_{eps}$ refers to the episode length, $clip\left(\varepsilon,A^{\pi_{\theta_k}}(s_{b,t},a_{b,t})\right)$ is the clipped value of the following terms $\frac{\pi_{\theta_{k-1}}(a_{b,t}|s_{b,t})}{\pi_{\theta_k}(a_{b,t}|s_{b,t})}A^{\pi_{\theta_k}}(s_{b,t},a_{b,t})$,
 with clipping parameter $\varepsilon$, and $V_{\phi_k}$ is the parameterized (i.e., $\phi$) value function at episode $k$. The details of the optimization procedure are shown in Algorithm \ref{alg:rlAlgo}. 


\subsection{Safe policy exploration and exploitation}
\lam{In order to enforce safe exploration during learning, and safe policy execution during testing, stochastic behaviors of the neural network in the \acrshort{drl} approach must be avoided. Therefore, we implemented an additional safety block on top of the \acrshort{drl} algorithms to transform the stochastic policy into a safe policy. Furthermore, the reward design should take into account the guarantee of safety requirement in addition to the delay optimization objective, to solve the problem \eqref{eq:P0}. In the following, we describe our reward design for the \acrshort{drl} algorithms, and the safety shield based on a \acrshort{cbf} function~\citep{amesControlBarrierFunctions2019}.}
 
\subsubsection{Reward design for safety-aware, delay optimization}
With respect to the design of the reward function, we note that the objective function of problem \eqref{eq:P0} is sufficient to minimize the average tunnel delay in the network. 
However, as mentioned in Section~\ref{sec:system}, when only  minimizing the average delay, the system may induce large delays for a small number of tunnels due to high link utilization or congestion, i.e. packet drops, to further reduce the average delay. Therefore, in order to make the \acrshort{drl} agent also aware of link utilization, we design our reward function to be the weighted sum of the average delay and the \acrshort{mlu} (i.e., $\mu$) as given by Equation \ref{eq:rew}. It should be noted that this reward function contains two terms: The former term reduces tunnel delay by directing the traffic load to the path with a lower propagation delay. It, however, can create congestion on the low propagation delay links. On the other hand, the second term attempts to minimize \acrshort{mlu} by steering toward the path with higher capacity. It, nevertheless, does not consider the delay of the link. Consequently, traffic can be forwarded on high capacity links with a high delay. Putting them together, \acrshort{drl} agents seek to achieve \textit{soft safety} during learning.

\begin{equation} \label{eq:rew}
    r_t(s_t,a_t)=-\sigma \frac{\sum_{k \in K} d_{k,t}}{\left|K \right|} - (1-\sigma )\mu
\end{equation}
where $\sigma \in [0,1]$ emphasizes the importance of the average tunnel delay over a low \acrshort{mlu}. While this reward cannot guarantee itself a hard safety, it guides the \acrshort{rl} agent in learning a policy which is both \acrshort{qos} optimal and safe after convergence. To ensure anytime and hard safety, we explain in the following how we incorporate a \acrshort{cbf} function on top of \acrshort{drl} algorithms to constrain the exploration and exploitation so that the \acrshort{mlu} does not exceed 100 $\%$ (i.e., $\mu\leq 100 \%$).

\subsubsection{Safe policy based on CBF function.}
\begin{figure}[b]
    \centering
    \includegraphics[scale=0.2]{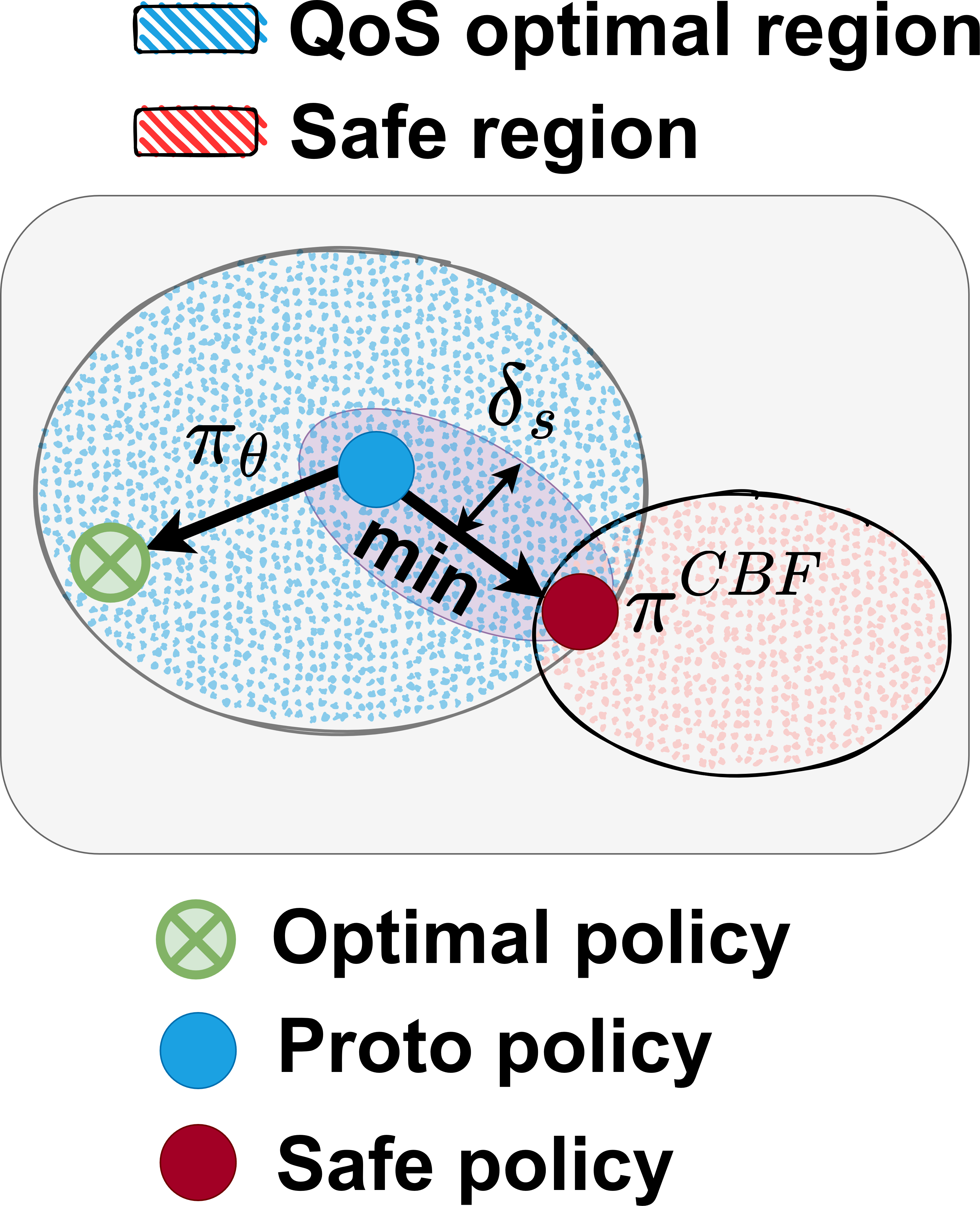}
    \caption{From proto-policy to safe policy with CBF.}
    \label{fig:cbf}
\end{figure}

The \acrshort{cbf} function serves as a projector to convert \textit{proto-policy} $\pi_\theta$, which is the parameterized actor network from which unsafe actions might cause congestion, into a \textit{safe policy} $\pi^{CBF}$. Its main objective is to guarantee that the \acrshort{mlu} $\mu$ remains below $1$ (i.e., $\mu(\gls{acbf}) \leq 1$, $\forall \gls{acbf} = \pi^{CBF}(.|s)$). 
As illustrated by Figure \ref{fig:cbf}, this  projection maintains the safe policy $\pi^{CBF}$ as close as possible to the proto policy and maintains the load-balancing system under the safety conditions. Safe policy projection is performed at the centralized controller in both learning and testing. Following the safe policy $\pi^{CBF}$, each \acrshort{cbf} action is determined according to the following optimization problem: 

\begin{align} \label{eq:safe}
    \begin{split}
       &\gls{acbf}= \underset{\gls{acbf}_t}{argmin}\left\|\gls{acbf}_t - \gls{arl}\right\|_1 \\
    \textit{s.t.} ~~~~~~~~~&  \gls{acbf} \in \mathcal{A} \\
    & \mu(\gls{acbf}) \leq 1 
    \end{split}
\end{align} 

The \acrshort{cbf} function is applied on top of the \acrshort{drl} algorithms to guarantee that any sampled action, which is stochastically derived from the parameterized policy, will be bounded in a safe region during learning. By targeting "bad" proto-action, which produces a high \acrshort{mlu} and overloaded links, the output \acrshort{cbf} action  results in a better or at least equal \acrshort{mlu} compared to the initial proto-action. For the projection of proto-actions into safe actions, we propose to use the local search algorithm detailed in Algorithm \ref{alg:Wapprox_exh} as the \acrshort{cbf} function .

In principle, given a proto-action, the \acrshort{cbf} function first attempts to generate \gls{ncbf} neighbor actions in the vicinity of the proto-action (i.e., within a radius $\delta_s$). The main goal is then to select among \gls{ncbf} target action that will be executed in the environment and will not violate the safety constraints (i.e., MLU). There is a trade-off to be made in the selection of $\delta_s$, that is, too small $\delta_s$ might lead to neighbor actions that are very close to the proto-action. In spite of helping the training become more stable when the origin policy only changes a little, it is more likely that no safe action can be found in that region when the proto-action is very bad. On the other hand, a high value of $\delta_s$ increases the probability that a safe action exists, but the safe action can be very different from the initial proto-action and the training might be unstable when the CBF function changes the proto-policy too rapidly.

\begin{algorithm}[tp]
\small
\footnotesize 
\caption{\acrshort{cbf} based on Local Search algorithm}\label{alg:Wapprox_exh}
Returned action from RL agent $a_{\theta}^{RL}$ \\
Local search radius $\delta_s$ \\
\acrshort{cbf} solutions \gls{ncbf} \\
Local search policy $\pi^{CBF}$\\
Local search max iteration \gls{mcbf}\\
Feasible solution $feas\_sol=\{\}$ \\
\eIf{$\mu(a_{\theta}^{RL}) \leq 100 \%$}{$\gls{acbf}=a_{\theta}^{RL}$}
{
\For{$m$ in \gls{mcbf}}{
\For {$n$ in \gls{ncbf}}{
Randomly sample $\epsilon _n \sim Uniform(0,\delta_s)$\\
Generation of stochastic actions $a_n^{CBF} \sim \pi^{CBF}$\\
Perform action update: $a_n^{CBF} = a_{\theta}^{RL} \pm \epsilon _n $ \\
\If{$\mu(a_n^{CBF})  \leq 100 \%$}{Append $a_n^{CBF}$ to $feas\_sol$}
}
}
\eIf{$feas\_sol \neq \oslash $}{$\gls{acbf}=\underset{a_i^{CBF} \in feas\_sol}{argmin}\left\|a_i^{CBF} - a_{\theta}^{RL} \right\|_1$}
{$\gls{acbf}= \underset{n}{argmin}\quad \mu\{a_n^{cbf}\}$}
}
\textbf{End}
\end{algorithm}

Once $\delta_s$ is properly determined, we identify any tunnel $k$ that has a path load utilization above a threshold of  $\eta$ (i.e., $ \exists p \in \gls{pk} ~ | ~\mu_p \geq \eta$, which is considered as unsafe). Then, we sampled a random value $\epsilon_k \sim Uniform(0,\delta_s)$ and it will be used to adjust the initial split-ratio of that tunnel, which is primarily decided by the proto-policy. As a consequence, an amount of traffic $\epsilon_k.T^k_{t,p}$ will be withdrawn from the path $p$ of tunnel $k$ with the highest path utilization, and added to the remaining paths. If the current evaluating action results in reducing \acrshort{mlu} of the network, it will be appended to a feasible set. Finally, a resulting CBF action (i.e., \acrshort{mlu} is below $\eta$) is selected from the feasible set in such a way that its L1 distance (Manhattan distance) to the original proto action is the smallest, as displayed in Equation \ref{eq:safe}. The returned action is heuristically safe and helps drive a stochastic policy toward a safe policy. As local search policies are heuristic, they may not be able to correct a proto-action for which a  safe action exists. In this case, the \acrshort{cbf} action, which returns the lowest \acrshort{mlu}, will be selected.

Our local search algorithm is important in both training and exploitation phases. In the training phase, it helps to ensure that the network learns safe policies by eliminating the possibility of taking actions that could cause safety issues (i.e., violation of link capacity constraint). In the exploitation phase, it helps to ensure that the network continues to operate safely even under unseen network conditions.

\section{Simulation campaign}
\label{sec:simulation}
In this section, we describe our simulation campaign to validate our safe load-balancing algorithm. Hereafter, we used the Abilene topology, which is an \acrshort{ip} network and it serves as a real-world topology \citep{knightInternetTopologyZoo2011}, to study load-balancing performance in terms of average delay (i.e., SLA minimization) and \acrshort{mlu} (i.e., safety requirement). 

In terms of simulation tools, we have developed two simulators. \textit{(1)} A flow-based simulator, which is based on Python’s Gym toolkit \citep{openaigym} and does not implement networking protocols. It will be used as a proof of concept and for benchmarking our algorithms in achieving high performance (i.e., delay) and guaranteeing safe load-balancing requirements (i.e., \acrshort{mlu}). 
\textit{(2)} A packet-based simulator which is based on \acrfull{ns3} \citep{ns3}, and takes into account network protocols and packet scheduling. Our goal is to prove the feasibility of our methods in a network setting that is \textit{close to} the real world. 
It is also worth noting that the link capacity and the real traffic demand in this packet-based simulator are down-scaled when compared to the real traffic rate and capacity links, so that the simulation runtime does not explode. 






\begin{figure}[ht]
    \centering
    \includegraphics[scale=0.42]{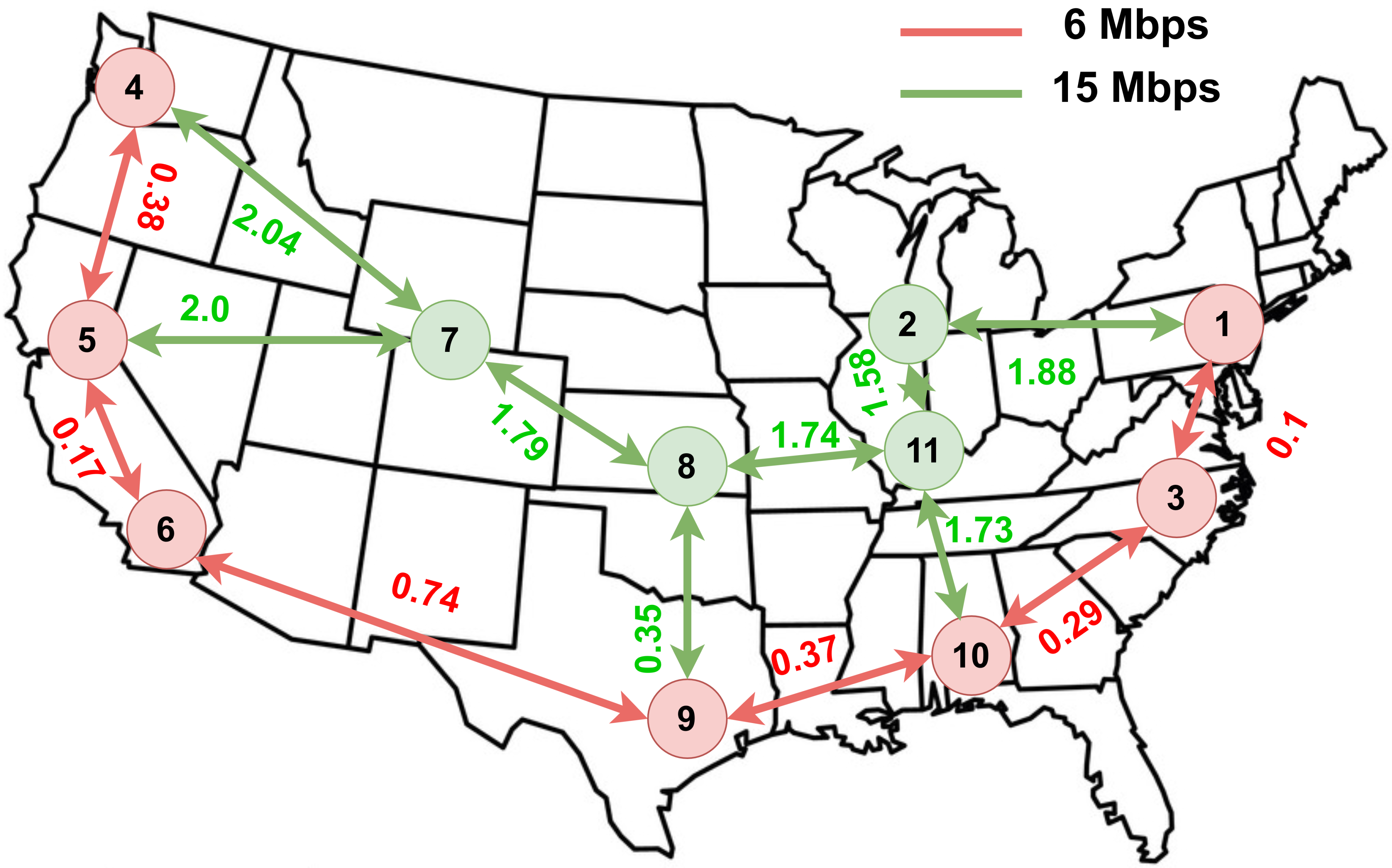}
    \caption{Abilene topology.}
    \label{fig:abilene}
\end{figure}

Figure~\ref{fig:abilene} presents the Abilene topology, which includes 11 Nodes. In this topology, we assume that three pairs of nodes (i.e. (1,5), (4,10) and (4,9)) establish six OD tunnels, in which each tunnel comprises two different paths with two different bandwidths, and path propagation delay, as illustrated in Table \ref{tab:path_abilene}. On the basis of the geographical positions (i.e., latitude and longitude) of two nodes, which are available in Zoo-topology \citep{knightInternetTopologyZoo2011}, we are able to compute the propagation delay of each link as shown in Figure~\ref{fig:abilene}. In addition, we further increase the propagation delay of each link that belongs to the path with higher capacity (green links). Each packet, that is sent to this path, results in a higher delay overall.  
On the other hand,  the remaining path contains links with a lower propagation delay (red links). At the same time, their capacity is limited when compared to green links. 

\lam{In this study, we selected the Abilene topology with three sender-receiver pairs to produce results in a reasonable amount of time with NS3. This scenario can also fit well with the use case of a small SME's business, in which the core operation, and management data are centralized in a single headquarter and connected to a few remotely business branches. }
\begin{table}[ht]
\centering
\caption{Paths in Abilene topology}
\label{tab:path_abilene}
\resizebox{250 pt}{!}{
\begin{tblr}{
  cells = {c},
  row{even} = {fg=JapaneseLaurel},
  row{3} = {fg=red},
  row{5} = {fg=red},
  row{7} = {fg=red},
  row{9} = {fg=red},
  row{11} = {fg=red},
  row{13} = {fg=red},
  hlines,
  vlines,
}
\textbf{Path idx} & \textbf{Tunnel OD} & \textbf{Nodes} & \textbf{Path delay} \\
0                 & 1-5                & 1,2,11,8,7,5   & 9,0 ms              \\
1                 & 1-5                & 1,3,10,9,6,5   & 1.67 ms             \\
2                 & 5-1                & 5,7,8,11,2,1   & 9.0 ms              \\
3                 & 5-1                & 5,6,9,10,3,1   & 1.67 ms             \\
4                 & 4-9                & 4,7,8,9        & 4.19 ms             \\
5                 & 4-9                & 4,5,6,9        & 1.28 ms             \\
6                 & 9-4                & 9,8,7,4        & 4.19 ms             \\
7                 & 9-4                & 9,6,5,4        & 1.28 ms             \\
8                 & 4-10               & 4,7,8,11,10    & 7.31 ms             \\
9                 & 4-10               & 4,5,6,9,10     & 1.65 ms             \\
10                & 10-4               & 10,11,8,7,4    & 7.31 ms             \\
11                & 10-4               & 10,9,6,5,4     & 1.65 ms             
\end{tblr}
}
\end{table}

\subsection{Algorithm implementation}
\begin{figure}[ht]
    \centering
    \includegraphics[scale=0.5]{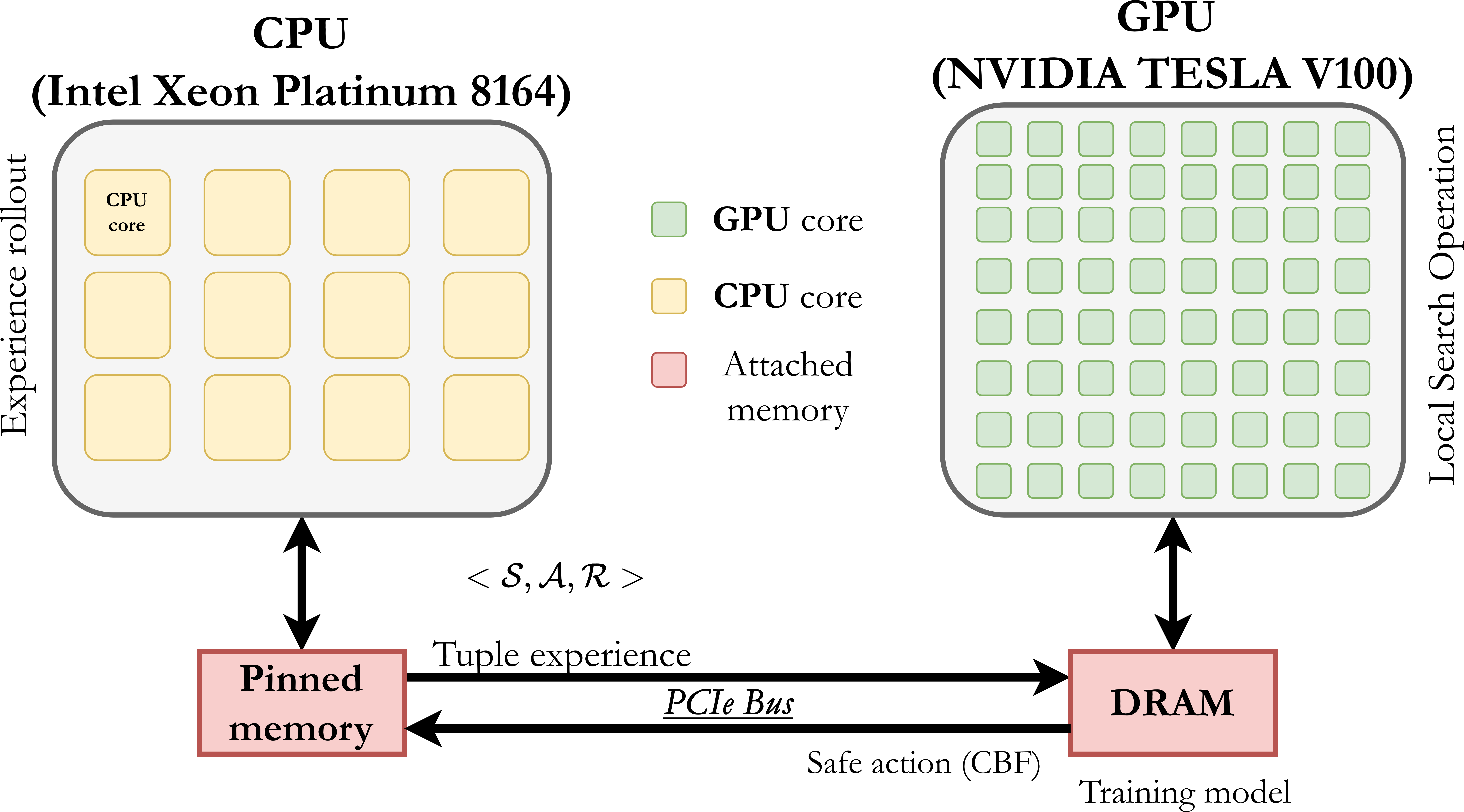}
    \caption{System architecture with 1) network environment on CPU and 2) safe RL algorithms on GPU.}
    \label{fig:flow}
\end{figure}

In this work, we used a server, which composed of a \acrshort{cpu} Intel\textsuperscript{\textregistered} Xeon\textsuperscript{\textregistered} Platinum 8164 (104 logical cores and 1024 GB of RAM memory) and 2x\acrshort{gpu} NVIDIA\textsuperscript{\textregistered}  Tesla V100 (each has 5120 \acrshort{cuda} cores and 16 GB of DRAM), to implement both flow- and packet-based simulations. \lam{We use multiple cores to parallelize network simulations in order to speed up data collection. Note that in practice we do not have to run these simulations because data can be collected from devices, as discussed in Section 4.2}. Because our local search and model training algorithms are computationally expensive, due to the large number of loops needed to find a safe solution at each gradient step, and huge arithmetic computations in model update, we implemented the solution over \acrshort{gpu} to fully benefit from all available CUDA cores, as shown in Figure \ref{fig:flow}. 
Therefore, model training and local search are fully performed at \acrshort{gpu} side. \lam{However, we use a simple neural network with only $3$ layers and $1024$ neurons in each layer (see Section 7.1.1). Therefore, model training and local search only take up to $20\%$ GPU computing capability. Note that the proposed algorithm can be deployed in less powerful and affordable systems.} Once a safe policy is found, it is transferred to the \acrshort{cpu}, where our \acrshort{sdn} environment is located, to perform a roll-out step. The resulting tuple experiences are then moved back to \acrshort{gpu} for further model training. 

\lam{Regarding flow-based simulations, the GPUs play a significant role in accelerating the computation of gradient-based updates and iterative local search algorithm during model training, while the CPU tasks are light.
On the other hand, the packet-based simulator requires heavy CPU computation to simulate a huge number of discrete events and is slower. By isolating each CPU core for an independent environment,  the use of our server is favorable to accelerate the packet-based simulation and reduce the simulation runtime.}


\subsection{Flow-based environment}
Our flow-based simulation is built  on the basis of Python's Gym toolkit \citep{openaigym}, which is a well-known \acrfull{api} for interfacing between our (safe) policy learning algorithms and our topologies (Abilene). Relying on Stable-Baseline3 \citep{stable-baselines3}, a well-known library of \acrshort{rl} algorithms, we have exploited two different types of policy-based algorithms: off-policy with \acrshort{ddpg}~\citep{lillicrapContinuousControlDeep2019} and on-policy with \acrshort{ppo}~\citep{schulmanProximalPolicyOptimization2017}. Given that both of them are not natively designed to guarantee  safety during training and testing phase, we additionally implemented our local search algorithm (Algorithm \ref{alg:Wapprox_exh}) on top of them to guarantee safety requirement on link capacity. 

\begin{figure}[ht]
    \centering
    \includegraphics[scale=0.45]{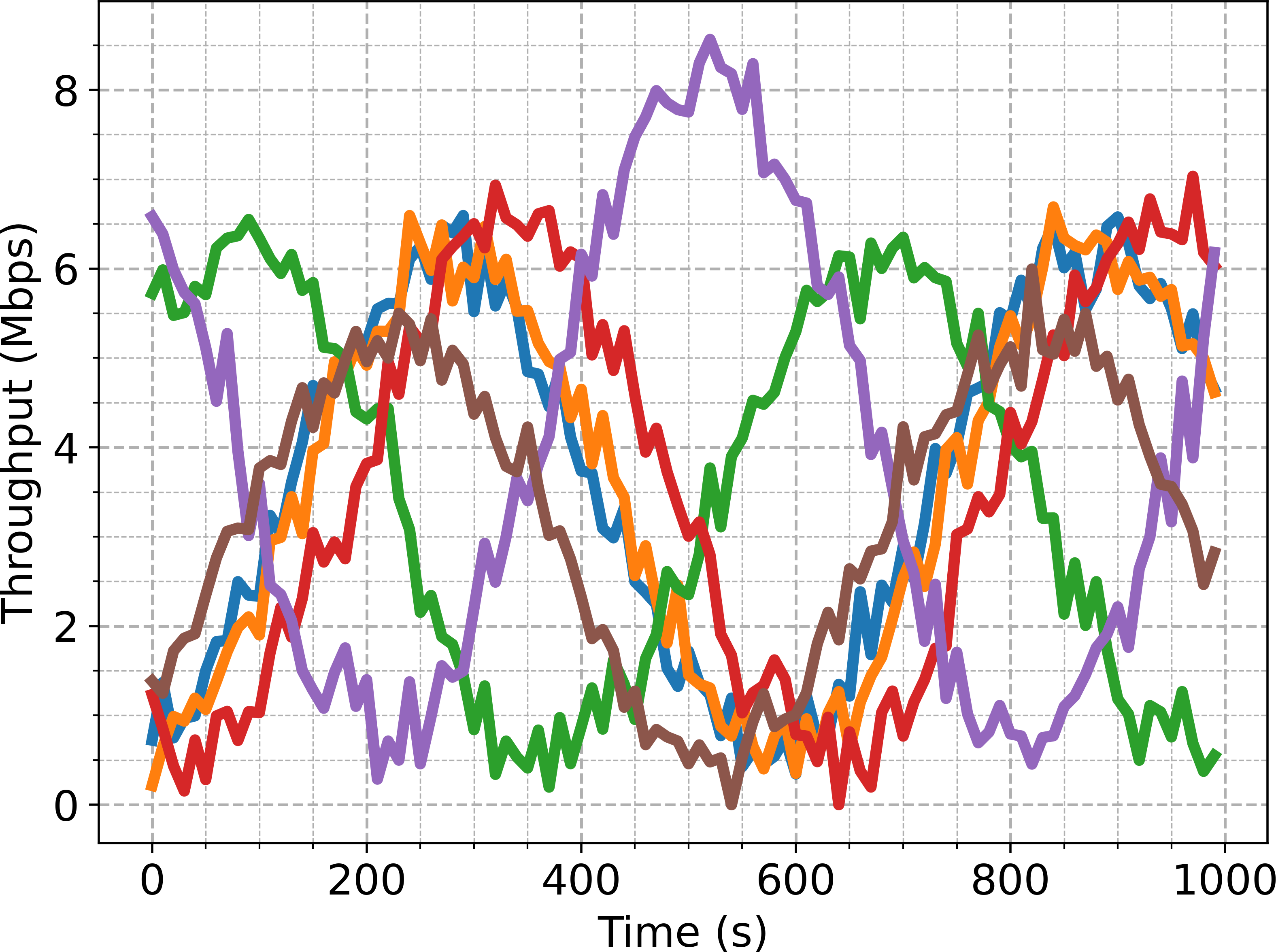}
    \caption{Tunnels' traffic samples in the flow-based simulation. 
    }
    \label{fig:traffic}
\end{figure}

In order to demonstrate the stochastic behavior at each \textit{OD tunnel}, 
we generate noisy sinusoidal traffic to model diurnal variations, as shown by Figure \ref{fig:traffic}. 
The phase of each OD flow traffic is also shifted to make traffic generation at each source somewhat different and likely to cause congestion at  bottleneck links without appropriate load-balancing. 

In order to simulate network behavior in the flow-based simulator, we adopt a simple M/M/1 queuing model to compute the delay on each \textit{link} $e$ and consider a propagation delay $\gls{dpe}$ in Table \ref{tab:path_abilene}. The link delay $\gls{de}$ is then derived as follows \citep{bertsekasDataNetworks1992}:
\begin{equation}
    \gls{de} = \gls{dpe} + \frac{1}{\gls{ce}-\gls{le}}
\end{equation}

The delay on each path $p$ of a tunnel $k$ (i.e. $\gls{dpk}$) is then calculated as the sum of the delay on the edge/link that is involved in that path as follows:
\begin{equation}
    \gls{dpk} = \sum_{e \in p, p\in \gls{pk}} \gls{de}
\end{equation}

Furthermore, in order to emulate the congestion control behavior of \acrshort{tcp} in the considered network, we adopted a min-max fairness rate allocation algorithm using a standard water-filling algorithm \citep{bertsekasDataNetworks1992} in case of congestion.

\subsection{Packet level environment}
Unlike the flow-based simulator as described above, which does not take into account any networking protocol, we developed a packet-based simulator using \acrshort{ns3} \citep{Riley2010} to empower sophisticated protocols into simulation for a better evaluation of the efficiency of algorithms. In particular, to demonstrate \acrfull{sdn} capabilities, we used the OFSwitch13 module \citep{chavesOFSwitch13EnhancingNs32016}, which supports the OpenFlow protocol version 1.3 \citep{openflow1_3} between switch devices and a centralized controller. 

\begin{figure}[ht]
    \centering
    \includegraphics[scale=0.28]{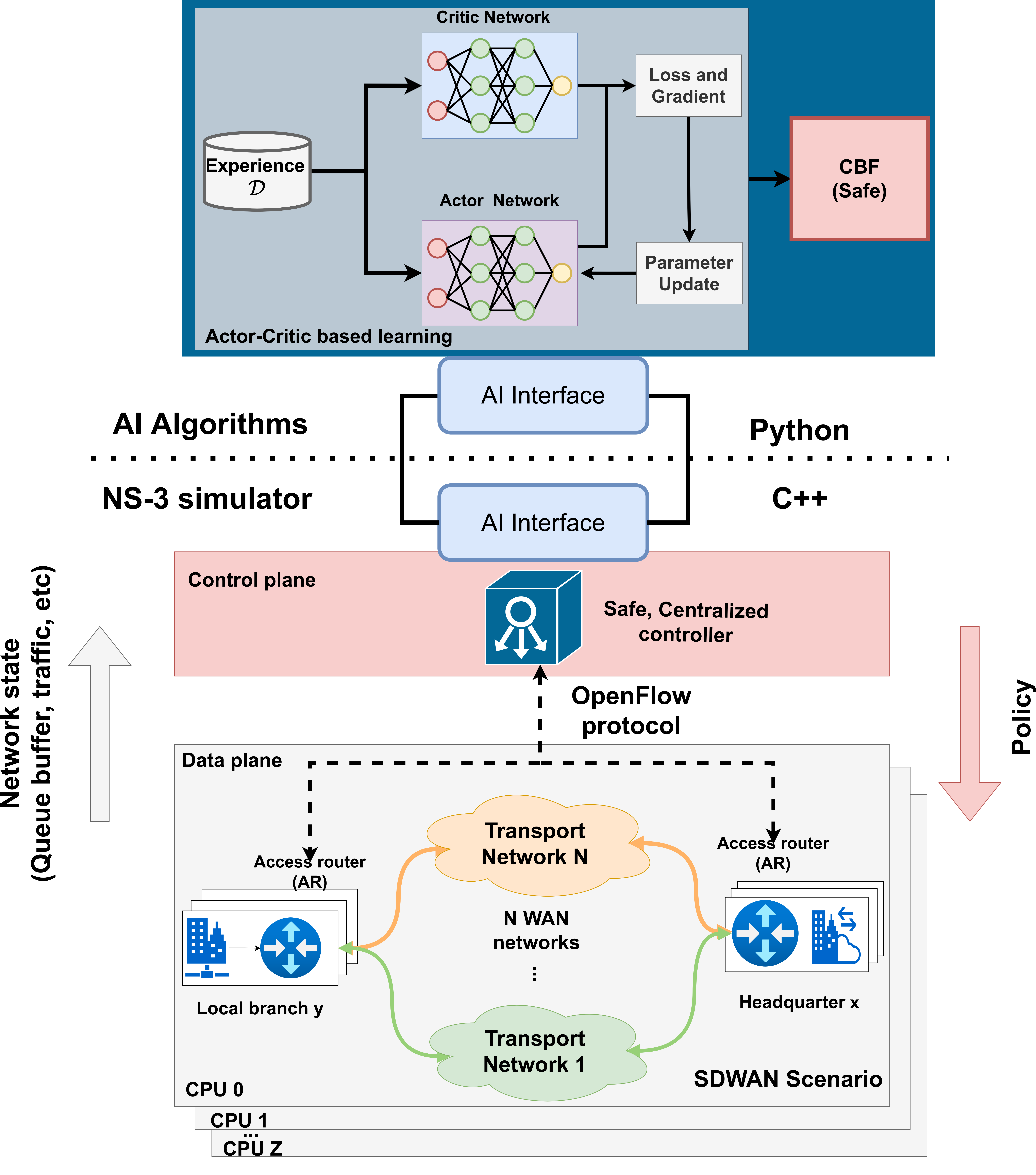}
    \caption{ML-based \acrshort{ns3} simulation.}
    \label{fig:ains3_simul}
\end{figure}

As \acrshort{ns3} does not yet natively support \acrshort{ml} features, we developed an \acrshort{ml}-based \acrshort{ns3} simulator which architecture is shown in Figure \ref{fig:ains3_simul}. In our \acrshort{ml}-based \acrshort{ns3} simulator, we implemented two major features compared with the native \acrshort{ns3} simulator:  \textit{(1)} parallelizing multiple \acrshort{ns3} instances on multiple \acrshort{cpu}s and \textit{(2)} an interface between \acrshort{ns3} instances and our \acrshort{drl} algorithms. The first feature is motivated by the fact that the simulation time required to collect each network measurement evolves exponentially as a function of the size of the network topology. As a result, policy training in a single \acrshort{ns3} instance is significantly slow and time-consuming, making the integration of \acrshort{drl} algorithms on the \acrshort{ns3} inappropriate. To address this issue, we allocated multiple \acrshort{ns3} instances on multiple  \acrshort{cpu}s to accelerate the training process.
In practice, to reduce frequent exchange between the controller and all routers, network measurements (e.g., delay, \acrshort{mlu}, packet loss, etc.) of each \acrshort{ns3} instance are reported to the centralized controller, after each \gls{DeltaT} seconds via OpenFlow protocol. Then, the controller and our safe \acrshort{drl}-\acrshort{cbf} algorithms must be interfaced to gather the simulation data and feed them to the models, as graphically depicted in Figure \ref{fig:lb_arch}. This interface is built on top of \citep{haons3ai2020} and it represents our second implementation contribution.

\begin{figure}[ht]
    \centering
    \includegraphics[scale=0.5]{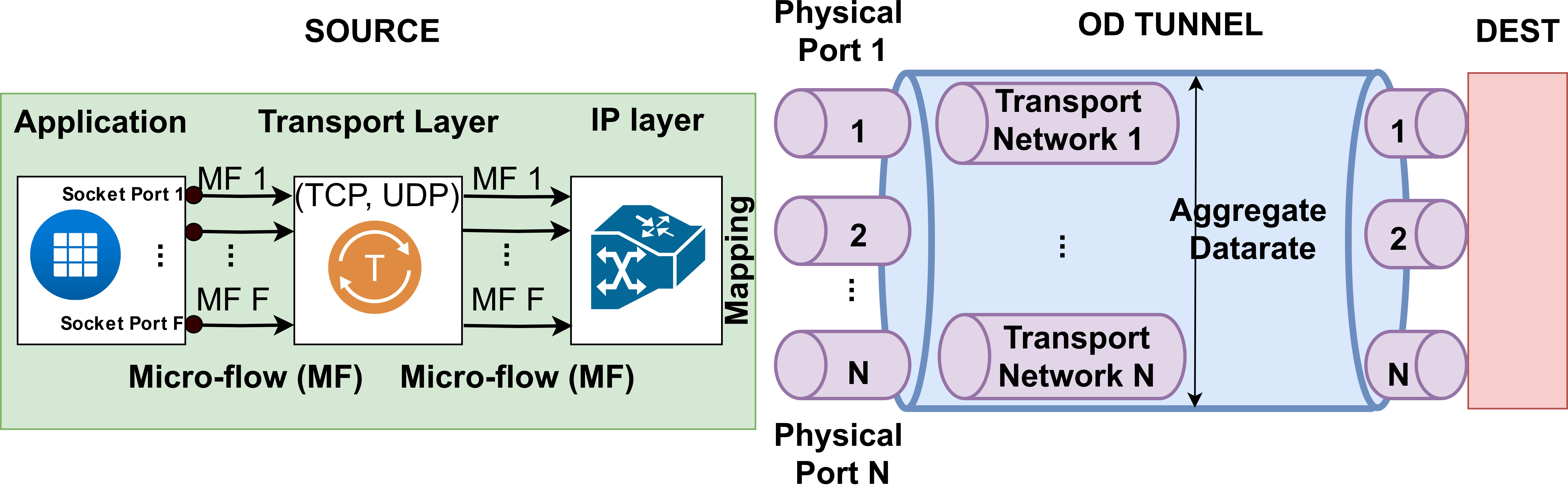}
    \caption{Traffic generation at source. 
    }
    \label{fig:traffic_sdwan}
\end{figure}

In order to produce a stochastic traffic behavior for each OD tunnel, we generated multiple microflows from a source to a destination as illustrated in Figure \ref{fig:traffic_sdwan}. Each microflow is defined as a 5-tuple flow, including the source/destination IP address, source/destination \acrshort{tcp}/\acrshort{udp} port and IP protocol. 
By assigning each destination IP port to each microflow, we are able to create multiple microflows between source and  destination. 

In the application layer, we create multiple microflows with an inter-arrival time randomly sampled according to an exponential distribution. The duration \gls{ttl} of each microflow is limited and the data rate of each microflow (e.g., \gls{dmf}) is fixed  to avoid  excessive data of a microflow injected to the transport network. By appropriately choosing the inter-arrival time between two microflows and controlling \gls{ttl}, \gls{dmf} of each microflow, we are able to control the aggregate data rate for each OD tunnel. At the source, each microflow will pass through the transport layer before reaching the IP layer. Here, the mapping between each microflow and a physical port is performed to forward the microflow traffic to a transport network. An optimal load-balancing policy solving \eqref{eq:P0} chooses the right path of each microflow such that the delay is minimized and link capacity along the path is not violated (i.e., congestion free).

\section{Results and discussions}
\label{sec:results}

In this section, we evaluate the performance of the \acrshort{drl}-\acrshort{cbf} algorithms in both flow-based simulation and packet-based simulation. 
Furthermore, we study how to transfer models, which can be quickly trained in flow-based simulations, to packet-based simulations, which are much slower.

\subsection{Simulation parameters}
\lam{For proof-of-concept purpose, the flow-based simulator is mainly designed to evaluate the proposed safe learning algorithm in a minimalist environment, in which networking protocols are neglected. It allows us to use simplistic models (i.e., M/M/1 queuing model) to simulate complex network behaviors and obtain a true optimal solution from the \acrshort{nlp} problem. It also plays a significant role in measuring how far our approach deviates from optimal, which can not be achieved in practice.  On the other hand, packet-based simulations provide us with a near real-time environment at the cost of a longer simulation. As we cannot compare to optimal in the packet-based simulator, we selected a number of practical baselines. The details of simulation (hyper-)parameters for both simulators are described below.} 
\subsubsection{(Hyper-) parameters for flow-based simulations} 
\begin{table}[ht]
\centering
\caption{(Hyper)parameters for flow-based simulations.}
\resizebox{330pt}{!}{%
\begin{tabular}{c|cc} \label{tab:simul}
                                                                                                    & \multicolumn{2}{c}{\textbf{Value}}                                                                                                                                       \\ \cline{2-3} 
\multirow{-2}{*}{\textbf{\begin{tabular}[c]{@{}c@{}}Simulation \\ (Hyper) parameters\end{tabular}}} & \multicolumn{1}{c|}{\textbf{\begin{tabular}[c]{@{}c@{}}Off-policy Algo\\ (\acrshort{ddpg} )\end{tabular}}} & \textbf{\begin{tabular}[c]{@{}c@{}}On-Policy Algo\\ (\acrshort{ppo} )\end{tabular}} \\ \hline
\multicolumn{1}{c|}{Gradient clipping maximum}                                                      & \multicolumn{1}{l|}{\cellcolor[HTML]{000000}}                                                  & 0.5                                                                     \\ \hline
Clipping parameter   $\varepsilon$                                                                               & \multicolumn{1}{c|}{\cellcolor[HTML]{000000}}                                                  & 0.2                                                                     \\ \hline
Target KL divergence  $\delta_{KL}$                                                                              & \multicolumn{1}{c|}{\cellcolor[HTML]{000000}}                                                  & 0.05                                                                  \\ \hline
Delayed network update rate                                                                   & \multicolumn{1}{c|}{0.05}                                                                      & \cellcolor[HTML]{000000}{\color[HTML]{000000} }                         \\ \hline
Learning rate                                                                                       & \multicolumn{2}{c}{1e-5}                                                                                                                                                 \\ \hline
Discounted factor  $\gamma$                                                                                 & \multicolumn{2}{c}{0.7}                                                                                                                                                  \\ \hline
Hidden layers                                                                                       & \multicolumn{2}{c}{3}                                                                                                                                                    \\ \hline
Dimension of hidden layer                                                                           & \multicolumn{2}{c}{1024}                                                                                                                                                  \\ \hline
Batch size  $|B|$                                                                                        & \multicolumn{2}{c}{64}                                                                                                                                                  \\ \hline
Frequency of model updates (steps)                                                                                        & \multicolumn{2}{c}{128}                                                                                                                                                  \\ \hline
Episode length $L_{eps}$ (steps)                                                                                          & \multicolumn{2}{c}{64}                                                                                                                                                  \\ \hline
Max local search iteration (\gls{mcbf})                                                                       & \multicolumn{2}{c}{20}                                                                                                                                               \\ \hline
Local search solutions (\gls{ncbf})                                                                       & \multicolumn{2}{c}{450}                                                                                                                                               \\ \hline
Local search radius ($\delta_s$)                                                                    & \multicolumn{2}{c}{0.3}                                                                                                                                                  \\ \hline
Reward parameter $\sigma$                                                                           & \multicolumn{2}{c}{0.8}                                                                                                                                                  \\ \hline
Training episodes                                                                          & \multicolumn{2}{c}{5000}                                                                                                                                                  \\ \hline
\end{tabular}
}
\end{table}

Table~\ref{tab:simul} enumerates the list of training parameters that we applied in the flow-based simulations. To stabilize the training process in \acrshort{ddpg}, delayed network updates are used so that the target actor/critic networks are updated less frequently than the main actor/critic by a factor of $0.05$. The clipped version of \acrshort{ppo}~ \citep{schulmanProximalPolicyOptimization2017} has been used with a clipping parameter $\varepsilon=0.2$ and a target $KL$ divergence  set at $\delta_{KL}=0.05$. To avoid destructively large weight updates, the gradient is also clipped at 0.5. In addition, a fully connected neural network with three layers of 1024 neurons is used to approximate the policy and value functions. Concerning local search, we take the best solution out of the $N*M=450\times20=90000$ points generated around each unsafe action returned by the \acrshort{drl} agent 
in a radius of $\delta_s=0.3$. Finally, our models are updated after each 128-step rollout, which is equivalent to two episodes, and the learning process is studied in $t_s=5000$ training episodes.


To obtain an optimal solution as a benchmark, the following constraint \eqref{eq:C21} is used in problem \ref{eq:P0} so that link delays are computed according to the M/M/1 queuing model. The resulting Non Linear Problem (\acrshort{nlp}) is solved with \acrshort{scip}~\citep{bestuzhevaSCIPOptimizationSuite2021}. Note that this constraint is crucial to  our benchmark, but, in practice, it cannot be applied.

\begin{align*}
     &\gls{de} \geq \frac{1}{c_e - \sum_{k\in K} \sum_{p \in \gls{pk}}\gls{lpk}} \quad \forall e \in E\tag{$\mathcal{C}_2$} \label{eq:C21} \\
\end{align*}

The main goal is to observe how close our algorithms are to the optimal policy in terms of delay minimization and safety.

\subsubsection{(Hyper-) parameters for packet-based simulations} 

Table \ref{tab:simul_packet} illustrates the network and training parameters that we used for our \acrshort{ml}-based \acrshort{ns3} environment. In particular, in the application layer, each microflow is generated with a fixed data rate \gls{dmf} of 0.3 $Mbps$ and a duration of 5 $[seconds]$. Each packet of a microflow is generated with a constant size of 350 [Bytes]. The inter-arrival time between microflows is exponentially distributed, and \acrfull{tcp} is used for the transport protocol. Then, Z=64 \acrshort{ns3} instances are executed in parallel to collect training tuples, as described in Figure \ref{fig:ains3_simul}. In each instance, the centralized controller collects training tuples every $\gls{DeltaT} = 5$ seconds.\lam{To avoid a long execution time of simulations, we consider the data collection period of $5$ seconds, which is assumed to be $5$ minutes in real system (traffic variations are scaled according to that also). In practice, the data collection period can be longer to further reduce the overhead on the control channel. }

\begin{table}[ht!]
\centering
\caption{(Hyper)parameters for packet-based simulations.}
\label{tab:simul_packet}
\resizebox{300pt}{!}{%
\begin{tabular}{c|c|c}
\begin{tabular}[c]{@{}c@{}}\textbf{Networking}\\\textbf{ parameters}\end{tabular}          & \multicolumn{2}{c}{\textbf{Value}}\\ 
\hline
Abilene's Tunnels & \multicolumn{2}{c}{ 6 } \\ 
\hline
Inter-arrival time distribution & \multicolumn{2}{c}{ EXP } \\ 
\hline
microflow data rate \gls{dmf} & \multicolumn{2}{c}{ 0.3 [Mbps] } \\ 
\hline
Time-To-Live of each microflow \gls{ttl} & \multicolumn{2}{c}{ 5 [Seconds] } \\
\hline
Application packet size  & \multicolumn{2}{c}{ 350 [Byte]} \\ 
\hline
Transport protocol & \multicolumn{2}{c}{ \acrshort{tcp} } \\ 
\hline
Queue buffer & \multicolumn{2}{c}{ FIFO, Max = 1000 [packets] } \\ 
\hline
Parallel NS3 instances Z & \multicolumn{2}{c}{ 64 } \\ 
\hline
Measurement period \gls{DeltaT} & \multicolumn{2}{c}{ 5 [Seconds] } \\ 
\hline
\begin{tabular}[c]{@{}c@{}}\textbf{Training}\\\textbf{ (Hyper) parameters}\end{tabular} & \multicolumn{1}{c|}{\begin{tabular}[c]{@{}c@{}}\textbf{Off-policy Algo}\\\textbf{ (\acrshort{ddpg} )}\end{tabular}} & \multicolumn{1}{l}{\begin{tabular}[c]{@{}c@{}}\textbf{On-Policy Algo}\\\textbf{ (\acrshort{ppo} )}\end{tabular}}  \\ 
\hline
\multicolumn{1}{c|}{Gradient clipping maximum}                                          & \multicolumn{1}{l}{{\cellcolor{black}}}                                                                       & 0.3                                                                                                   \\ 
\hline
Clipping parameter $\varepsilon$                                                        & {\cellcolor{black}}                                                                                           & 0.2                                                                                                   \\ 
\hline
Target KL divergence $\delta_{KL}$                                                        & {\cellcolor{black}}                                                                                           & 0.05                                                                                                  \\ 
\hline
Delayed network update rate  $\tau$                                                       & 0.05                                                                                                          & {\cellcolor{black}}                                                                                   \\ 
\hline
Learning rate  training                                                                         & \multicolumn{2}{c}{1e-5}                                                                                                                                                                                        \\ 
\hline
Learning rate  fine-tuning                                                                         & \multicolumn{2}{c}{1e-6}                                                                                                                                                                                        \\
\hline
Discounted factor $\gamma$                                                              & \multicolumn{2}{c}{0.7}                                                                                                                                                                                               \\ 
\hline
Hidden layers                                                                           & \multicolumn{2}{c}{3}                                                                                                                                                                                                 \\ 
\hline
Dimension of a hidden layer                                                               & \multicolumn{2}{c}{1024}                                                                                                                                                                                               \\ 
\hline
Batch size $|B|$                                                                        & \multicolumn{2}{c}{64}                                                                                                                                                                                               \\ 
\hline
Frequency of model updates                                                       & \multicolumn{2}{c}{128 [steps]}                                                                                                                                                                                               \\ 
\hline
Episode length $L_{eps}$                                                           & \multicolumn{2}{c}{64 [steps]}                                                                                                                                                                                               \\ 
\hline
Max local search iteration (\gls{mcbf})                                                       & \multicolumn{2}{c}{20}                                                                                                                                                                                                \\ 
\hline
Local search solutions (\gls{ncbf})                                                           & \multicolumn{2}{c}{450}                                                                                                                                                                                             \\ 
\hline
Local search radius ($\delta_s$)                                                        & \multicolumn{2}{c}{0.3}                                                                                                                                                                                               \\ 
\hline
Reward parameter $\sigma$                                                               & \multicolumn{2}{c}{0.8}                                                                                                                                                                                               \\ 
\hline
Training episode                                                                          & \multicolumn{2}{c}{600 [Episodes]}                                                                                                                                                                                                  \\
\hline
\end{tabular}
}
\end{table}
In terms of training parameters, we kept the hyper-parameters unchanged compared to the flow-based simulator, as described in Table \ref{tab:simul}, except for two  minor differences: (1) we added the learning rate for fine-tuning of pre-trained models, which is $1e-6$, and (2) the number of training episodes is reduced to 600 instead of 4000 due to long and extensive simulations. 
In this simulation, we use the CSMA full-duplex  \citep{tobagiPerformanceAnalysisCarrier1980} to model Ethernet links between nodes. 

\begin{figure*}[ht!]
        \centering
        \begin{subfigure}[b]{0.24\textwidth}
            \centering
    \includegraphics[width=\textwidth]{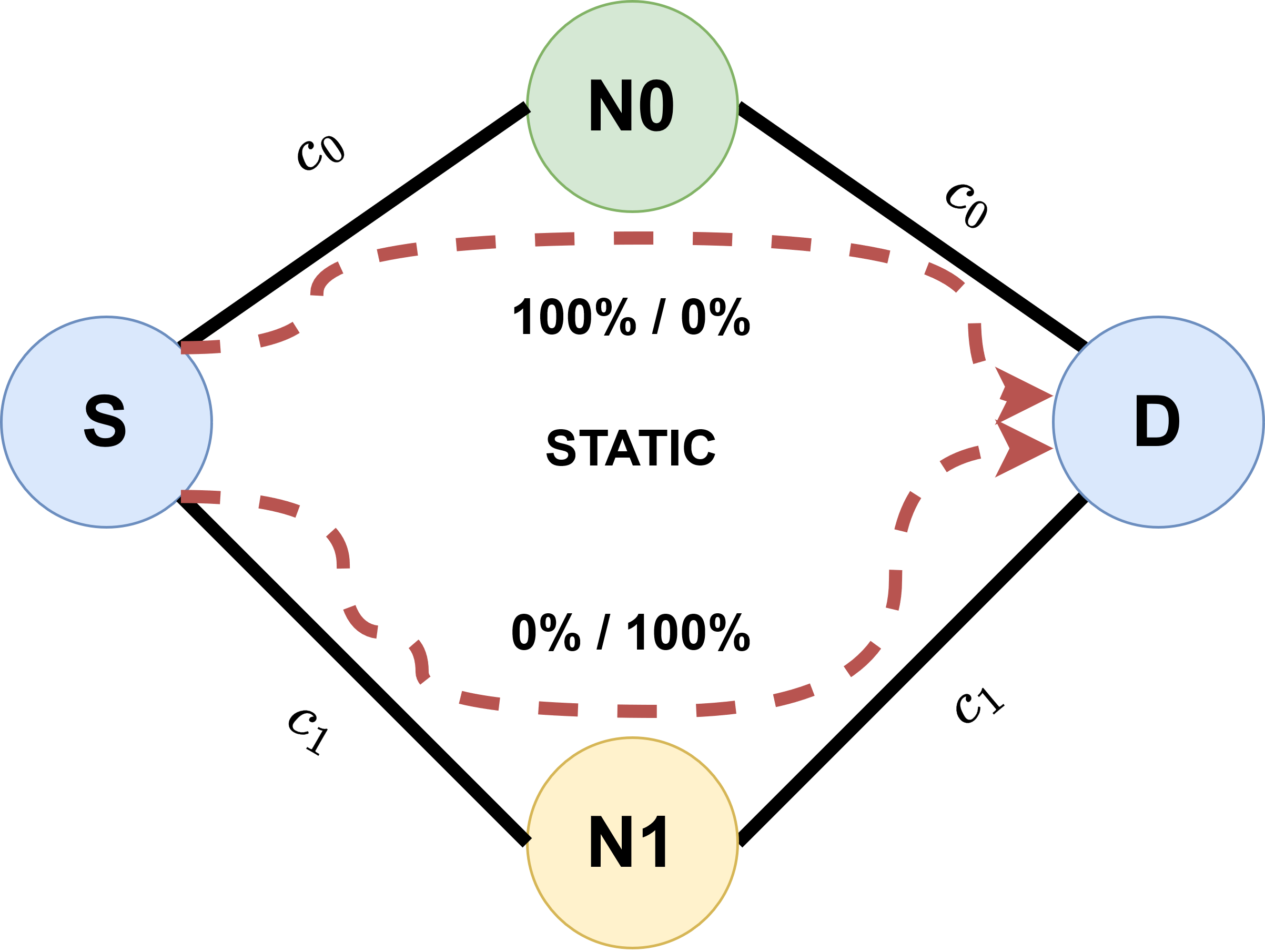}
    \caption{No Load Balancing}
    \label{fig:lb_static}
        \end{subfigure}
        \begin{subfigure}[b]{0.24\textwidth} 
            \centering
    \includegraphics[width=\textwidth]{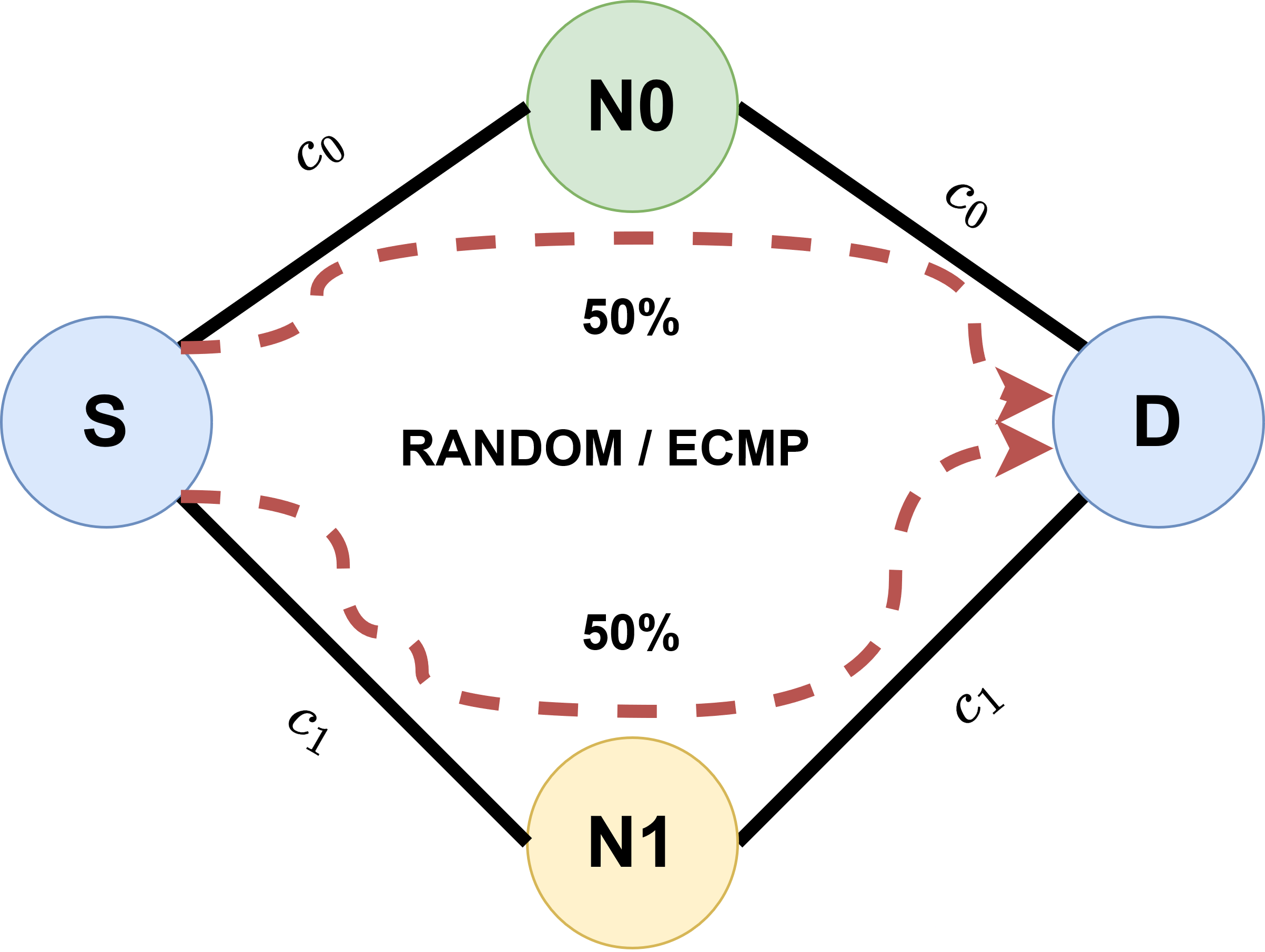}
    \caption{\acrshort{ecmp} Load Balancing}
    \label{fig:lb_ecmp}
        \end{subfigure}  
        \begin{subfigure}[b]{0.24\textwidth} 
           \centering
    \includegraphics[width=\textwidth]{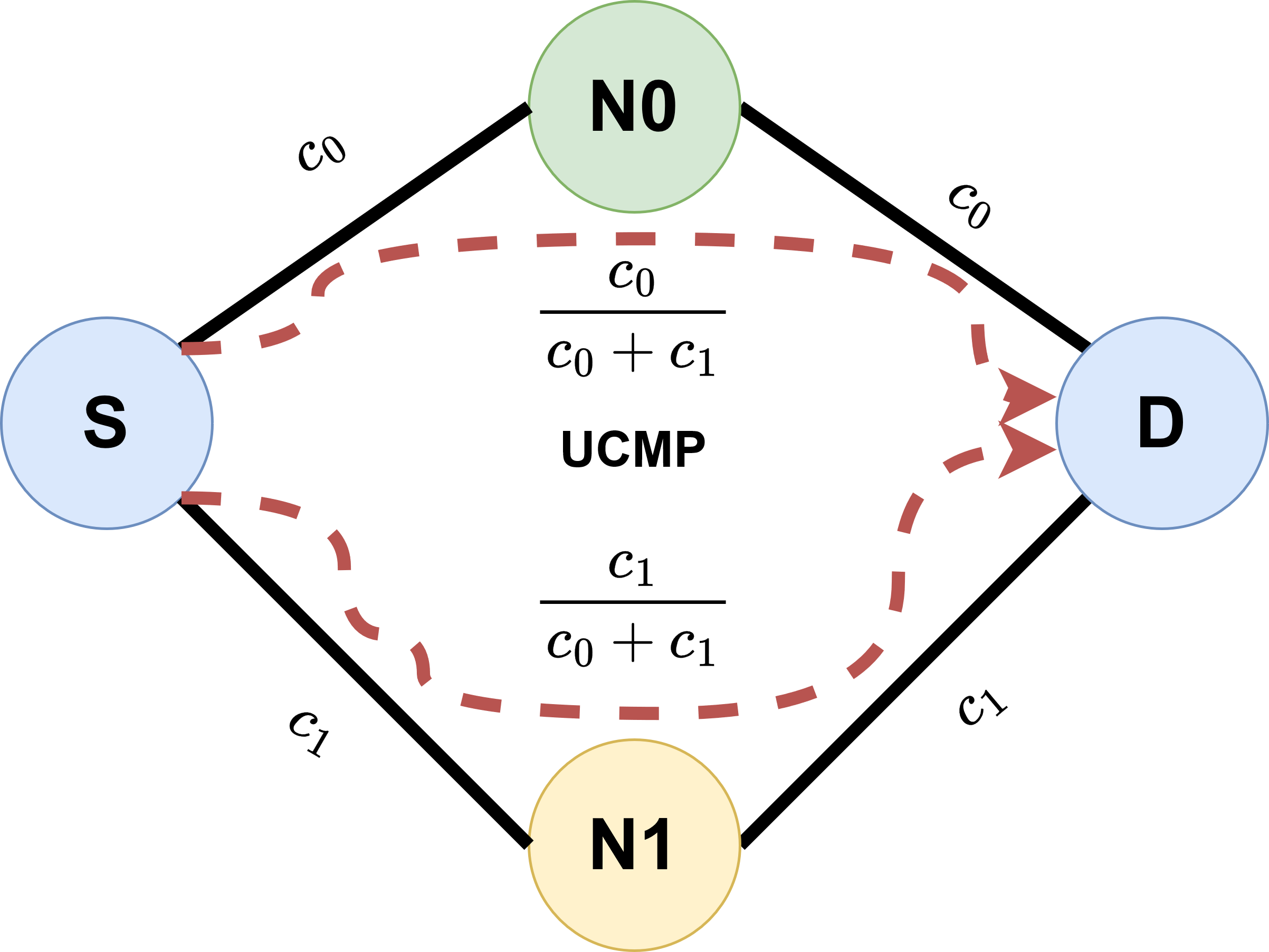}
    \caption{\acrshort{ucmp} Load Balancing}
    \label{fig:lb_ucmp}
        \end{subfigure}
        \begin{subfigure}[b]{0.24\textwidth} 
           \centering
    \includegraphics[width=\textwidth]{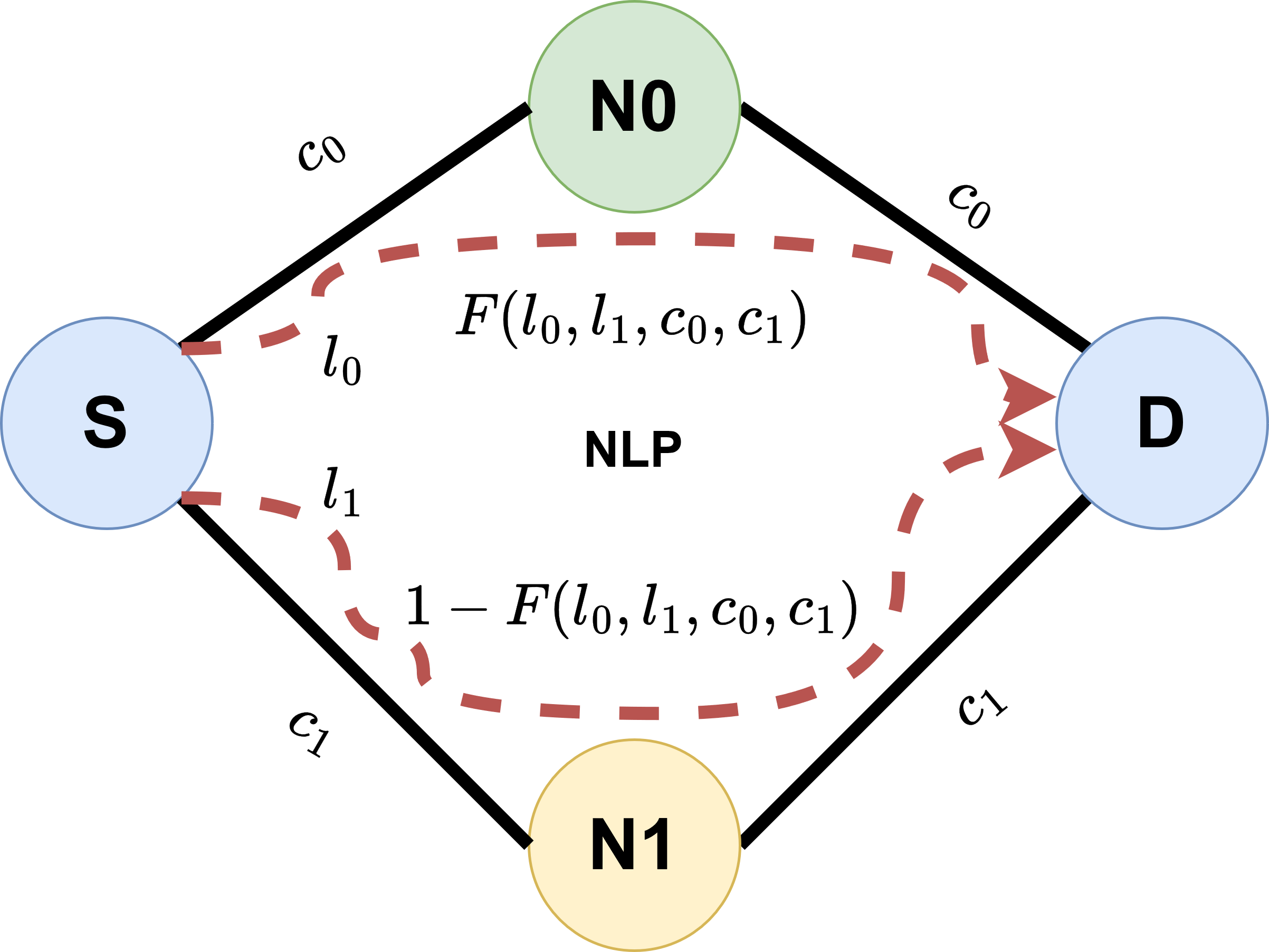}
    \caption{\acrshort{nlp} Load Balancing}
    \label{fig:lb_nlp}
        \end{subfigure}
        \vskip\baselineskip
        \caption
        {\small Load balancing algorithms used as benchmarks in packet-based simulations.} 
        \label{fig:bechhmark}
\end{figure*}

In order to compare the gains brought by our solution to existing baselines, we will use the following policy:\textbf{ STATIC}, \textbf{RANDOM}, \textbf{\acrfull{ecmp}} \citep{rhamdaniEqualCostMultipathRouting2018a}, \textbf{\acrshort{ucmp}} \citep{medagliani2016global} and \textbf{\acrshort{nlp}} solutions \citep{dinhSafeLoadBalancing2024a} for bench-marking. For illustration purpose, Figure \ref{fig:bechhmark} illustrates these solutions when traffic from source S is sent to destination D via two paths (i.e., $N0$ and $N1$) on a topology where links have two different link capacities (i.e., $c_0$ and $c_1$).

Figure \ref{fig:lb_static} shows that traffic is forwarded over only one path with the \textbf{STATIC} approach. As this approach constantly steers traffic on the path with minimum capacity, it represents the worst-case scenario when load-balancing is not applied. It easily causes congestion, which heavily deteriorates \acrshort{qos} performance. Therefore, load-balancing approaches are required to avoid congestion and optimize \acrshort{sla}s.

Figure \ref{fig:lb_ecmp} depicts  load-balancing with \textbf{\acrshort{ecmp}}/\textbf{RANDOM}. 
The traffic load is equally distributed over the available equal cost paths, regardless of their capacity. 
As shown in the figure, $50\%$ of the traffic load is forwarded to path $N0$, and the remaining is sent to $N1$. As for \textbf{RANDOM} method, it randomly picks a path to forward a microflow, and an equal distribution of data traffic on the available paths is also obtained.  However, it cannot guarantee zero congestion as it does not take into account link capacities, and in general, the equal distribution of traffic over paths is likely to introduce congestion on links with the lowest capacities.

Figure \ref{fig:lb_ucmp} demonstrates \textbf{\acrshort{ucmp}} load-balancing, which takes into account link capacities over different paths. Specifically, \textbf{\acrshort{ucmp}} properly divides source traffic so that paths with higher capacity receive a higher amount of traffic and vice versa. This approach reduces the risks of link saturation because the traffic placed on a path is proportional to its capacity. However, the end-to-end delay from source S to destination D might be high if the path with higher capacity has a higher propagation delay.

Finally, Figure \ref{fig:lb_ucmp} shows the use of \textbf{\acrshort{nlp}}, which is based on \acrshort{scip} solver \citep{bestuzhevaSCIPOptimizationSuite2021}, to solve the load-balancing problem. The \acrshort{nlp} solution will be derived based on the current load on each path (i.e., $l_0$, $l_1$) and the link capacity (i.e., $c_0$, $c_1$) to estimate the delay according to M/M/1. This estimation can be inaccurate as it cannot model complex scheduling behaviors. 



\subsection{Results on flow-based simulations}

\subsubsection{GPU acceleration}
\begin{figure}[ht!]
    \centering
    \includegraphics[scale=0.55]{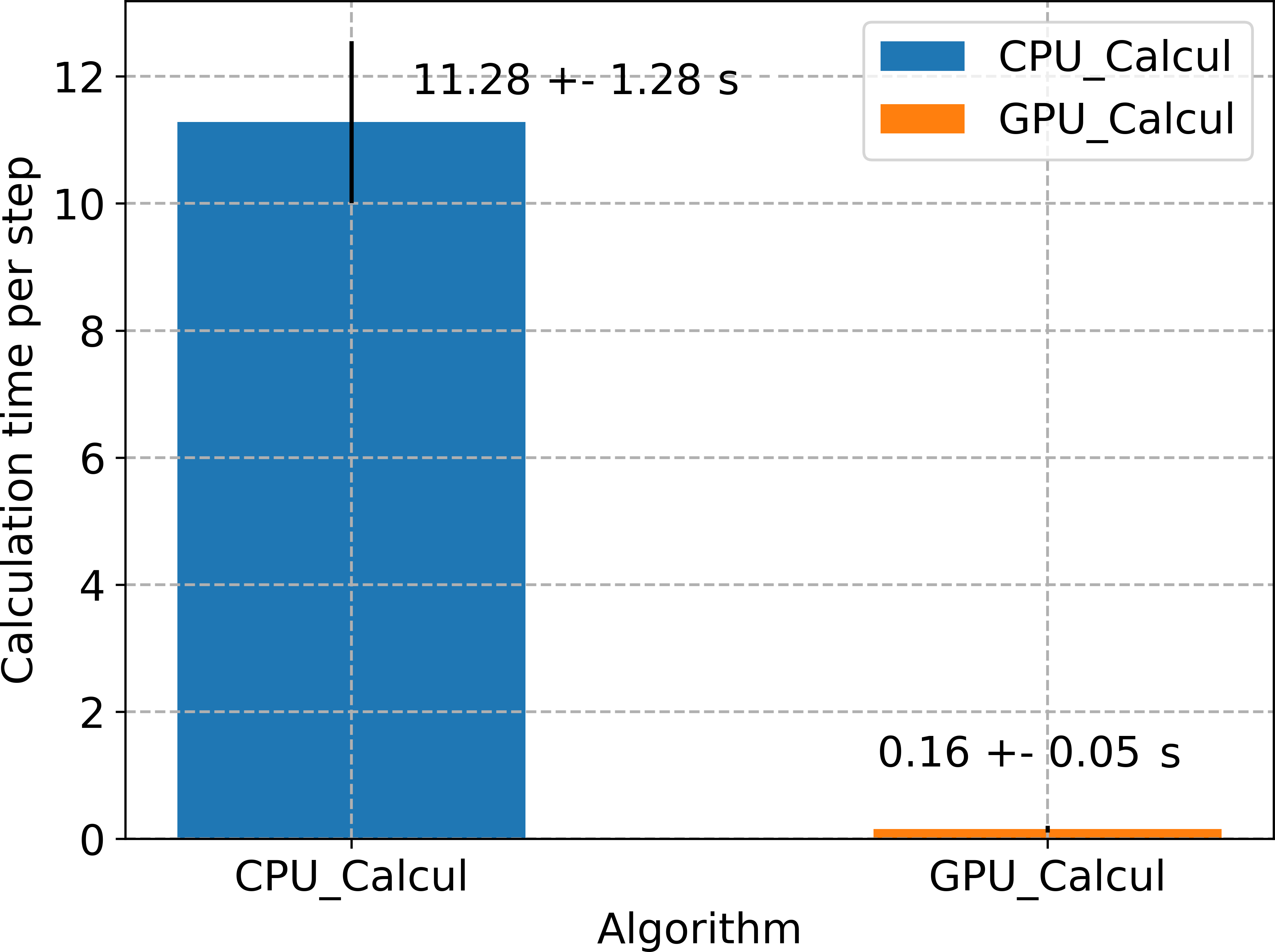}
    \caption{Computation time per training step using CPU and GPU.}
    \label{fig:calcul_time}
\end{figure}
Figure \ref{fig:calcul_time} demonstrates the benefits of our hardware architecture, as illustrated in Figure \ref{fig:flow} for training acceleration. 
We compare the average execution time required to successfully perform one iteration of Algorithm \ref{alg:rlAlgo}, which includes local search and model update algorithms, with and without (i.e. \acrshort{cpu}) using \acrshort{gpu}. It can be observed that the \acrshort{gpu} implementation significantly speeds up the calculation and training time. In particular, feasible (safe) action is found in roughly $0.16(s)$ when using \acrshort{gpu}, compared to more than $11(s)$ using \acrshort{cpu}. This acceleration favors the use of our methods in practice, since the model can also be updated every 128 steps in roughly 20.4 seconds, rather than 1445 seconds using only \acrshort{cpu}. As a result, our model can be fully trained in nearly 13 hours using 300.000 training steps instead of 41 days when using \acrshort{cpu}. This time acceleration has a significant impact on the possibility of deploying our solution in practice. Indeed, since network measurements (e.g., delay, traffic, loss, etc.) can be collected every $5$s, on-policy learning models can be updated every $128*5=640$s in $0.16*128=20.5$s,
which is reasonable.

\subsubsection{Training performance}
\begin{figure*}[ht!]
        \centering
        \begin{subfigure}[b]{0.23\textwidth}
            \centering
  \includegraphics[width=\textwidth]{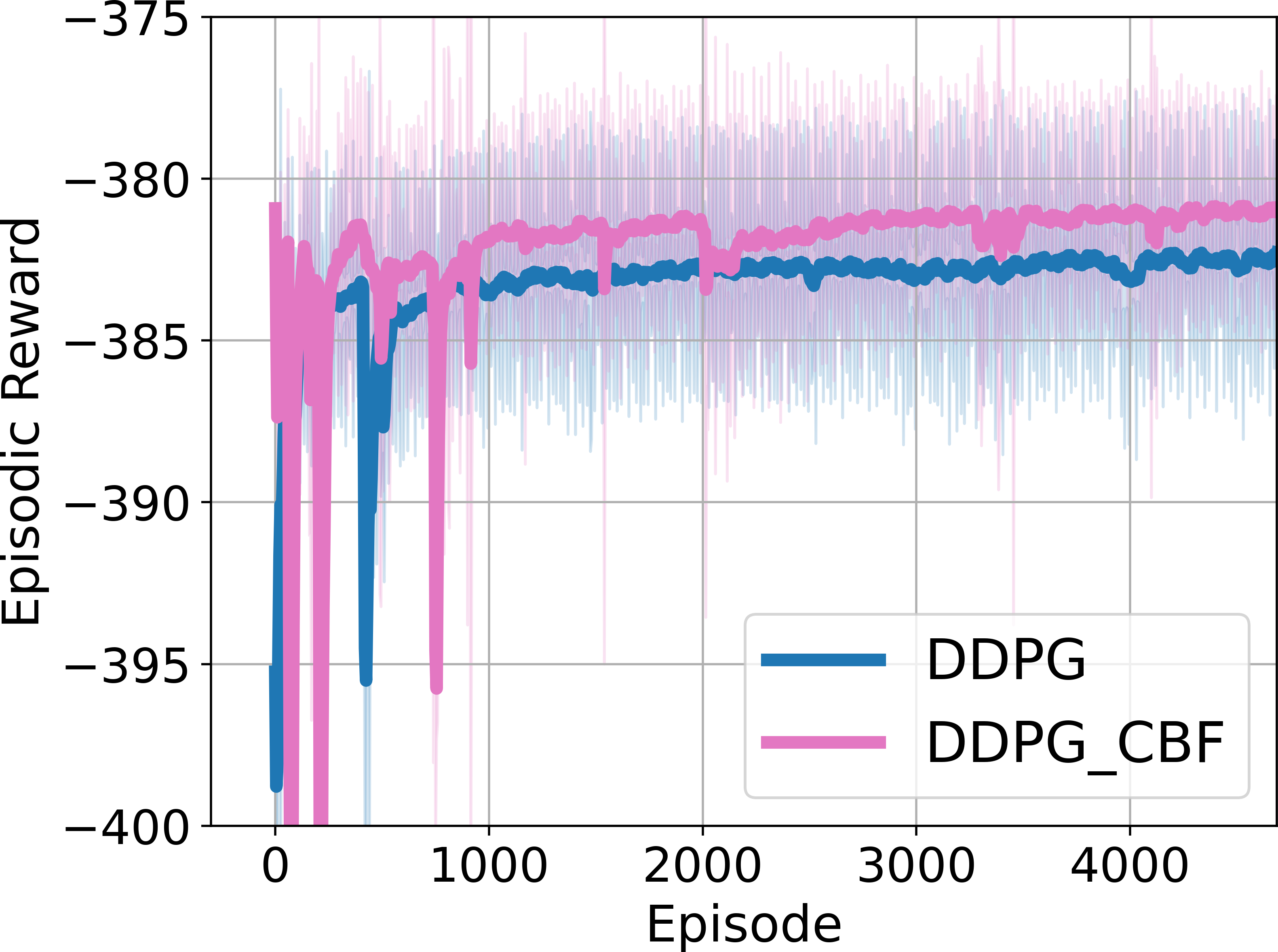}
  \caption{Reward of \acrshort{ddpg}-CBF}
  \label{fig:rew_ddpg}
        \end{subfigure}
        \begin{subfigure}[b]{0.23\textwidth}  
            \centering
  \includegraphics[width=\textwidth]{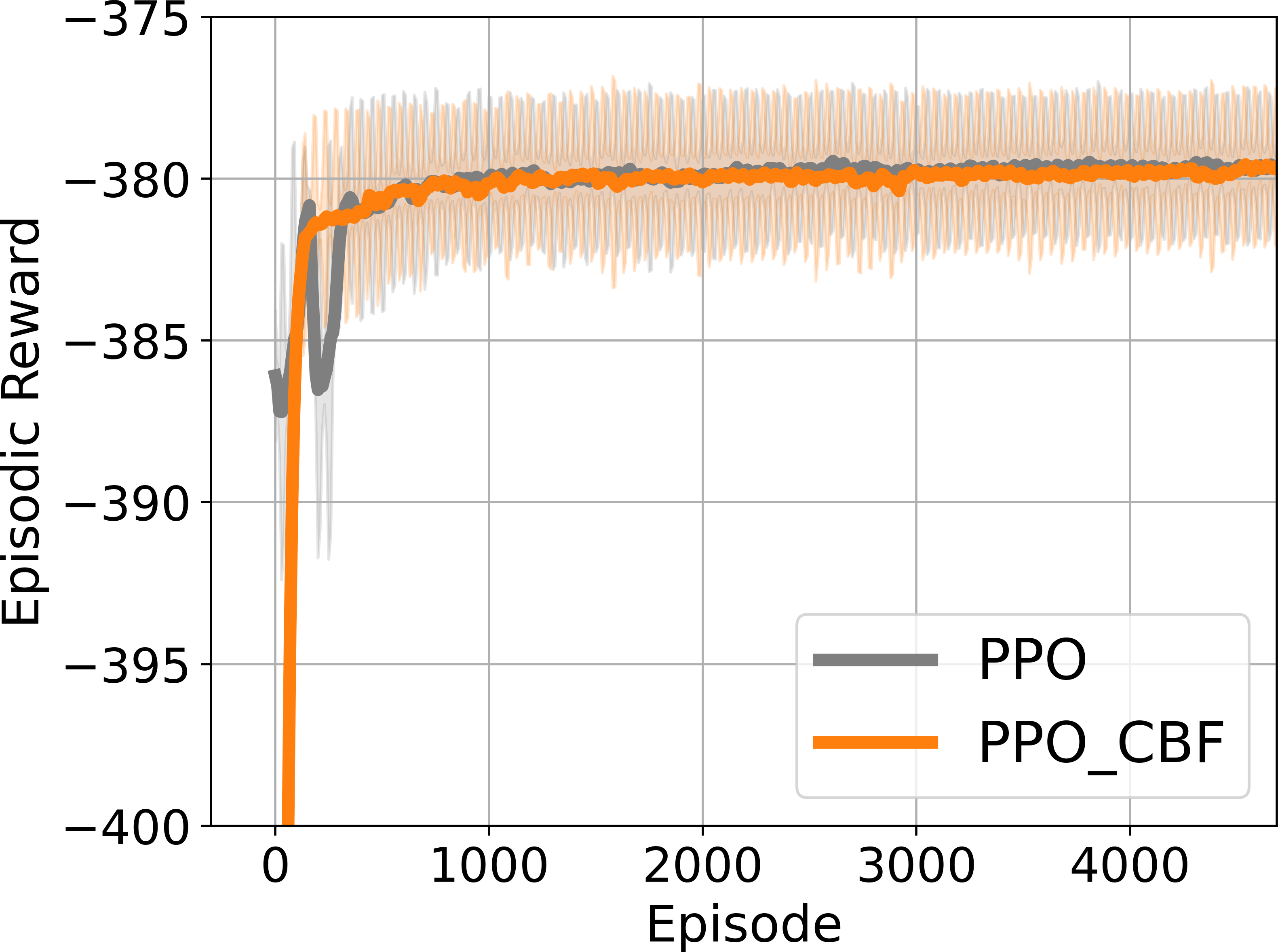}
  \caption{Reward of \acrshort{ppo}-CBF}
  \label{fig:rew_ppo}
        \end{subfigure}
        \begin{subfigure}[b]{0.23\textwidth} 
           \centering
    \includegraphics[width=\textwidth]{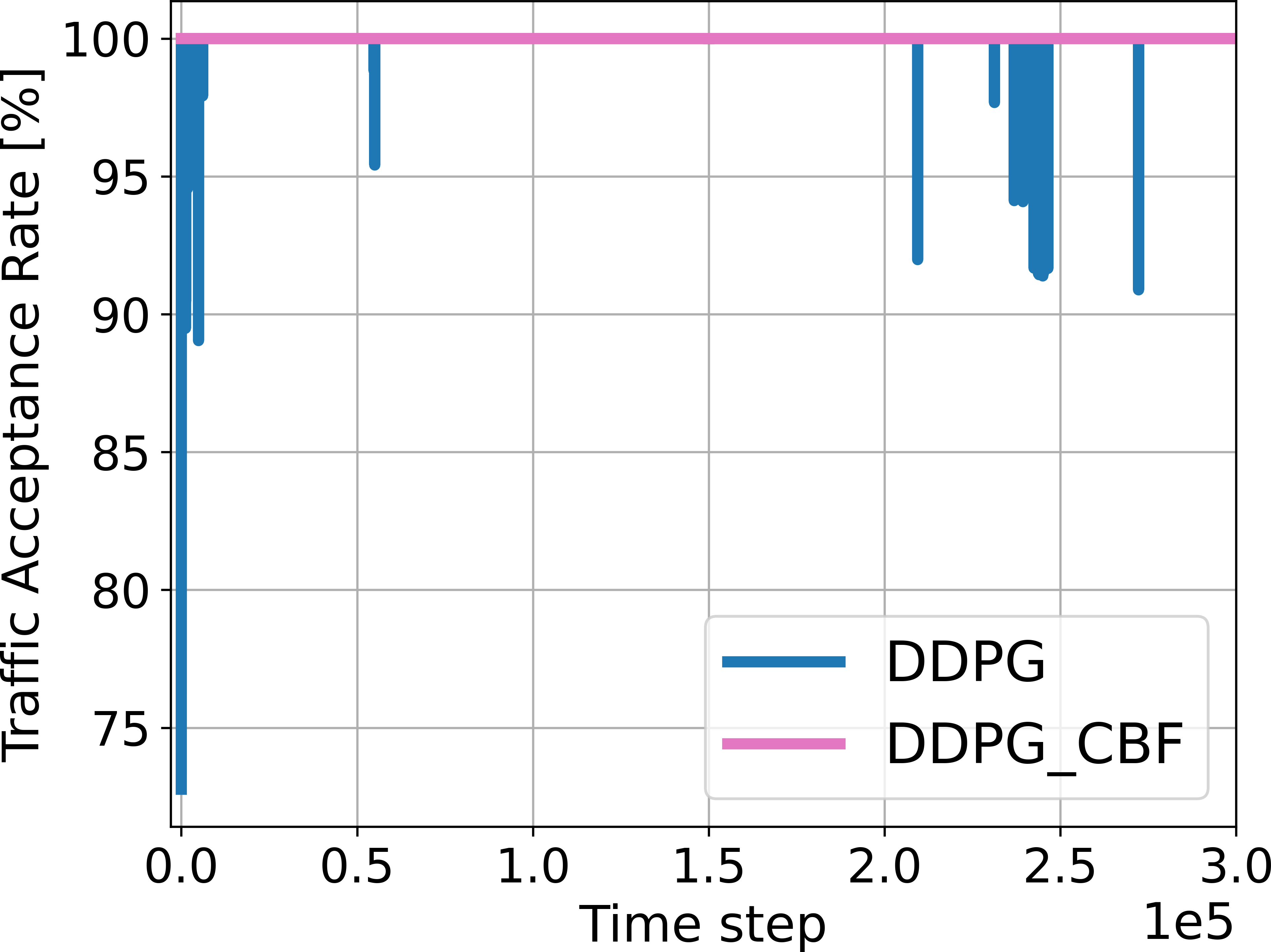}
    \caption{Acceptance  \acrshort{ddpg}-CBF}
    \label{fig:acc_rate_ddpg}
        \end{subfigure}
        \begin{subfigure}[b]{0.23\textwidth}  
            \centering
  \includegraphics[width=\textwidth]{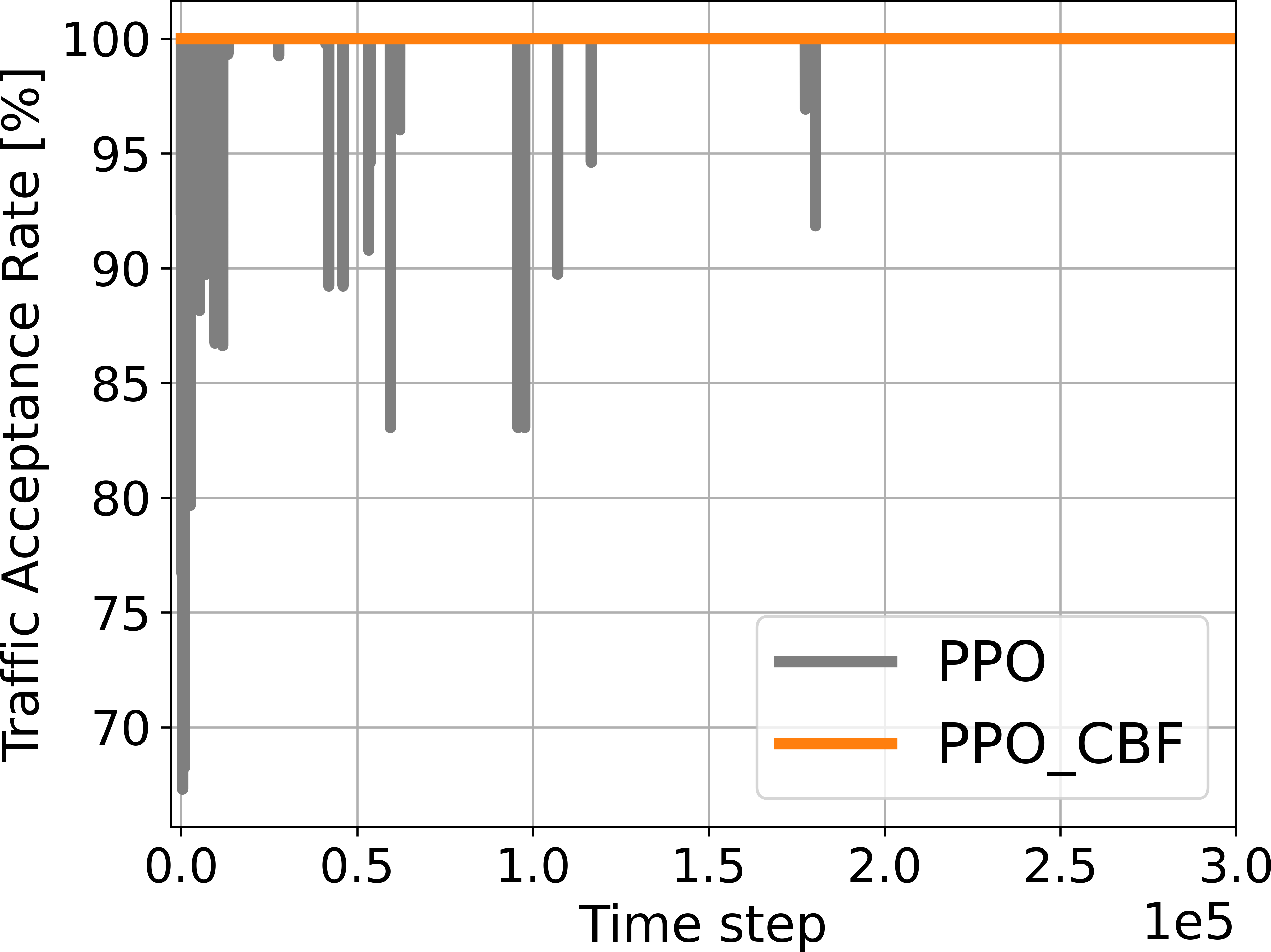}
  \caption{Acceptance  \acrshort{ppo}-CBF}
  \label{fig:acc_rate_ppo}
        \end{subfigure}
        \caption
        {Training performance: (a,b) average episodic training reward, (c,d) traffic acceptance rate of flow-based simulator.} 
        \label{fig:training_cbf}
\end{figure*}

Figures \ref{fig:rew_ddpg} and \ref{fig:rew_ppo} compare the episodic rewards of \acrshort{ddpg}, \acrshort{ppo}, and their \acrshort{cbf} version, respectively. These figures highlight that all learning progress has converged and \acrshort{ppo}  typically achieves smoother convergence than \acrshort{ddpg}. In terms of safety, Figures \ref{fig:acc_rate_ddpg} and \ref{fig:acc_rate_ppo} illustrate the percentage of the total traffic  that is accepted due to link capacity constraints with/without \acrshort{cbf} applied on top of the \acrshort{ddpg} and \acrshort{ppo}  algorithms. When \acrshort{cbf} is not applied, the algorithms encourage policy exploration at early stages in training, which accidentally causes heavy link congestion and traffic rejection. Essentially, the \acrshort{ddpg} algorithm also faces an unstable acceptance rate in the course of training as the traffic acceptance rate is suddenly reduced around time step 200000, whereas this is not the case for \acrshort{ppo}  training. This observation is explained as the learning process of \acrshort{ddpg} algorithm stochastically samples experience from its long replay buffer to update the current policy. It introduces a non-negligible probability that most of the sampling experiences are bad and turns the model update to a worse one. As a result, the acceptance rate drops. On the other hand, the \acrshort{ppo} algorithm constantly improves the policy, and the acceptance rate improves during  training. When \acrshort{cbf} is applied on top of the learning algorithms, the models have also converged while no traffic rejection (i.e., traffic acceptance rate is 100 $\%$) is seen during training. 


\subsubsection{Testing performance}
\begin{figure*}[ht!]
        \centering
        \begin{subfigure}[b]{0.3\textwidth}
            \centering
    \includegraphics[width=\textwidth]{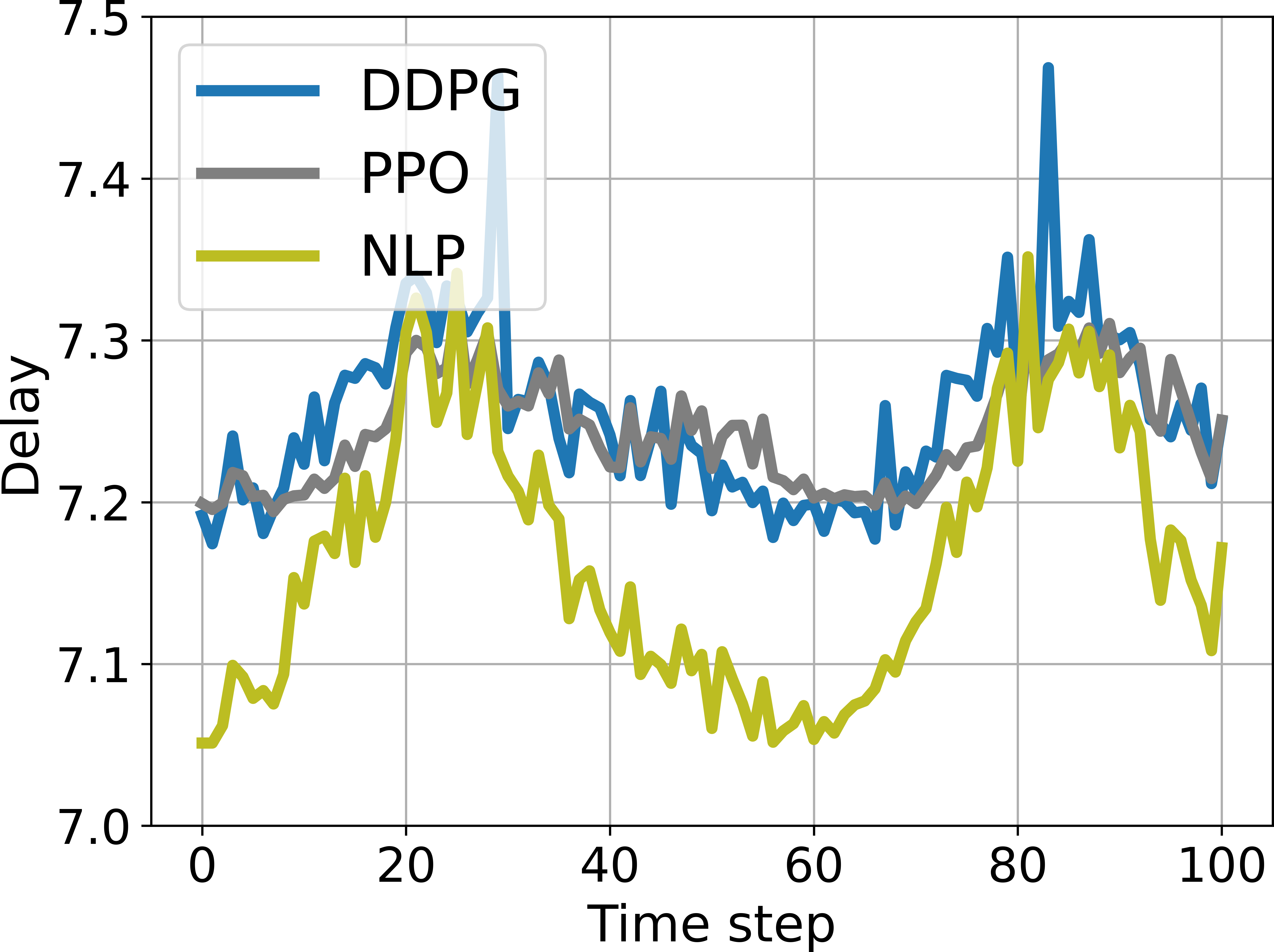}
    \caption{Delay results of \acrshort{ddpg}, \acrshort{ppo},   \acrshort{nlp}}
    \label{fig:del}
        \end{subfigure}
        \begin{subfigure}[b]{0.3\textwidth} 
            \centering
    \includegraphics[width=\textwidth]{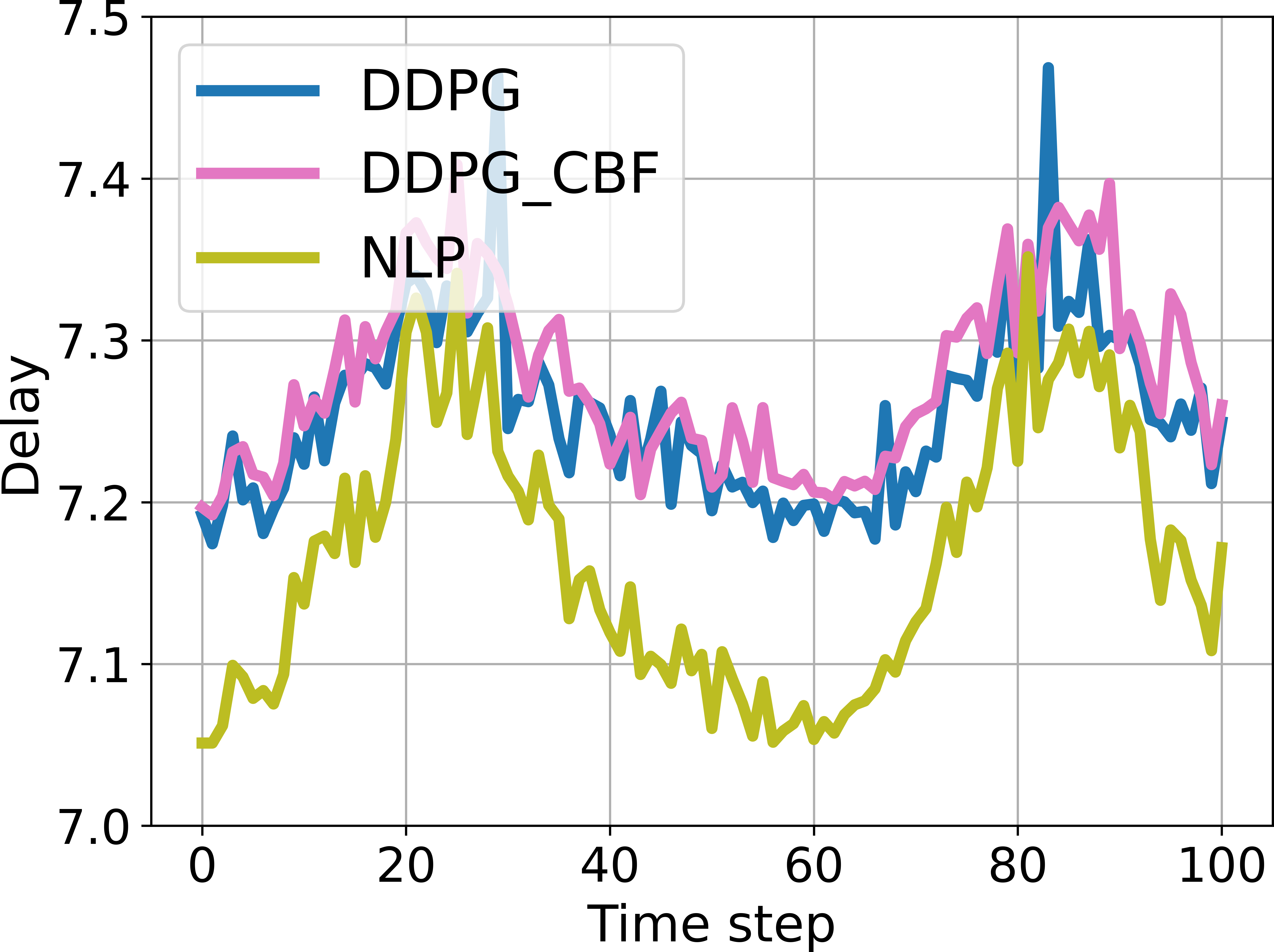}
    \caption{Delay results of  \acrshort{ddpg}-\acrshort{cbf},  \acrshort{nlp}}
    \label{fig:delddpg}
        \end{subfigure}  
        \begin{subfigure}[b]{0.3\textwidth} 
           \centering
    \includegraphics[width=\textwidth]{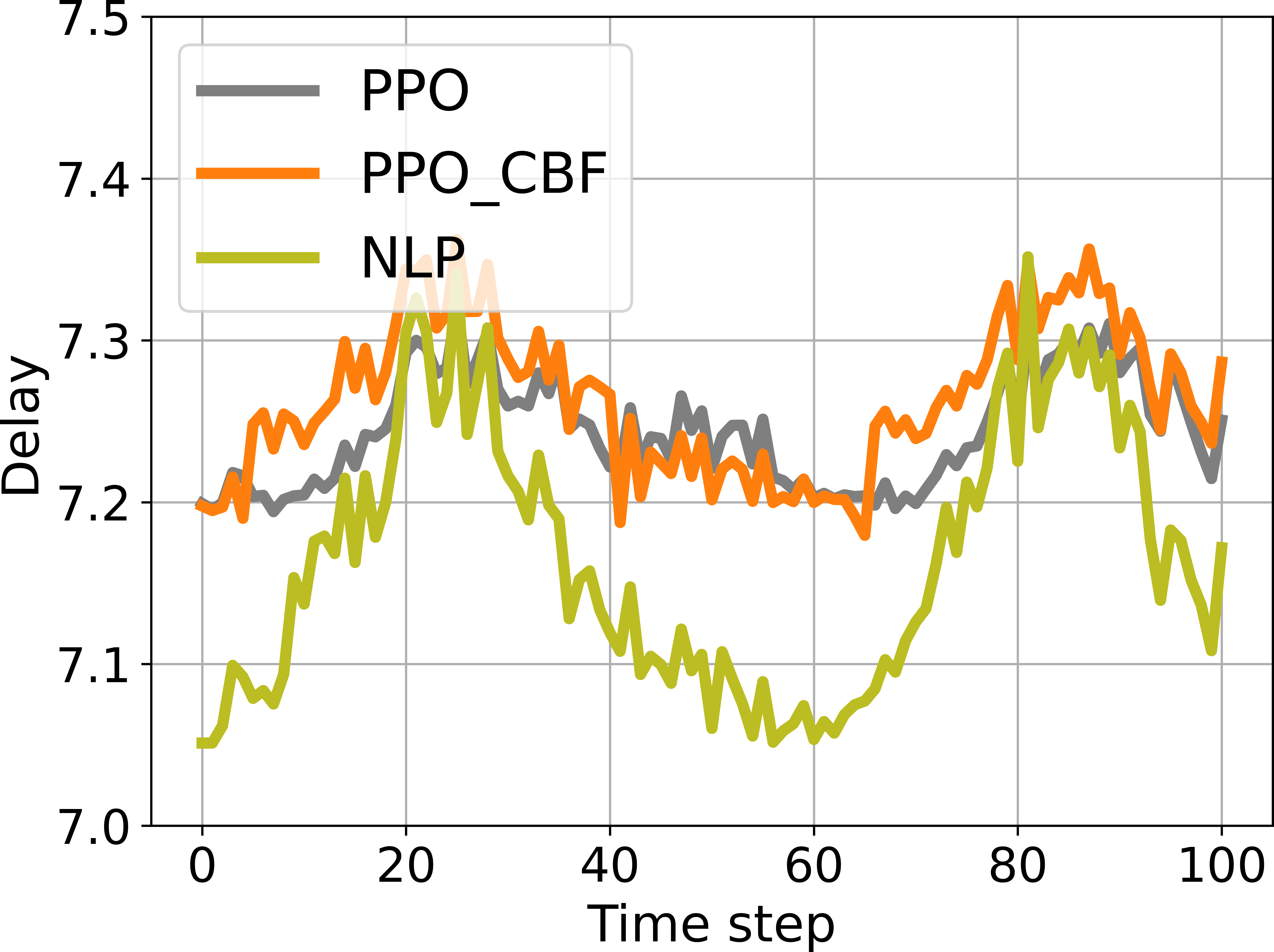}
    \caption{Delay results of \acrshort{ppo}-\acrshort{cbf},  \acrshort{nlp}}
    \label{fig:delppo}
        \end{subfigure}
        \vskip\baselineskip
        \begin{subfigure}[b]{0.3\textwidth} 
            \centering
    \includegraphics[width=\textwidth]{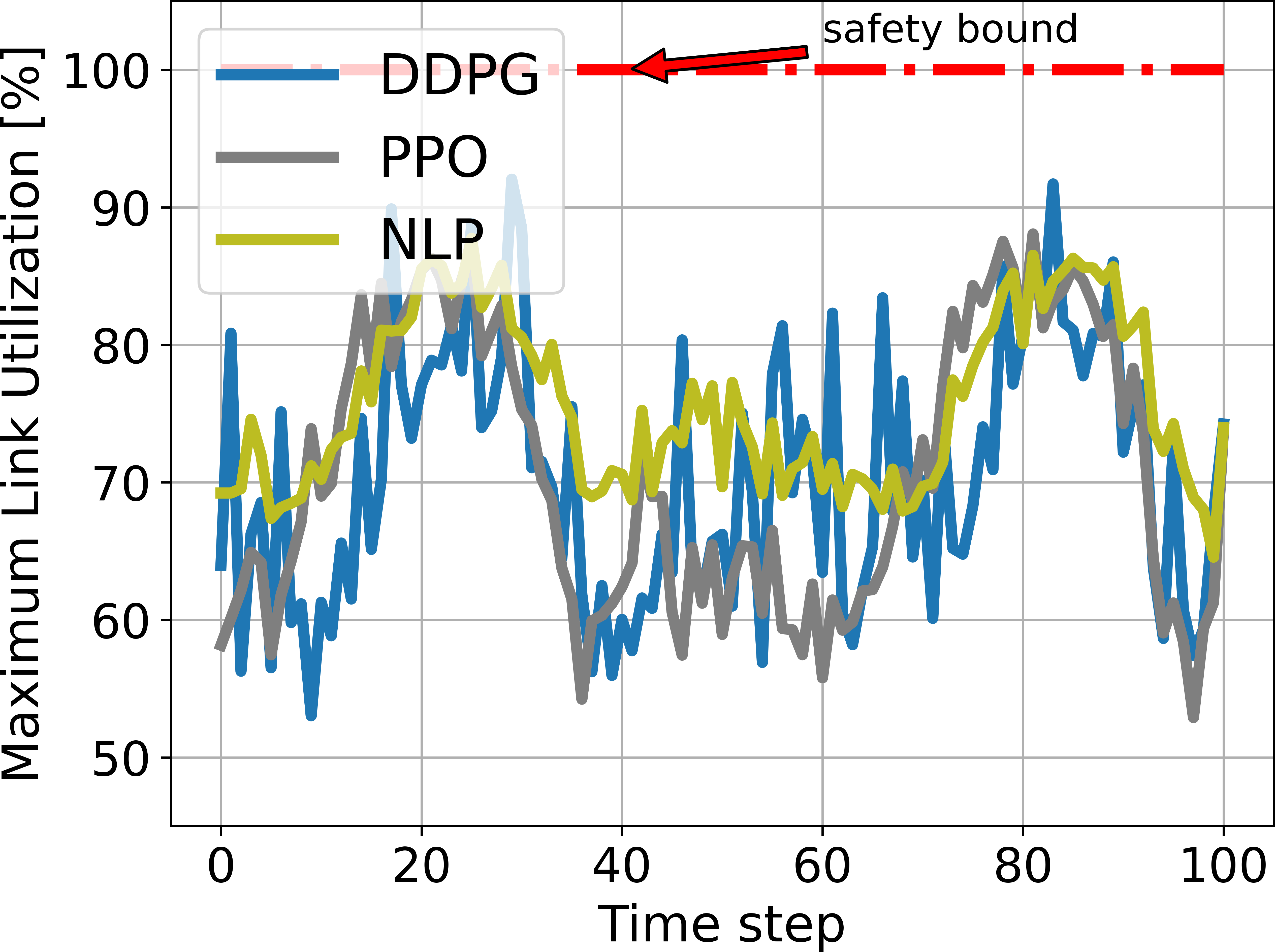}
    \caption{\acrshort{mlu} results of \acrshort{ddpg}, \acrshort{ppo}, \acrshort{nlp}}
    \label{fig:mlu}
        \end{subfigure}  
        \begin{subfigure}[b]{0.3\textwidth} 
           \centering
    \includegraphics[width=\textwidth]{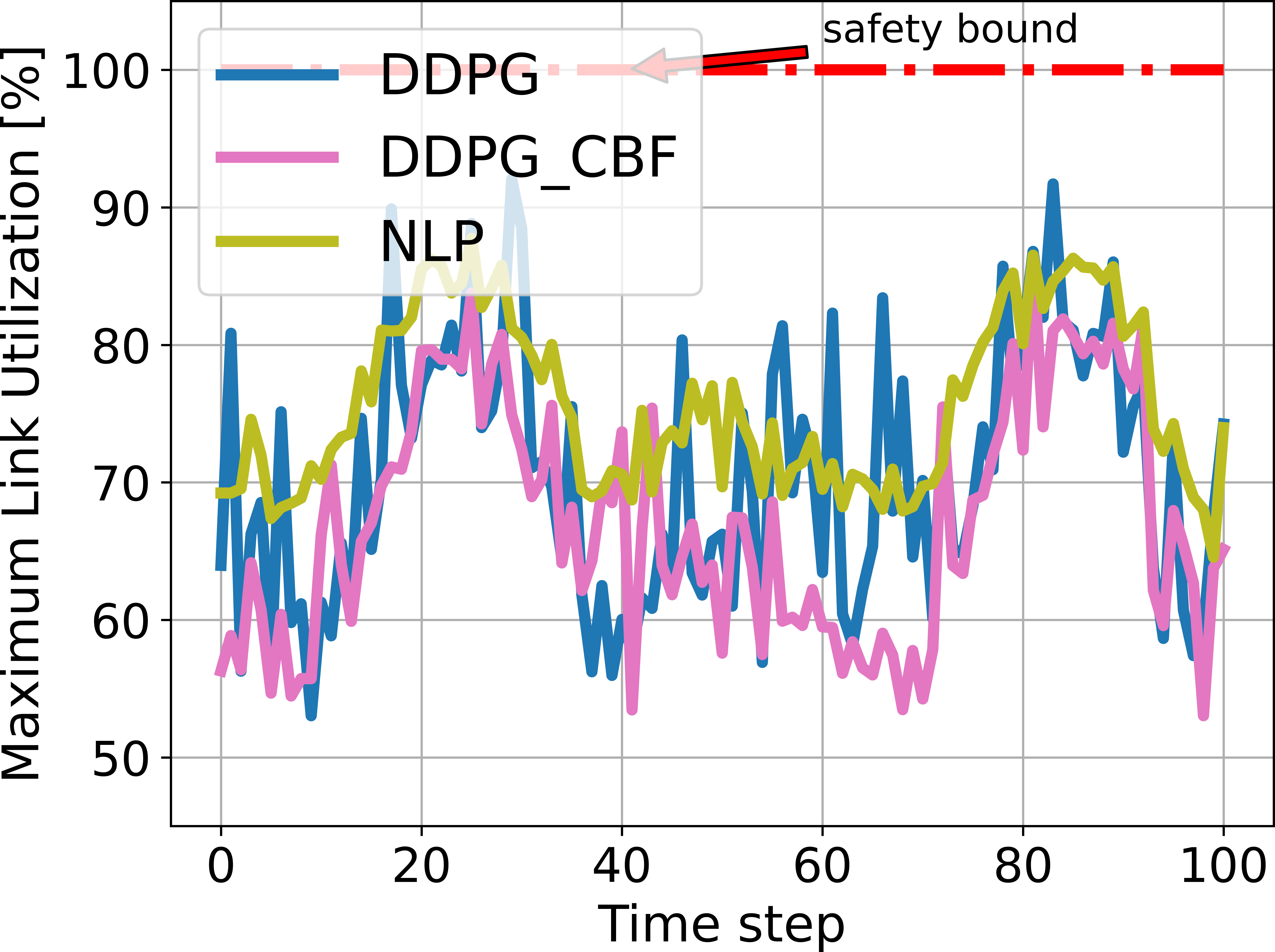}
    \caption{\acrshort{mlu} results of \acrshort{ddpg}-\acrshort{cbf},  \acrshort{nlp}}
    \label{fig:mluddpg}
        \end{subfigure}       
        \begin{subfigure}[b]{0.3\textwidth} 
            \centering
  \includegraphics[width=\textwidth]{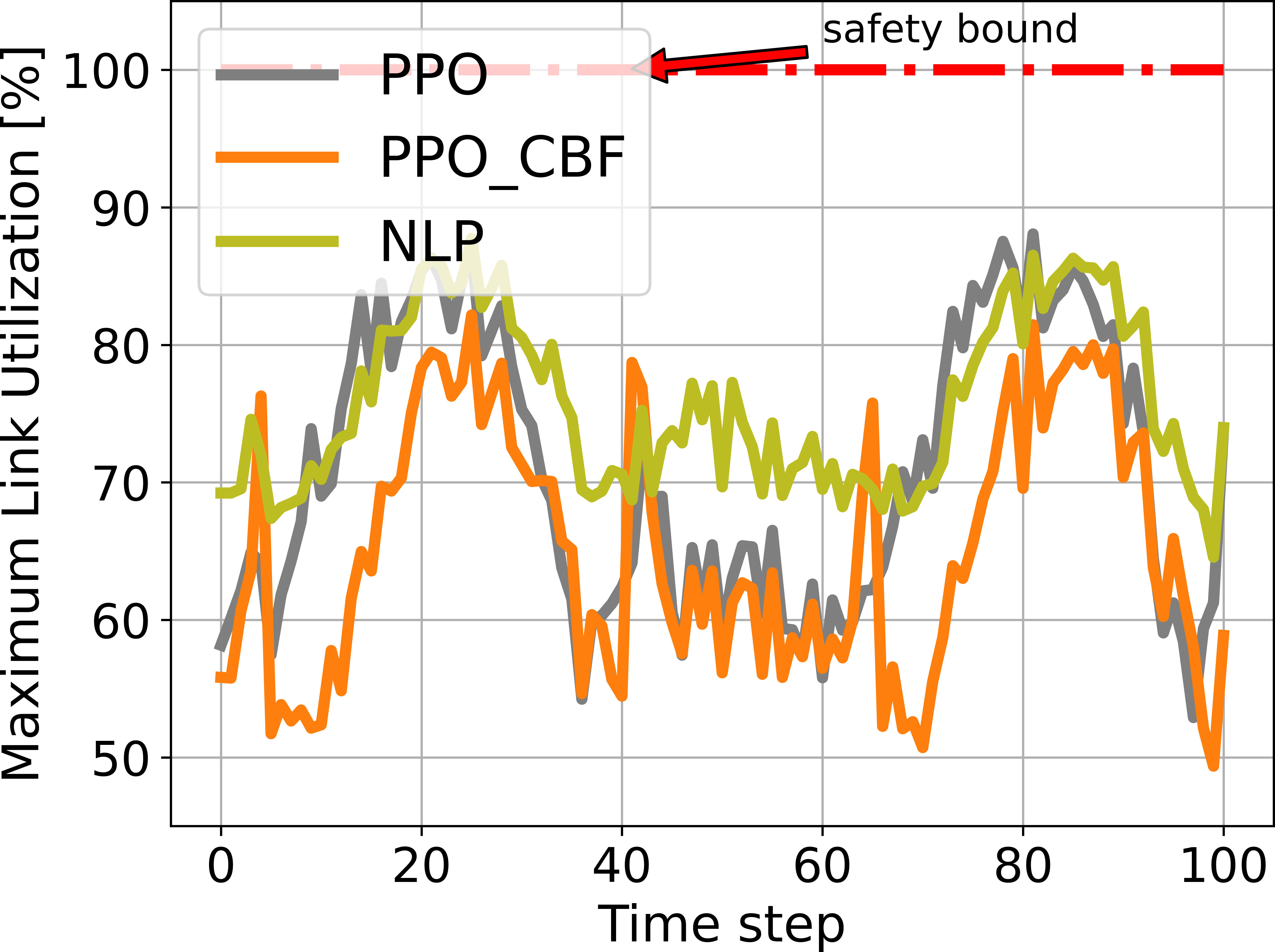}
  \caption{\acrshort{mlu} results of \acrshort{ppo}-\acrshort{cbf}, \acrshort{nlp}}
  \label{fig:mluppo}
        \end{subfigure}        
        \caption[ Local traffic and corresponding shaping rate, loss of each solution ]
        {\small Testing performance on flow-based simulator for average tunnel delay and \acrshort{mlu}.} 
        \label{fig:train_test}
\end{figure*}
To compare the performance during testing, we selected learning models for the \acrshort{ddpg} and \acrshort{ppo} algorithms after training in the last episode. At this point, the training models are well aware of delay and \acrshort{mlu} minimization in the objective function. Therefore, traffic rejection due to link capacity violations is unlikely to occur. To define a test scenario, we generated 100 traffic samples for each OD tunnel following the traffic pattern presented in Figure \ref{fig:traffic}. 

As illustrated in Figures \ref{fig:del} and \ref{fig:mlu}, the delay and \acrshort{mlu} of the learning-without-safety models are compared to the optimal \acrshort{nlp} solution obtained with \acrshort{scip}. They reveal that near-optimal delays are obtained using conventional \acrshort{ddpg} and \acrshort{ppo} learning algorithms. Besides, the \acrshort{mlu} during testing is safely kept below 100 $\%$, resulting in no traffic rejection. Furthermore, network delay is better for \acrshort{ppo} compared to \acrshort{ddpg}  since the local optimum trap is not observed during training. 

When the safety \acrshort{cbf} layer is applied on top of the current off/on policy learning algorithms, the testing performance results are depicted in Figures \ref{fig:delddpg}, \ref{fig:mluddpg}, \ref{fig:delppo} and \ref{fig:mluppo}, respectively. With respect to safety, Figures \ref{fig:delddpg} and \ref{fig:mluddpg} demonstrate that off-policy learning with \acrshort{ddpg}-\acrshort{cbf} does not handle delay well, especially when the total traffic demand is high. In particular, many high delay peaks are observed during testing regardless of \acrshort{cbf} policies, although safety is always respected (\acrshort{mlu} belows 100 $\%$). 
On-policy learning with \acrshort{ppo}-\acrshort{cbf}  significantly improves delay and better controls the \acrshort{mlu} in testing as illustrated by Figures \ref{fig:delppo} and \ref{fig:mluppo}. Furthermore, no capacity violations occur during learning, as indicated in Figure \ref{fig:acc_rate_ppo}. These results show that, with safety, on-policy learning outperforms off-policy learning as near-optimal performance is obtained and hard safety requirements are met.



\begin{table}[htp!]
\centering
\caption{Computation time between NLP and DRL-based models to solve 100 traffic samples}\label{tab:comp-lat}
\resizebox{\linewidth}{!}{%
\begin{tblr}{
  row{1} = {c},
  cell{1}{1} = {c=5}{},
  cell{1}{6} = {c=5}{},
  cell{3}{1} = {black},
  hlines,
  vlines,
}
\textbf{Pre-computation (Training)  [s]} &      &          &     &         & \textbf{Computation (Inference)  [s]} &      &          &     &         \\
NLP                                             & DDPG & DDPG-CBF & PPO & PPO-CBF & NLP                                          & DDPG & DDPG-CBF & PPO & PPO-CBF \\
                                                &  $\approx$ 2.5 $\times$10$^3$    &    $\approx$ 4.8$\times$10$^5$      &  $\approx$ 3$\times$10$^3$    &   $\approx$ 4.8$\times$10$^5$       &  $\approx$ 81.34                                             &  $\approx$ 0.34    & $\approx$ 7.07          &  $\approx$ 0.39    &  $\approx$ 7.97 
\end{tblr}
}
\end{table}

\lam{Table \ref{tab:comp-lat} presents the total runtime (in seconds) required to compute  the optimal load balancing for a window of 100 traffic samples with the optimal solution (i.e., \acrshort{nlp}), and \acrshort{drl}-based algorithms.  In this table, \textit{pre-computation} time is the training time required for the models to converge (i.e., 300.000 time steps are reached), while \textit{computation} times measure the inference time. Pre-computation is not necessary for the \acrshort{nlp} because it directly solves problem \eqref{eq:P0} in an online fashion. As can be observed, \acrshort{cbf}  increases the training time by a factor of 19, from around 3$\times$10$^3$ [s] to 4.8$\times$10$^5$[s]. For the inference test, \acrshort{nlp} takes approximately 81.34 [s] to produce 100 split ratios. It has the highest computation time because it recomputes the optimal split for each traffic matrix. Without \acrshort{cbf}, both \acrshort{ddpg} and \acrshort{ppo} models produce near-optimal solutions in 210x faster than the NLP solver, because the output splitting ratio is simply derived from the input matrix of the converged models. When \acrshort{cbf} is applied, computation latency increases to 7-8 [s] because local search operations are also required to maintain safety. However, it remains 10 times faster than the \acrshort{nlp} calculations. Furthermore, this inference time of our CBF solution is shown to be viable in practice, because each splitting ratio output, in which the safety bound is already taken into account, can be computed in approximately 0.07-0.08 [s]. }
\subsection{Results on packet-based simulation}
\lam{This section presents graphical results produced by our packet-based simulation, which is based on \acrshort{ns3} simulator, and incorporates real protocols for real time evaluation. Firstly, we present the benefits of using the multi-CPU architecture, which is displayed in Figure \ref{fig:ains3_simul}, to generate the training dataset and accelerate the training progress. Secondly, the training results,  along with training delay and training MLU,  are shown in a window of 40.000 steps. Finally, the converged models are used for testing compared with non-AI baselines. }
\subsubsection{Training acceleration using multiple CPU cores}
\begin{figure}[ht!]
    \centering
    \includegraphics[scale=0.45]{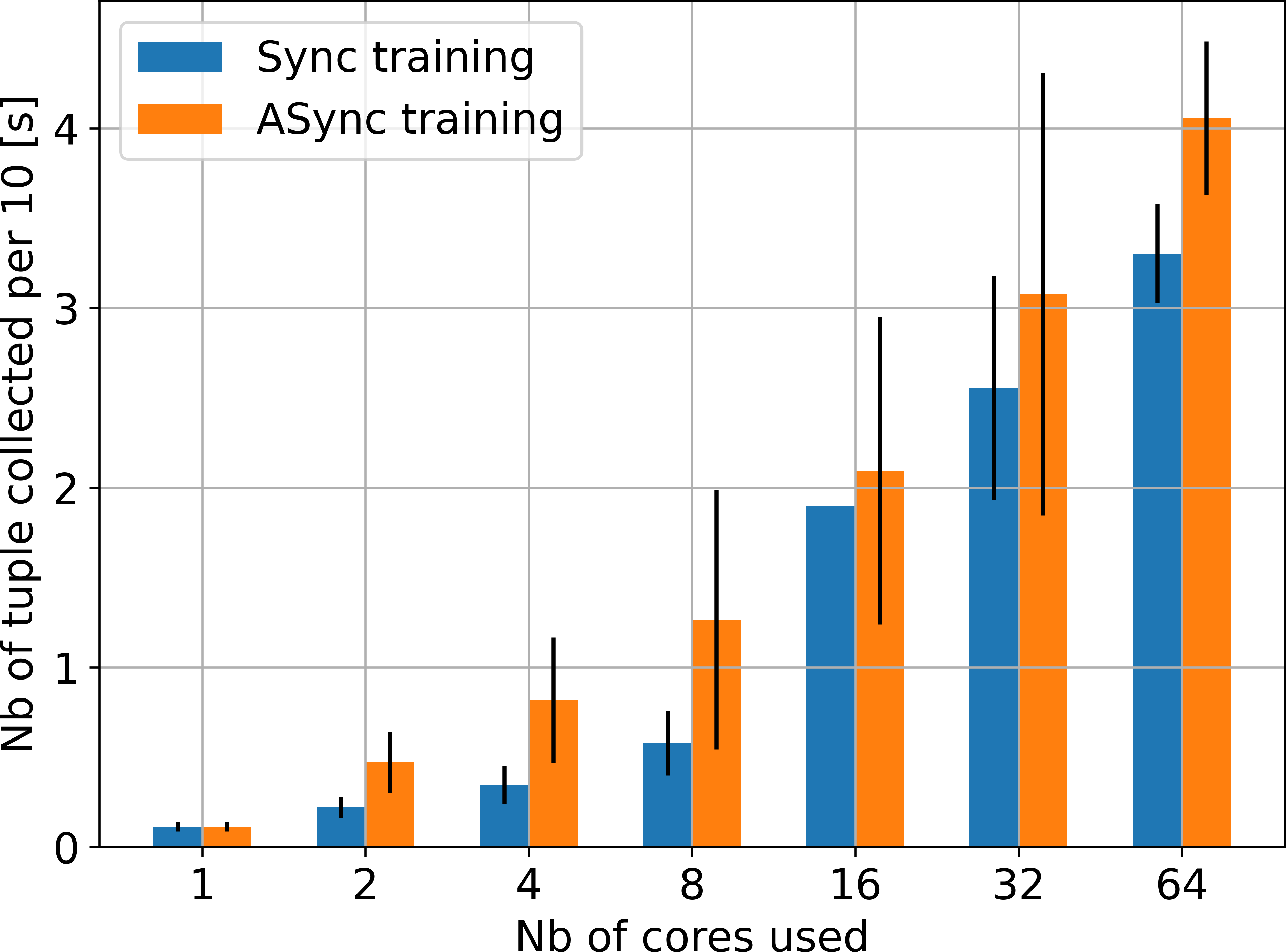}
    \caption{Average training tuples collected on multiple CPU cores.}
    \label{fig:ns3_parallel}
\end{figure}

Figure \ref{fig:ns3_parallel} shows the benefits of using multiple cores to generate training samples for our learning algorithms. 
The \acrshort{ml}-based \acrshort{ns3} architecture is depicted in Figure \ref{fig:ains3_simul}. By running different \acrshort{ns3} instances on independent CPU cores, we are able to collect more training experience (i.e., state, action, reward) and feed them to learning algorithms. 
As we can see from this figure, a higher number of parallel \acrshort{ns3} instances results in more training tuples collected per 10 [seconds]. In the \acrshort{ddpg}-based algorithm, we can store experience, which can be achieved from other policies in the past, in the replay buffer. Then, a batch of experience is randomly taken from the buffer to update the current policy (i.e., off-policy). In this sense, each \acrshort{ns3} instance can use different policies to generate training samples, thus, their environment is not necessarily synchronized with the other running instances. This way of training sample generation is called asynchronous training (i.e., \textbf{Async training}) because each \acrshort{ns3} instance can run at different speeds and does not depend on others. On the other hand, each training sample in the \acrshort{ppo}-based algorithm must be related to the same training policy (i.e., on-policy). Therefore, every \acrshort{ns3} instance has to be synchronized with all available instances, and it defines a synchronous way of collecting the training samples (i.e., \textbf{Sync training}). As we can observe in Figure \ref{fig:ns3_parallel}, \textbf{Async training} is able to collect more samples than \textbf{Sync training} using the same number of CPU cores, because each instance in \textbf{async training} does not need to wait for other instances to reach a certain point, as in \textbf{sync training}. 


\subsubsection{Training Performance}
\begin{figure}[ht!]
        \centering
    \begin{subfigure}[b]{0.44\textwidth} 
    \includegraphics[width=\textwidth]{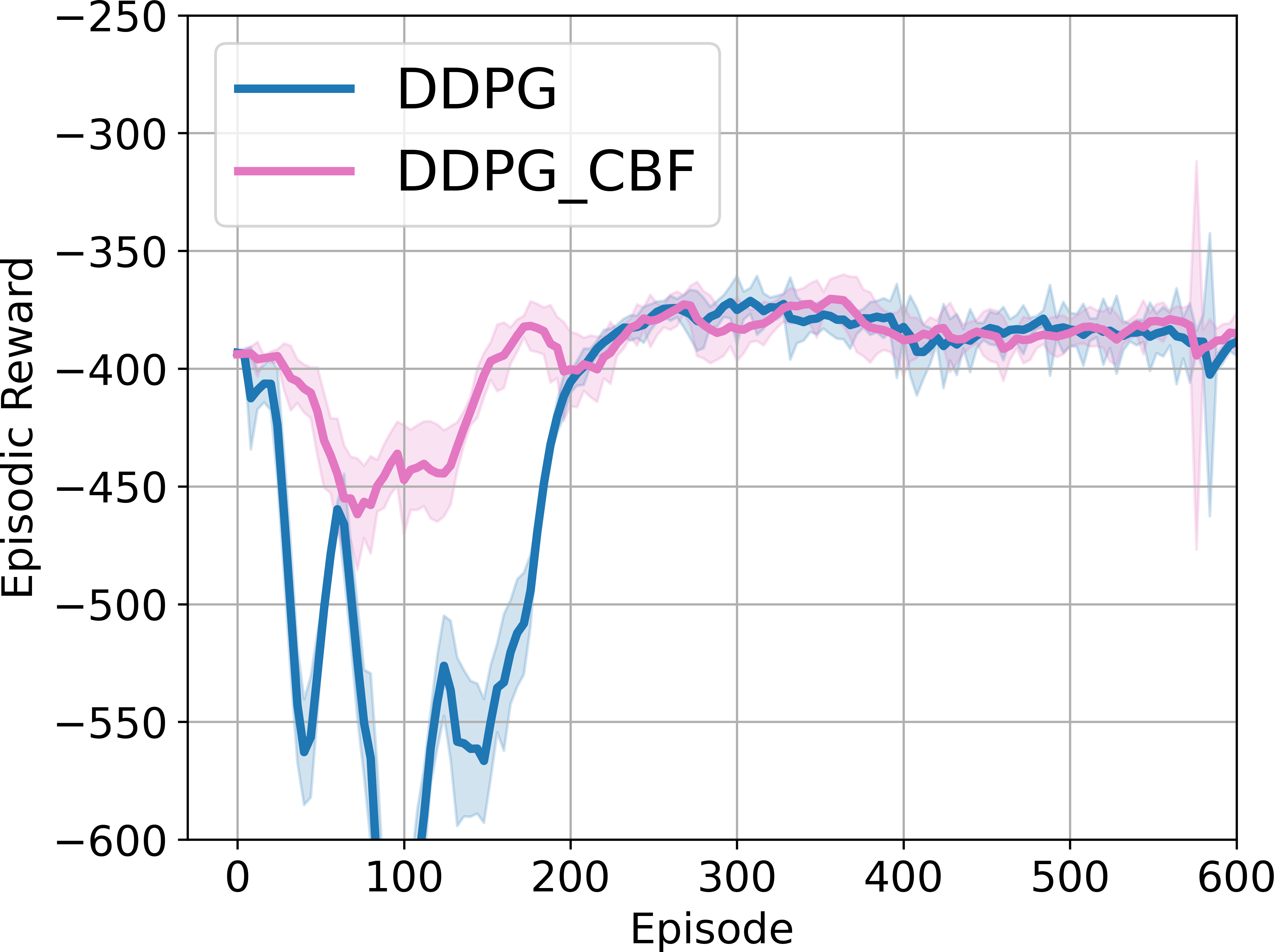}
    \caption{ \acrshort{ddpg}  and \acrshort{ddpg}-CBF}
    \label{fig:eps_training_ns3_ddpg}
        \end{subfigure}
        \begin{subfigure}[b]{0.44\textwidth}  
            \centering
    \includegraphics[width=\textwidth]{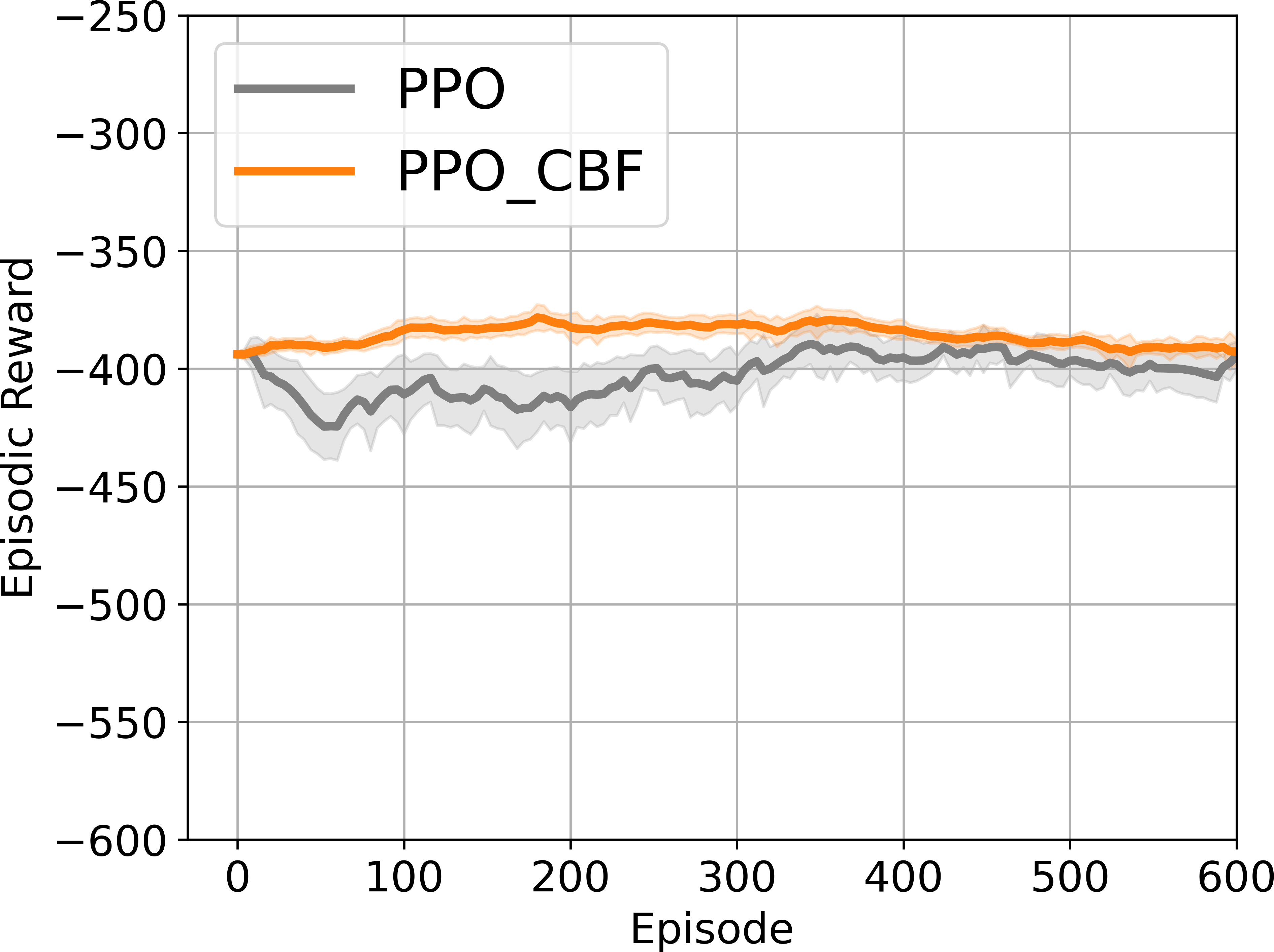}
    \caption{\acrshort{ppo}  and \acrshort{ppo}-CBF}
    \label{fig:eps_training_ns3_ppo}
        \end{subfigure}
        \caption{Episodic training reward} 
        \label{fig:ns3_eps_rew}
\end{figure}

Figures \ref{fig:eps_training_ns3_ddpg} and \ref{fig:eps_training_ns3_ppo} illustrate the episodic training reward of \acrshort{ddpg}, \acrshort{ddpg}-\acrshort{cbf}, \acrshort{ppo}  and \acrshort{ppo}-\acrshort{cbf}, respectively. In general, \acrshort{ddpg}-\acrshort{cbf} and \acrshort{ppo}-\acrshort{cbf} reach stable convergence and higher training rewards than their non-safe versions. Particularly, the \acrshort{ddpg}  algorithm does not behave well at the start of the training progress when the episodic reward significantly drops before uprising at episode 100. This is not seen in \acrshort{ddpg}-\acrshort{cbf} and is achieved thanks to the inclusion of the \acrshort{mlu} part in the reward (Equation \ref{eq:rew}), which is heuristically improved by the \acrshort{cbf} function.

\begin{figure}[ht!]
        \centering
    \begin{subfigure}[b]{0.44\textwidth} 
    \includegraphics[width=\textwidth]{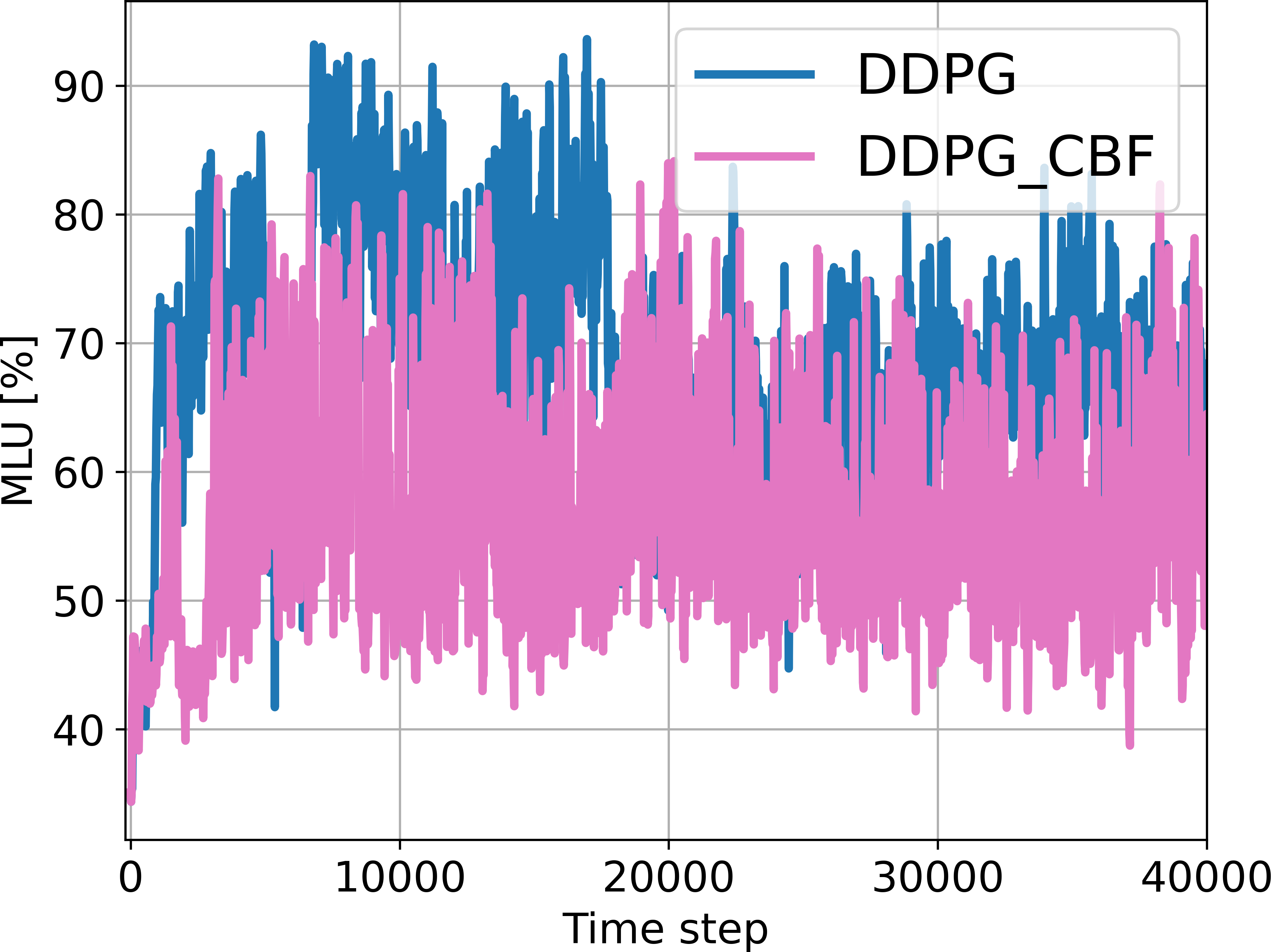}
    \caption{ \acrshort{ddpg}  and \acrshort{ddpg}-CBF}
    \label{fig:mlu_training_ns3_ddpg}
        \end{subfigure}
        \begin{subfigure}[b]{0.44\textwidth}  
            \centering
    \includegraphics[width=\textwidth]{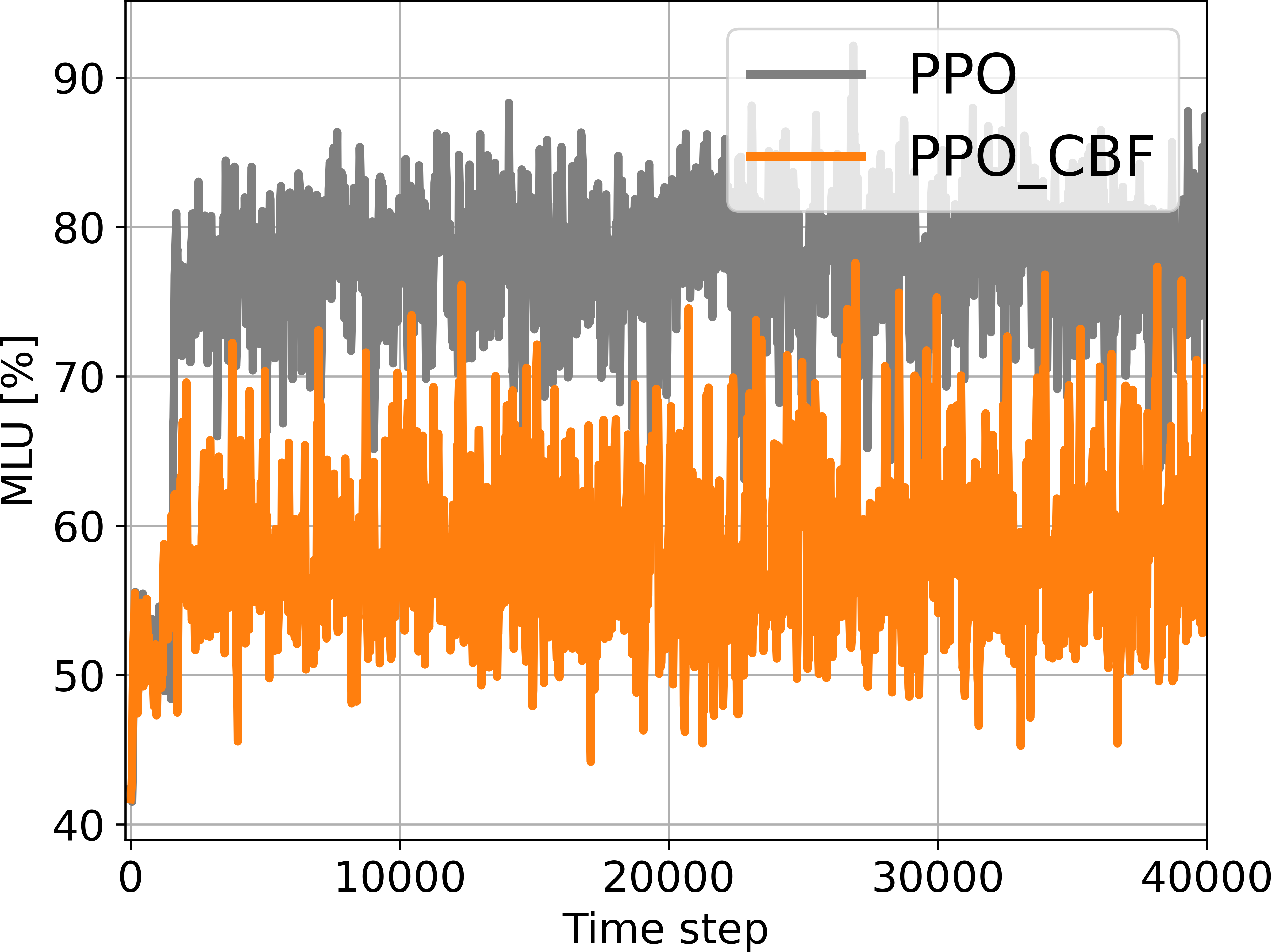}
    \caption{\acrshort{ppo}  and \acrshort{ppo}-CBF}
    \label{fig:mlu_training_ns3_ppo}
        \end{subfigure}
        \caption{\acrshort{mlu} during training} 
        \label{fig:mlu_training}
\end{figure}

Concerning \acrshort{mlu} during training, Figure \ref{fig:mlu_training} demonstrates how \acrshort{mlu} is improved when using \acrshort{cbf} function in addition to the learning algorithms (i.e., \acrshort{ddpg}  and \acrshort{ppo}). As shown in Figures \ref{fig:mlu_training_ns3_ddpg} and \ref{fig:mlu_training_ns3_ppo}, \acrshort{ddpg}-\acrshort{cbf} and \acrshort{ppo}-\acrshort{cbf} attain a lower \acrshort{mlu} during training compared to their non-safety version, respectively. It means that our \acrshort{cbf} effectively corrects the \acrshort{mlu} during training and makes learning safer for the networking system as link congestion is hardly seen.

\begin{figure}[ht!]
        \centering
    \begin{subfigure}[b]{0.44\textwidth} 
    \includegraphics[width=\textwidth]{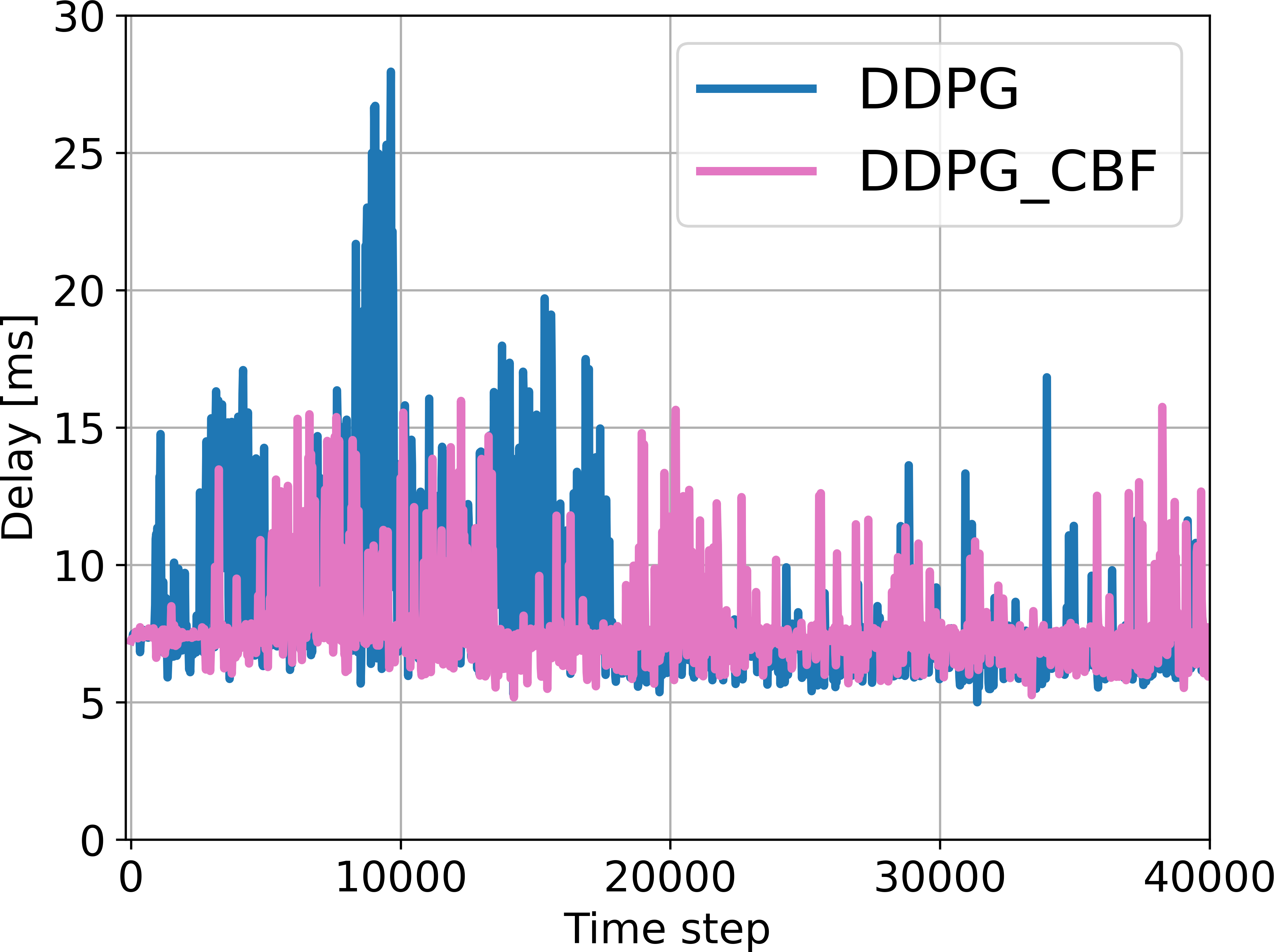}
    \caption{ \acrshort{ddpg}  and \acrshort{ddpg}-CBF}
    \label{fig:delay_training_ns3_ddpg}
        \end{subfigure}
        \begin{subfigure}[b]{0.44\textwidth}  
            \centering
    \includegraphics[width=\textwidth]{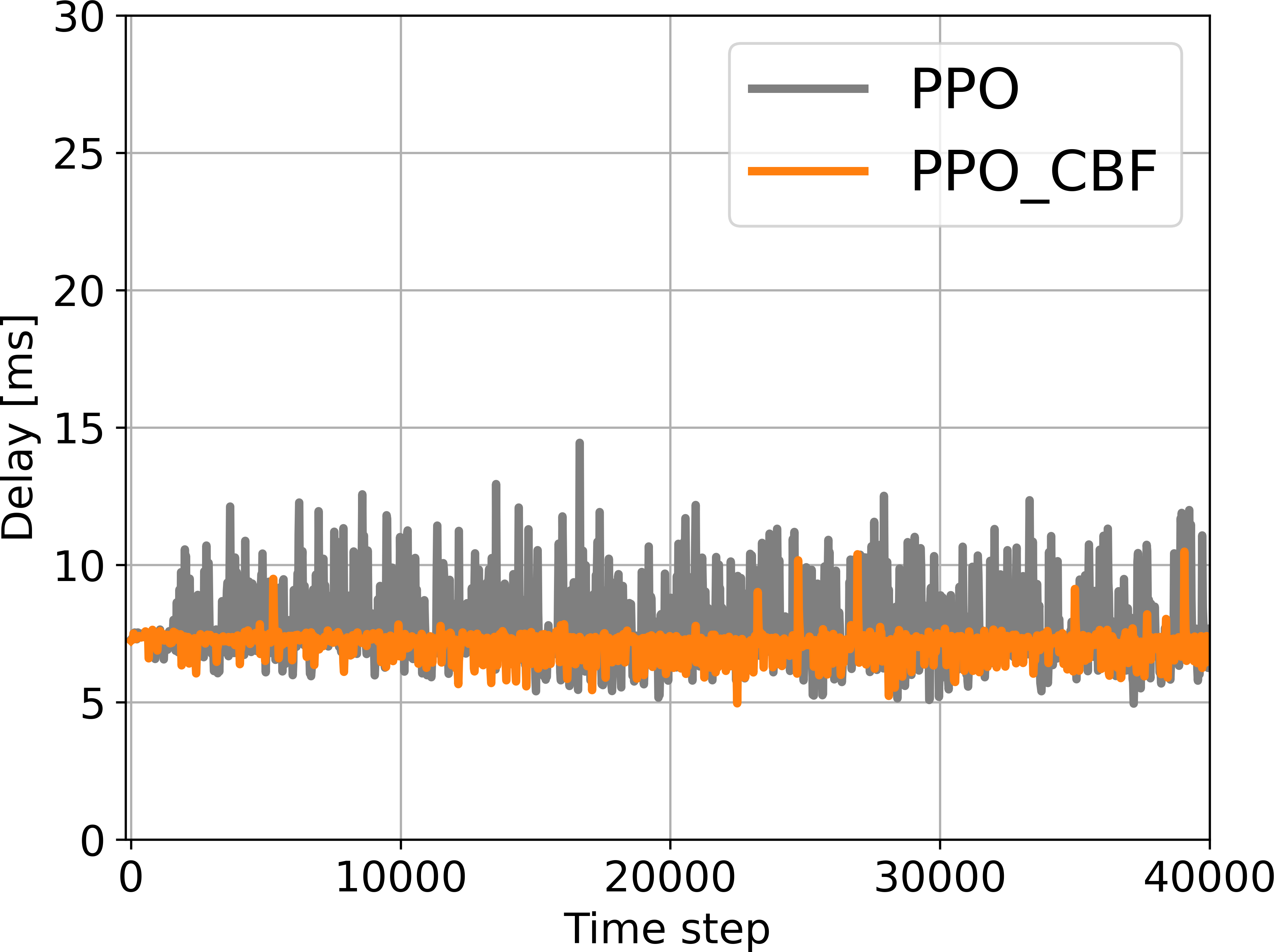}
    \caption{\acrshort{ppo}  and \acrshort{ppo}-CBF}
    \label{fig:delay_training_ns3_ppo}
        \end{subfigure}
        \caption{Delay during training} 
        \label{fig:delay_training}
\end{figure}

Then, Figure \ref{fig:delay_training} displays the delay of the learning algorithms during training. As we can see, \acrshort{ddpg}   
results in a more noisy and high delay during training, especially at the beginning of training. The extremely high delay is caused by the link congestion (i.e., high MLU) as evidenced in Figure \ref{fig:mlu_training}. 

\subsubsection{Testing Performance}

\begin{figure}[ht!]
        \centering
        \begin{subfigure}[b]{0.45\textwidth}
            \centering
  \includegraphics[width=\textwidth]{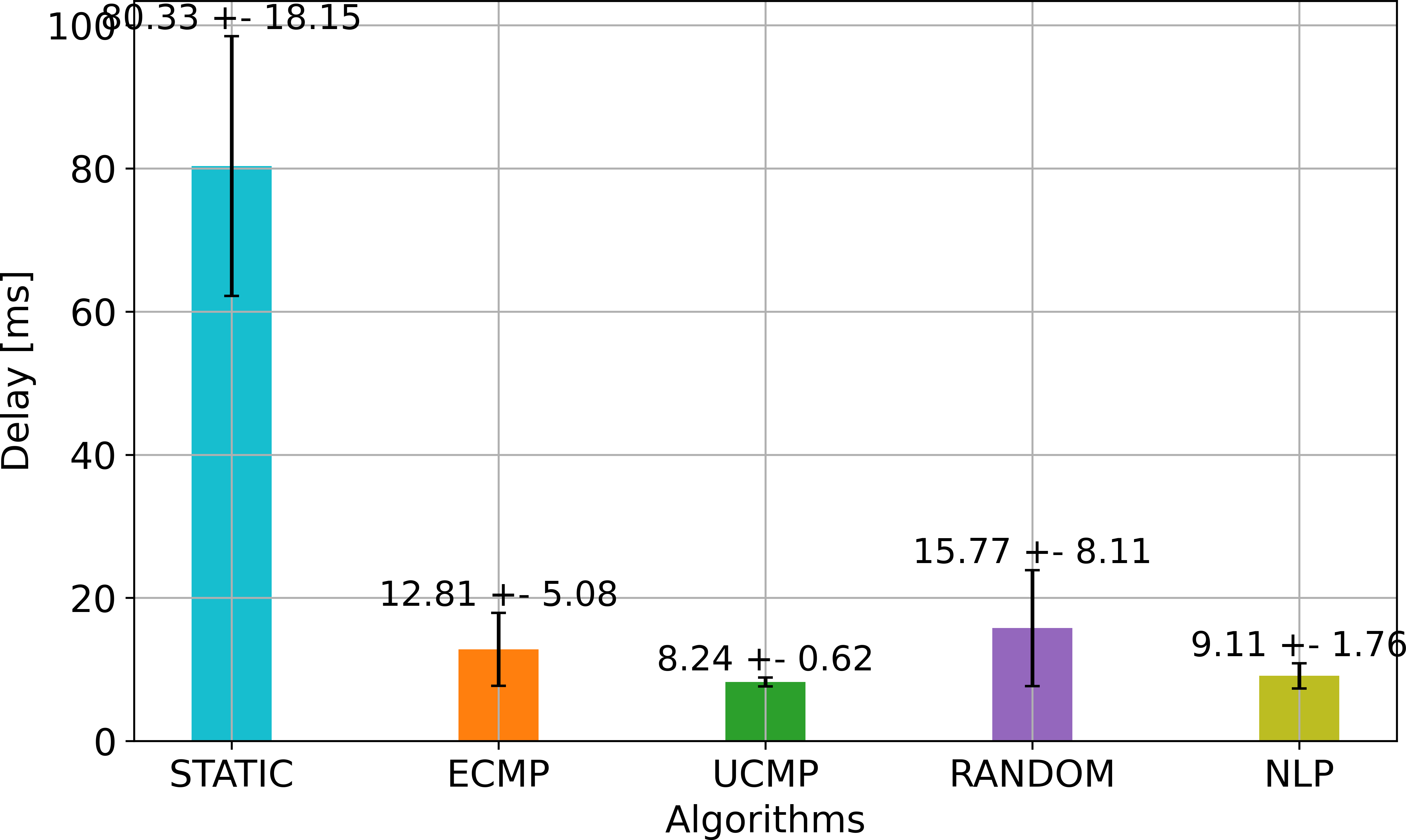}
  \caption{Average delay}
  \label{fig:ave_b_delay}
        \end{subfigure}
        \begin{subfigure}[b]{0.45\textwidth}  
            \centering
  \includegraphics[width=\textwidth]{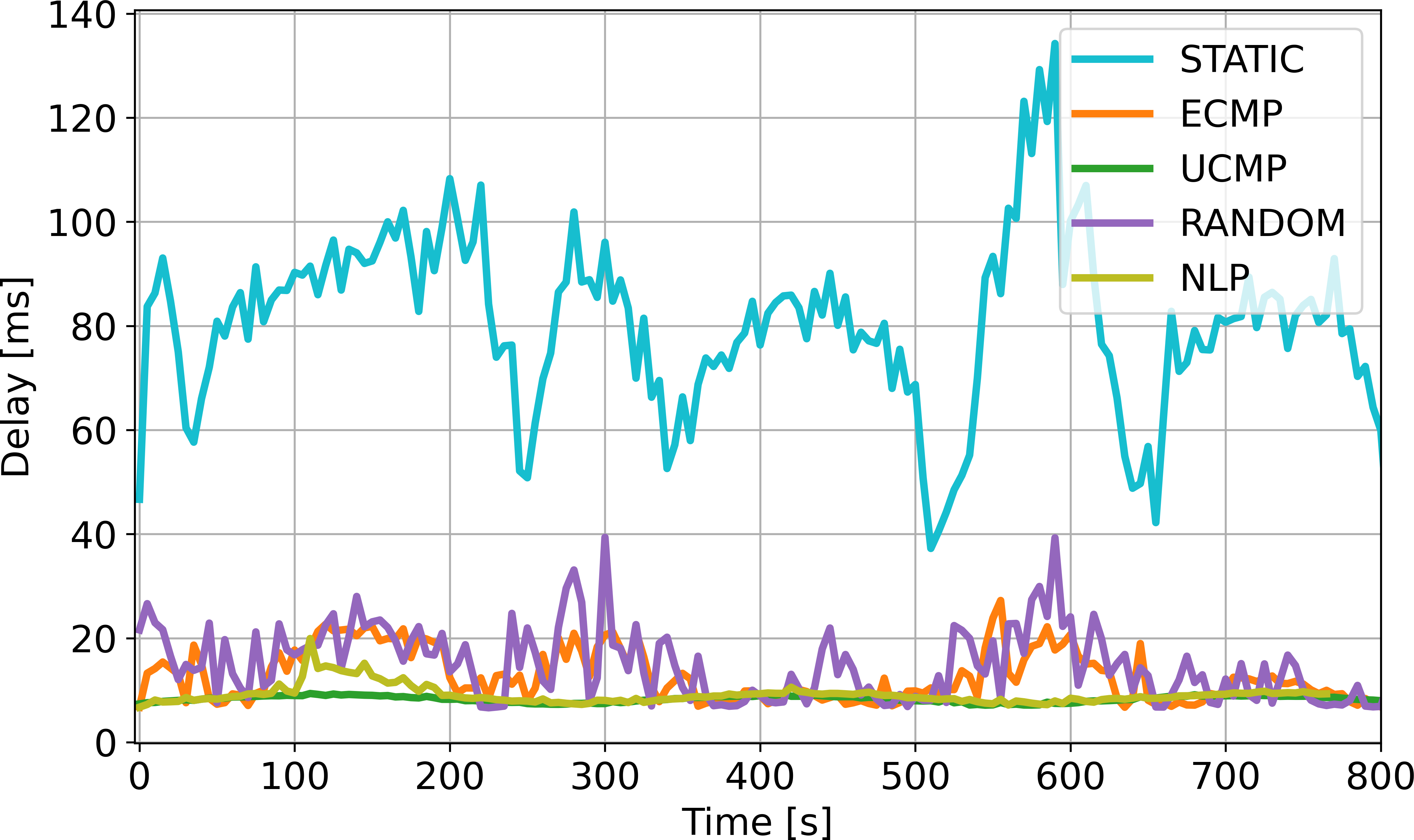}
  \caption{Delay evolution}
  \label{fig:temp_b_delay}
        \end{subfigure}
        \caption{Average delay (a) and delay evolution (b) of the baselines} 
        \label{fig:bench_delay}
\end{figure}
\begin{figure}[ht!]
        \centering
        \begin{subfigure}[b]{0.45\textwidth}
            \centering
  \includegraphics[width=\textwidth]{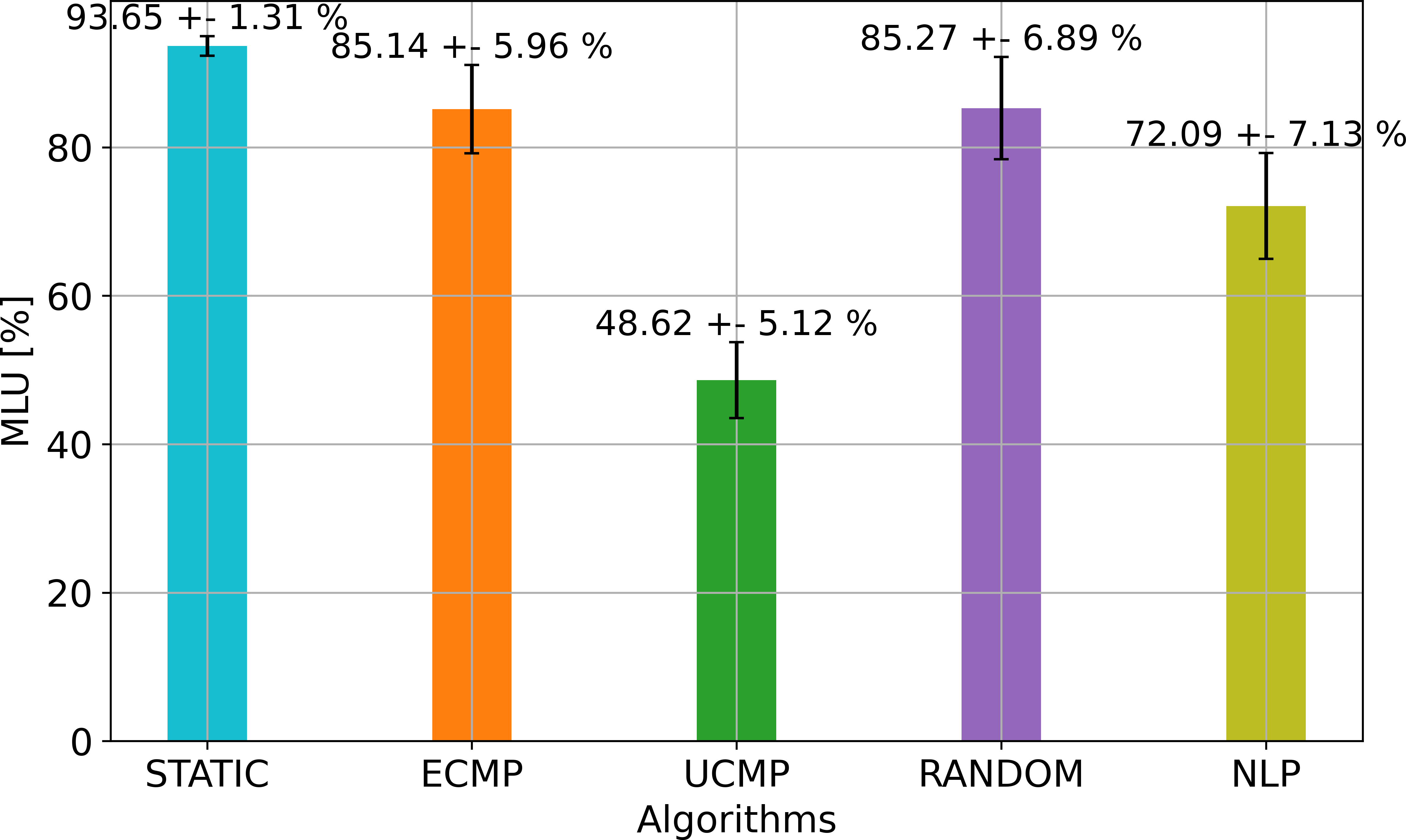}
  \caption{Average \acrshort{mlu}}
  \label{fig:ave_b_mlu}
        \end{subfigure}
        \begin{subfigure}[b]{0.45\textwidth}  
            \centering
  \includegraphics[width=\textwidth]{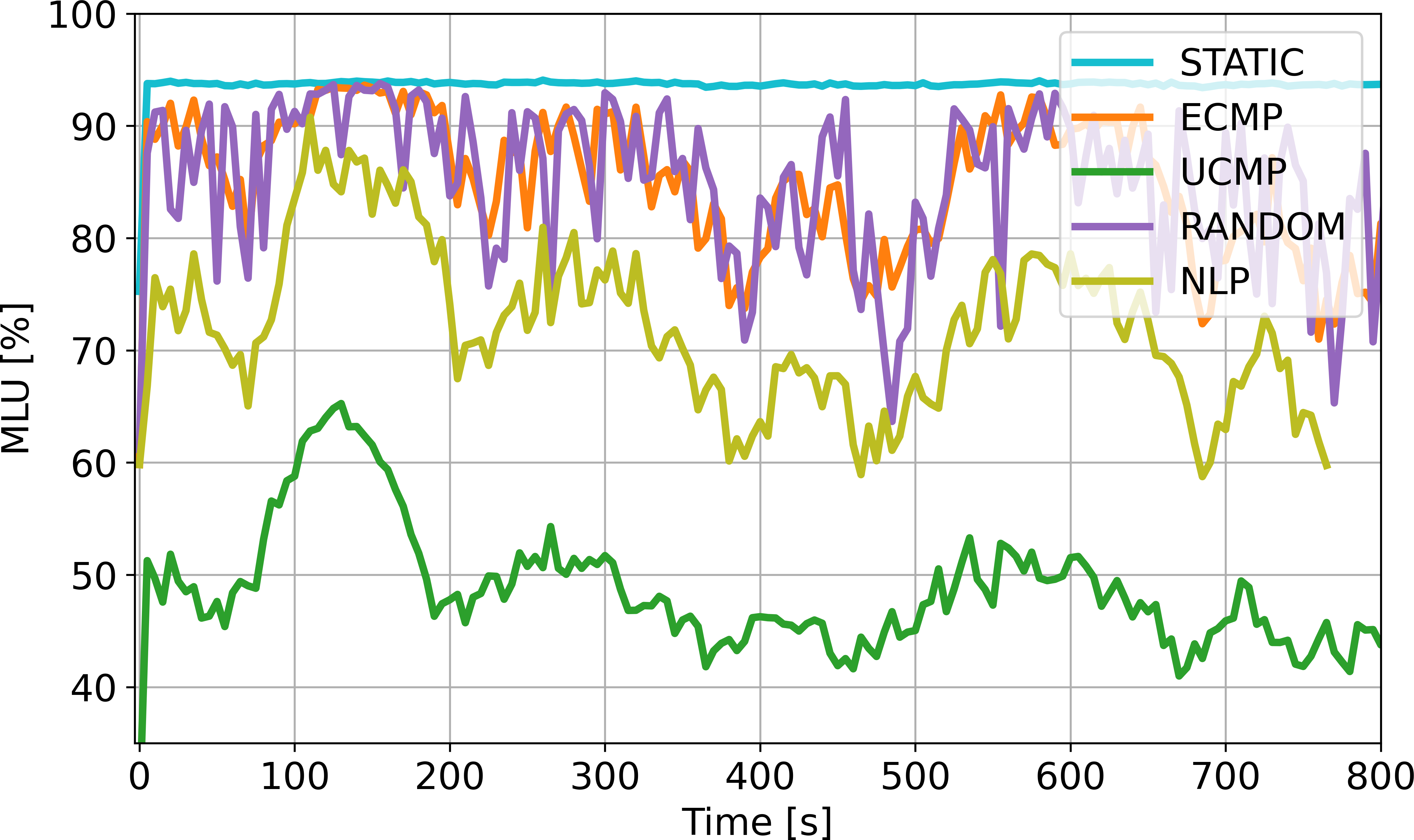}
  \caption{\acrshort{mlu} evolution}
  \label{fig:temp_b_mlu}
        \end{subfigure}
        \caption{Average \acrshort{mlu} (a) and \acrshort{mlu} evolution (b) of the baselines} 
        \label{fig:bench_mlu}
\end{figure}

\begin{figure}[ht!]
        \centering
        \begin{subfigure}[b]{0.45\textwidth}
            \centering
  \includegraphics[width=\textwidth]{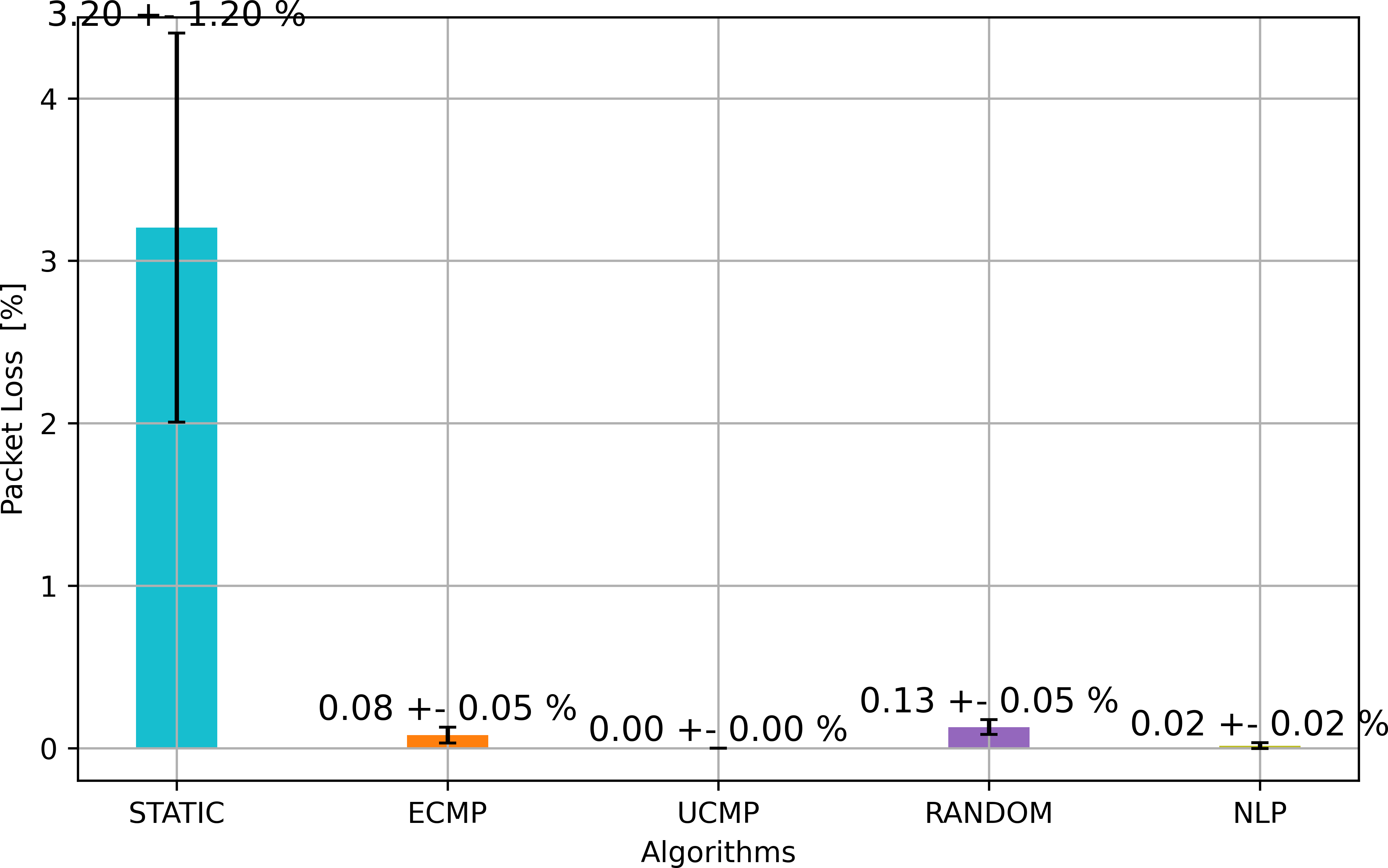}
  \caption{Average PKLoss}
  \label{fig:ave_b_pkl}
        \end{subfigure}
        \begin{subfigure}[b]{0.45\textwidth}  
            \centering
  \includegraphics[width=\textwidth]{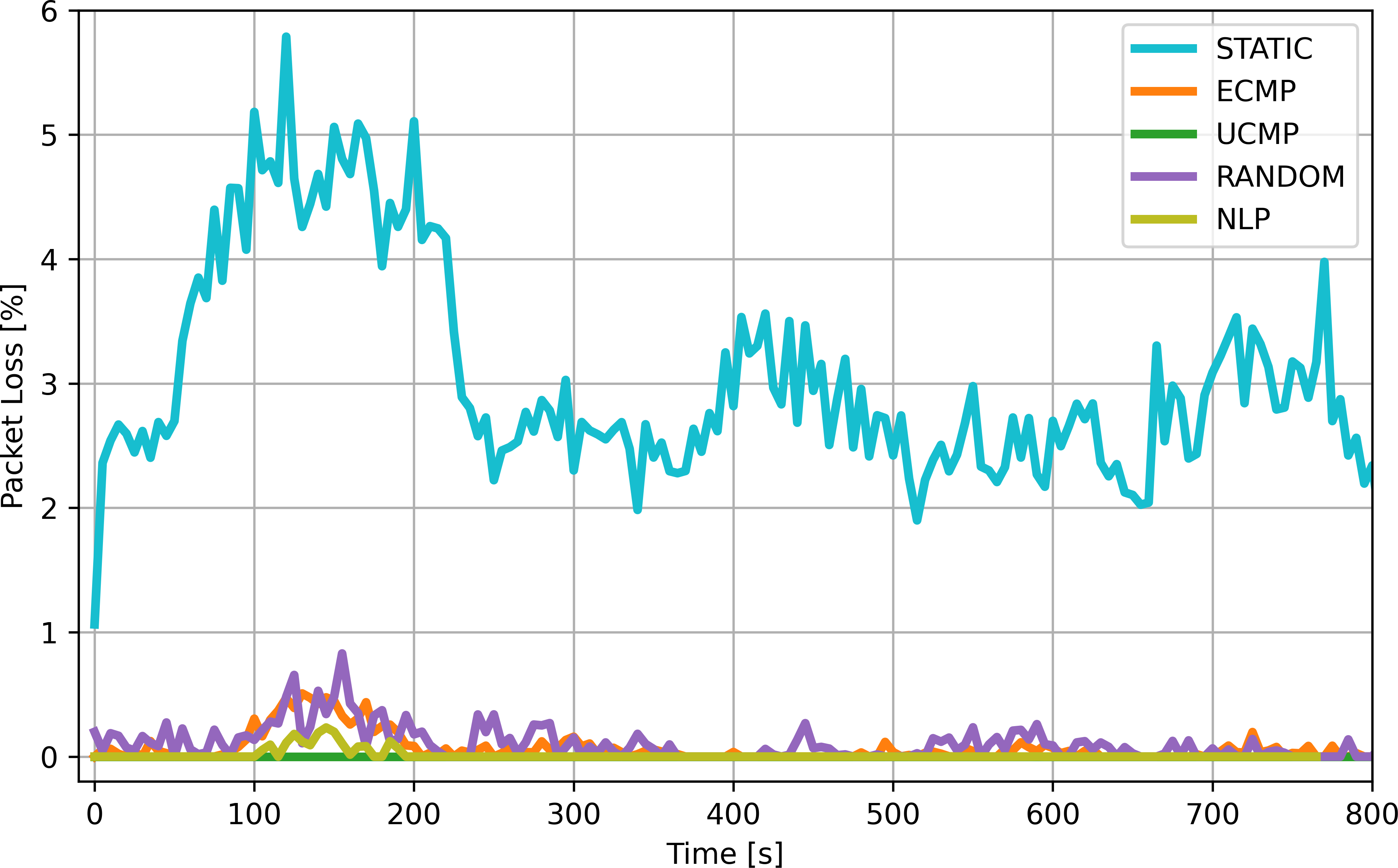}
  \caption{PKLoss evolution}
  \label{fig:temp_b_pkl}
        \end{subfigure}
        \caption{Average (a) and PKLoss evolution (b) of the baselines} 
        \label{fig:bench_pkl}
\end{figure}

Figures \ref{fig:bench_delay}, \ref{fig:bench_mlu} and \ref{fig:bench_pkl} compare the delay, \acrshort{mlu}, and packet loss performance of the baselines in our packet-based simulation. When load-balancing is not applied (\textbf{STATIC}), Figure \ref{fig:temp_b_mlu} suggests that traffic on a single path always causes congestion. As a consequence, delay and packet loss are worse compared to the other benchmarks, and load-balancing is needed to attain better \acrshort{qos} and \acrshort{mlu}. \textbf{\acrshort{ecmp}} and \textbf{RANDOM} slightly increase performance when traffic is equally distributed over available paths. Although the issue related to link saturation and packet loss is partly resolved, 50$\%$ of the traffic load is sent to a path where link capacity is limited, which results in a higher \acrshort{mlu} overall. \textbf{\acrshort{ucmp}} load-balancing significantly reduces \acrshort{mlu} by proportionally splitting  traffic according to the path capacity. In this method, higher traffic is steered in the path with higher bandwidth, but it may induce a higher end-to-end delay. Using \textbf{\acrshort{nlp}}, we can observe that delay and packet loss are nearly identical to \acrshort{ucmp}, although \acrshort{mlu} is higher. It is due to the fact that it prioritizes paths with lower propagation delays. 
It implies that there is a better load-balancing policy than \acrshort{ucmp}, which can further reduce delay. Hereafter, we will use \acrshort{ucmp} and \acrshort{nlp} solutions as the main benchmarks for comparison with our learning-based algorithms.


\begin{figure}[ht!]
        \centering
        \begin{subfigure}[b]{0.43\textwidth}
            \centering
  \includegraphics[width=\textwidth]{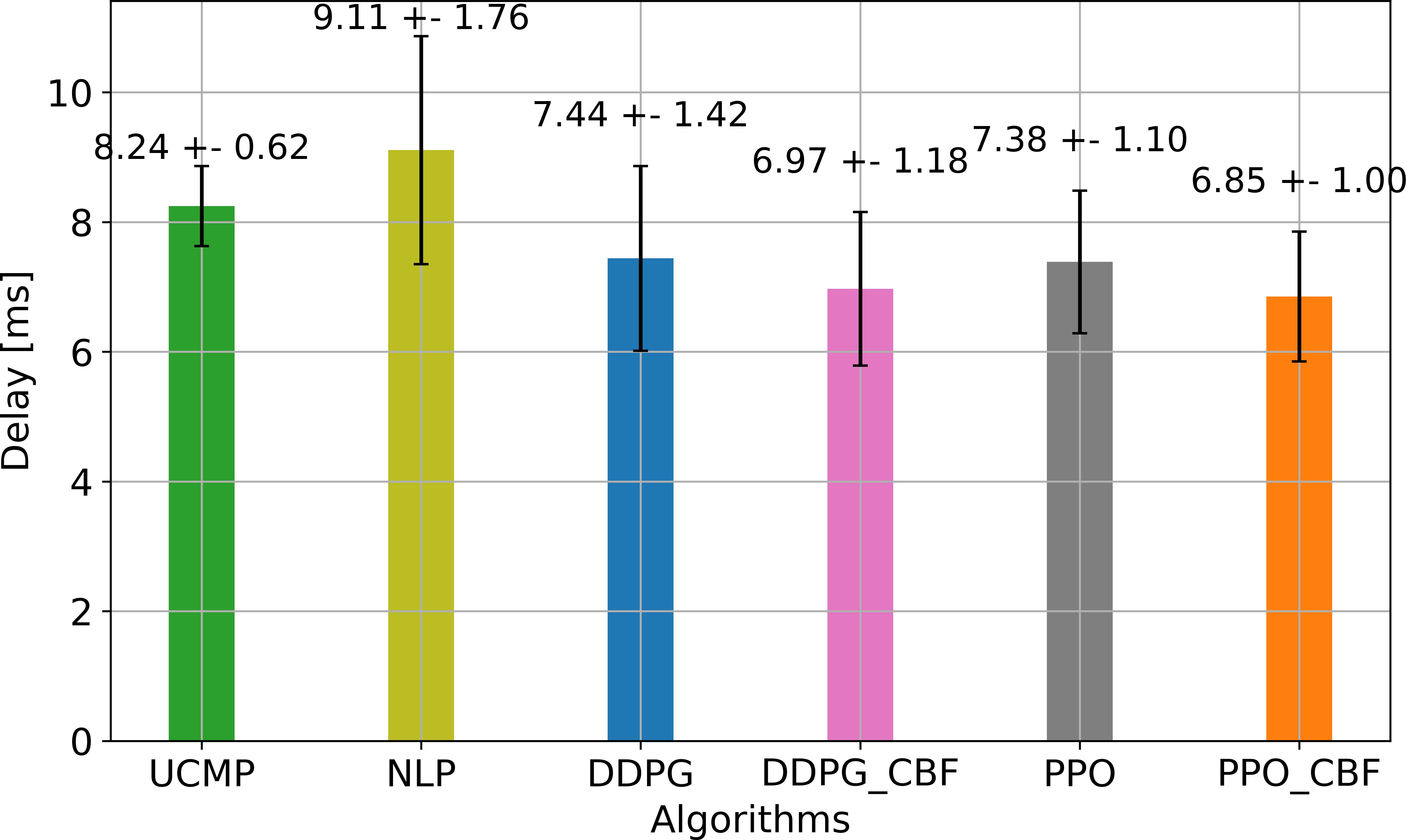}
  \caption{Average Delay}
  \label{fig:b_del_dp}
        \end{subfigure}
        \begin{subfigure}[b]{0.43\textwidth}  
            \centering
  \includegraphics[width=\textwidth]{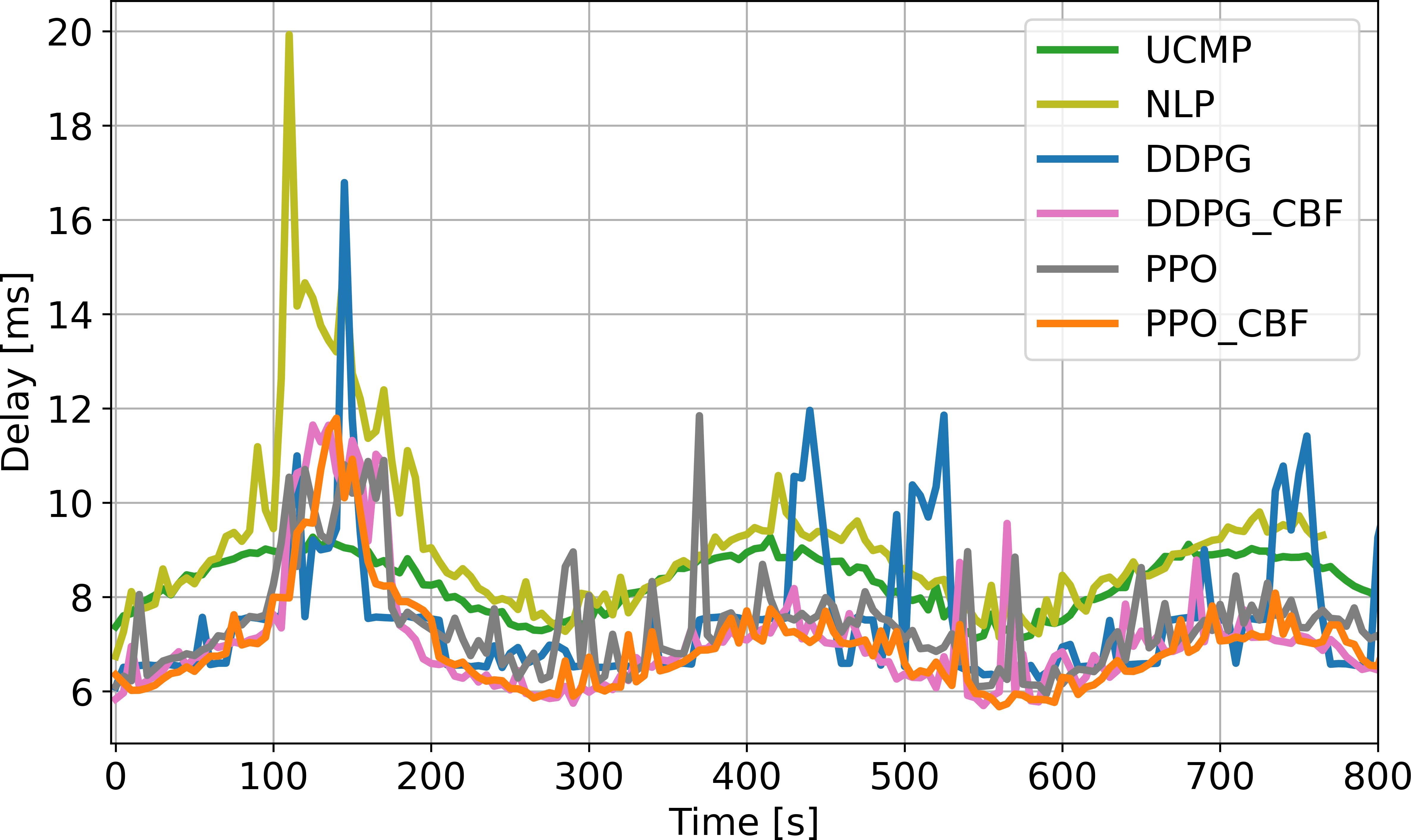}
  \caption{Delay evolution}
  \label{fig:t_del_dp}
        \end{subfigure}
        \caption{Average (a) and delay evolution (b) in testing.} 
        \label{fig:delay_learning}
\end{figure}

Figure \ref{fig:delay_learning} illustrates the average delay and  delay evolution of our algorithms with respect to the \acrshort{ucmp} and \acrshort{nlp} baselines during testing phase, which lasts 800 seconds. As we can see from Figure \ref{fig:b_del_dp}, both \acrshort{ddpg}-\acrshort{cbf} and \acrshort{ppo}-\acrshort{cbf} obtain better end-to-end delays, which are around 6.97 ms and 6.73 ms, respectively. Without CBF, the \acrshort{ppo}  and \acrshort{ddpg}  models reach a slightly higher delay compared with their CBF versions, because they caused high \acrshort{mlu} in some cases and accidentally increase queuing delay. Effectively, our learning algorithms outperform \acrshort{ucmp} and \acrshort{nlp} in terms of end-to-end delay. 

\begin{figure}[ht!]
        \centering
        \begin{subfigure}[b]{0.43\textwidth}
            \centering
  \includegraphics[width=\textwidth]{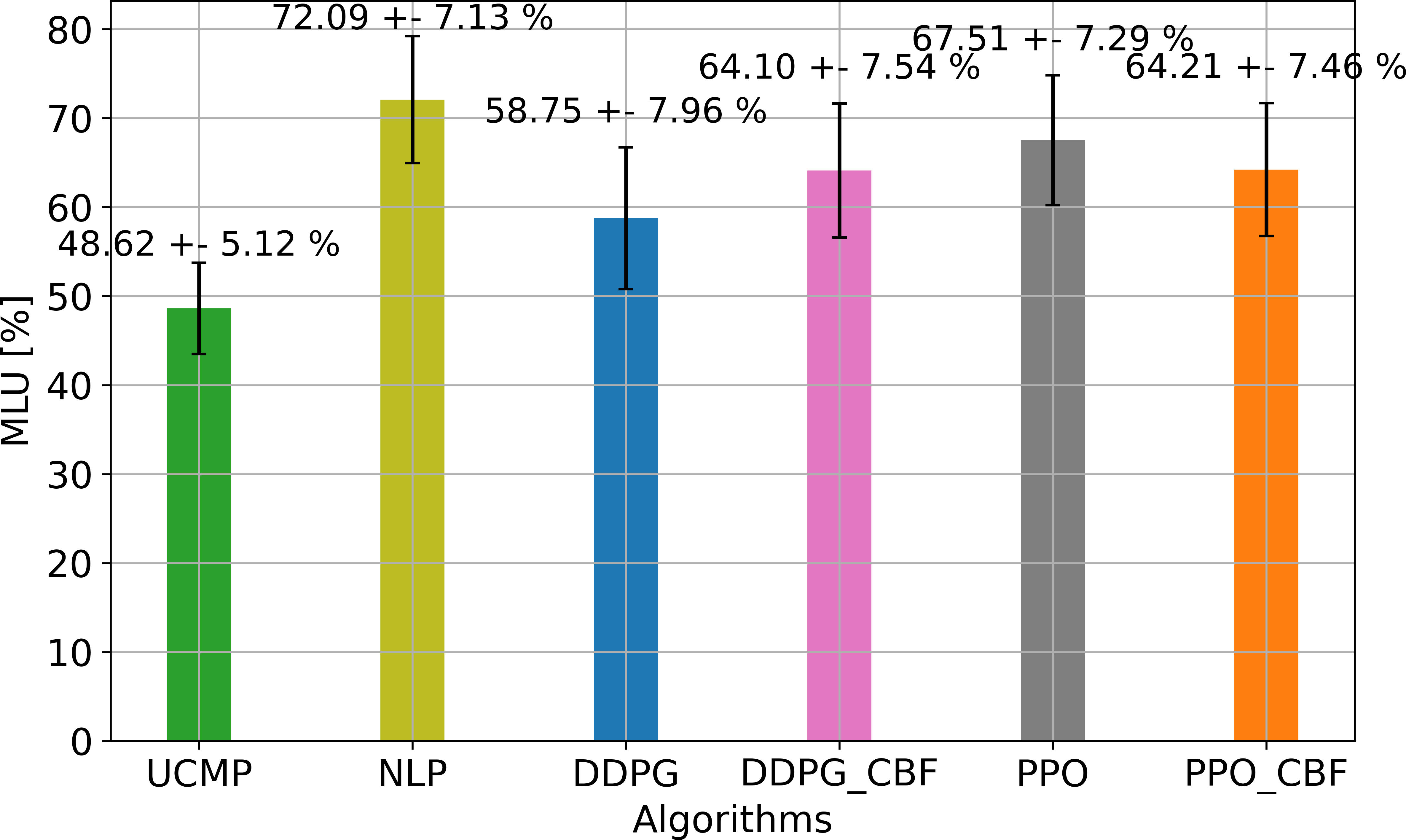}
  \caption{Average \acrshort{mlu}}
  \label{fig:b_mlu_dp}
        \end{subfigure}
        \begin{subfigure}[b]{0.43\textwidth}  
            \centering
  \includegraphics[width=\textwidth]{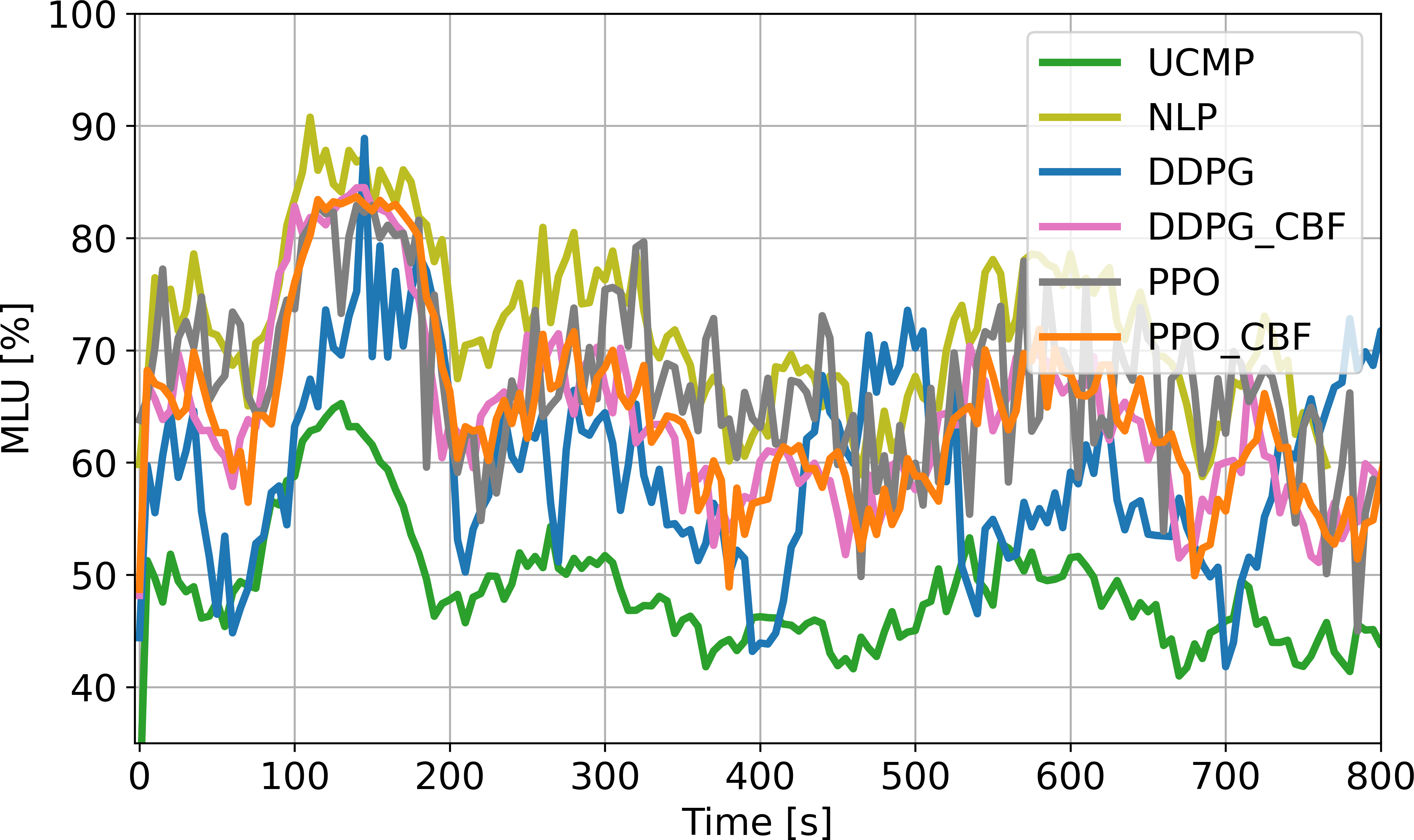}
  \caption{\acrshort{mlu} evolution}
  \label{fig:t_mlu_dp}
        \end{subfigure}
        \caption{ Average (a) and evolution (b) \acrshort{mlu} in testing. } 
        \label{fig:mlu_testing}
\end{figure}

The better delay performance can be explained by observing their \acrshort{mlu} during the testing phase. As mentioned above, better policies than \acrshort{ucmp} and \acrshort{nlp} should prioritize the path with a lower propagation delay and without overloading that path as \acrshort{nlp} to achieve a better delay. Figure \ref{fig:mlu_testing} shows that our learning algorithms achieve the \acrshort{qos} and safety objectives when the paths with high propagation delays are not prioritized and the paths with lower capacity are not overloaded. As shown in Figure \ref{fig:b_mlu_dp}, \acrshort{ucmp} reaches the lowest \acrshort{mlu} as most of the traffic is sent to the path with higher capacity and higher propagation delay. On the other hand, \acrshort{nlp} steers traffic on a path with lower propagation delay and limited capacity, making queuing delay dominated. Our proposals properly orchestrate the traffic on the available paths to (1) avoid the path with higher propagation delay and (2) maintain a good level of queuing delay in the path with limited capacity.  This explains why the learning algorithms reach better performance than \acrshort{ucmp} and \acrshort{nlp}.

\begin{figure}[ht!]
        \centering
        \begin{subfigure}[b]{0.43\textwidth}
            \centering
  \includegraphics[width=\textwidth]{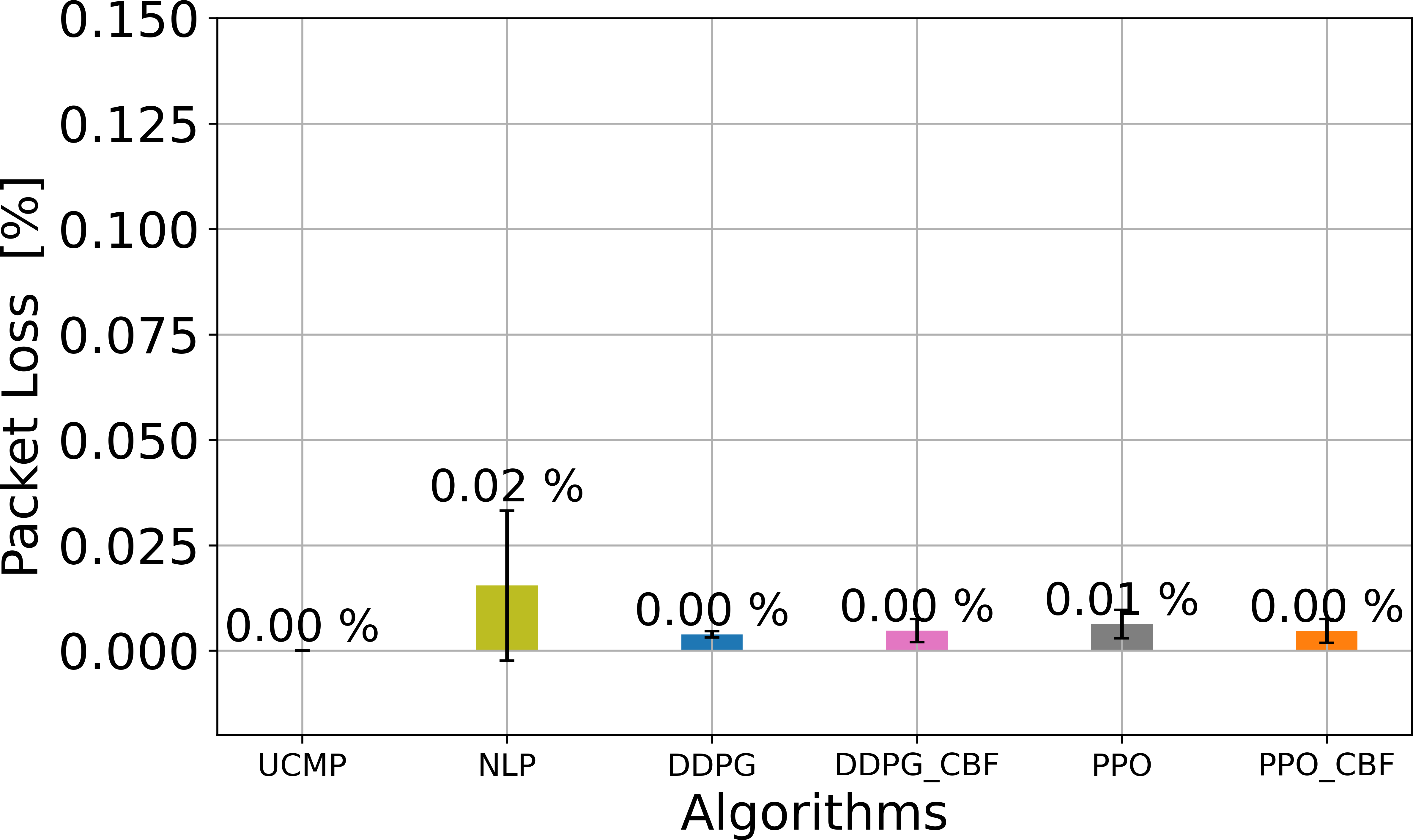}
  \caption{Average PKLoss}
  \label{fig:ave_pkl_t}
        \end{subfigure}
        \begin{subfigure}[b]{0.43\textwidth}  
            \centering
  \includegraphics[width=\textwidth]{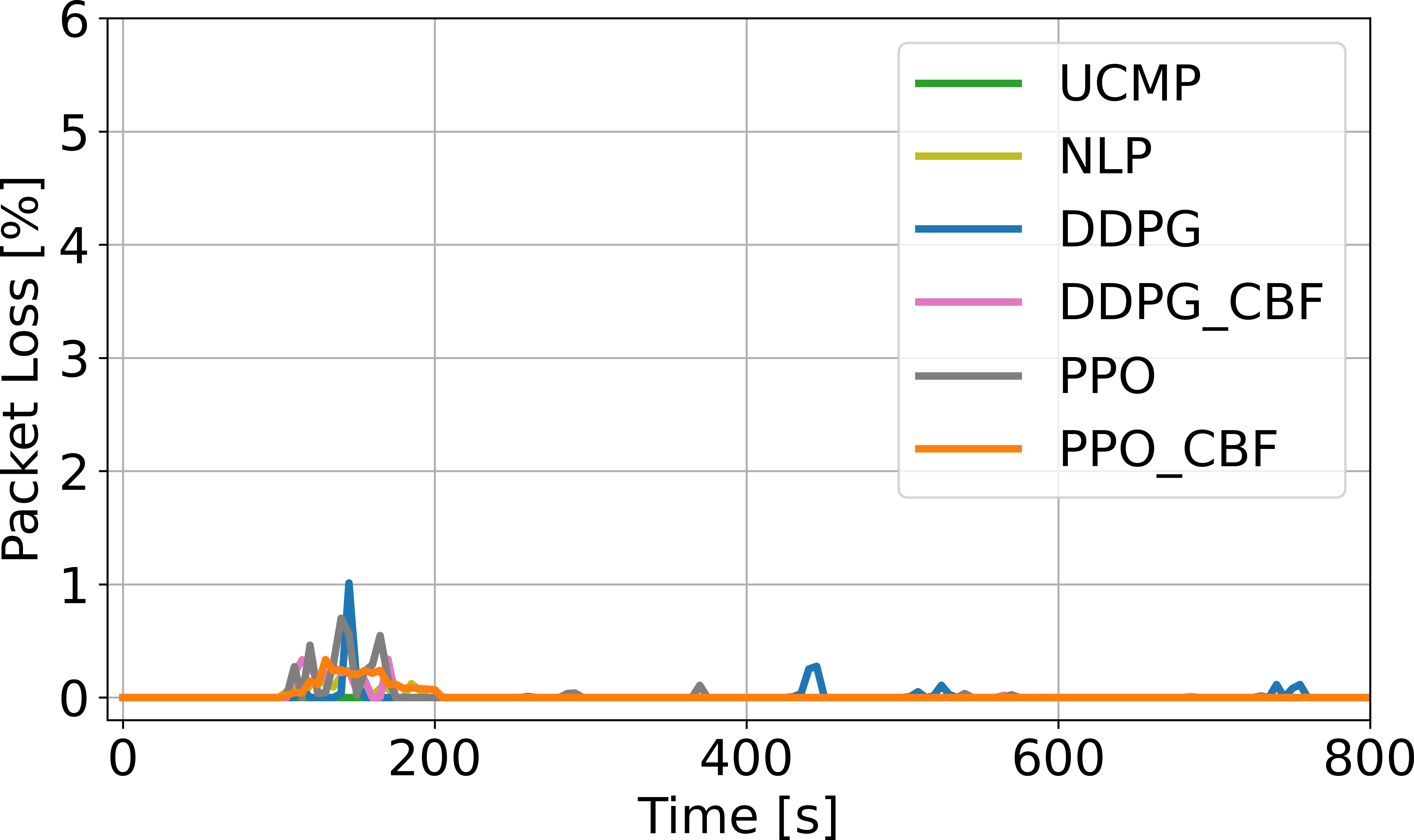}
  \caption{PKLoss evolution}
  \label{fig:te_pkl_t}
        \end{subfigure}
        \caption{Average (a) and evolution (b) PKLoss } 
        \label{fig:pkl_testing}
\end{figure}

Lastly, Figure \ref{fig:pkl_testing} shows the packet loss of our algorithms compared to  \acrshort{ucmp} and \acrshort{nlp}. Overall, all approaches reach very low packet loss during testing.

\subsection{Transfer learning from flow-based simulator to packet-based simulator}
So far, we have seen our simulation results in two independent simulators: flow-based and packet-based. 
In this section, we examine how models, which have been trained in a flow-based simulator, can also be applied to the packet-based simulator, using transfer learning \citep{zhuangComprehensiveSurveyTransfer2020}. Following this hypothesis, we consider the model that has been well trained in the flow-based simulator as a \textit{pre-trained model}. Afterward, this model will be fine-tuned in our packet-based simulator within a few episodes.

\subsubsection{Without fine-tuning}
In this study, we demonstrate how training models in the flow-based simulator behave in the packet-based simulator without fine-tuning. In this sense, those models will be used to perform load-balancing in an environment which is more complex than the one for which they have been trained.

\begin{figure}[ht!]
    \centering
    \includegraphics[scale=0.55]{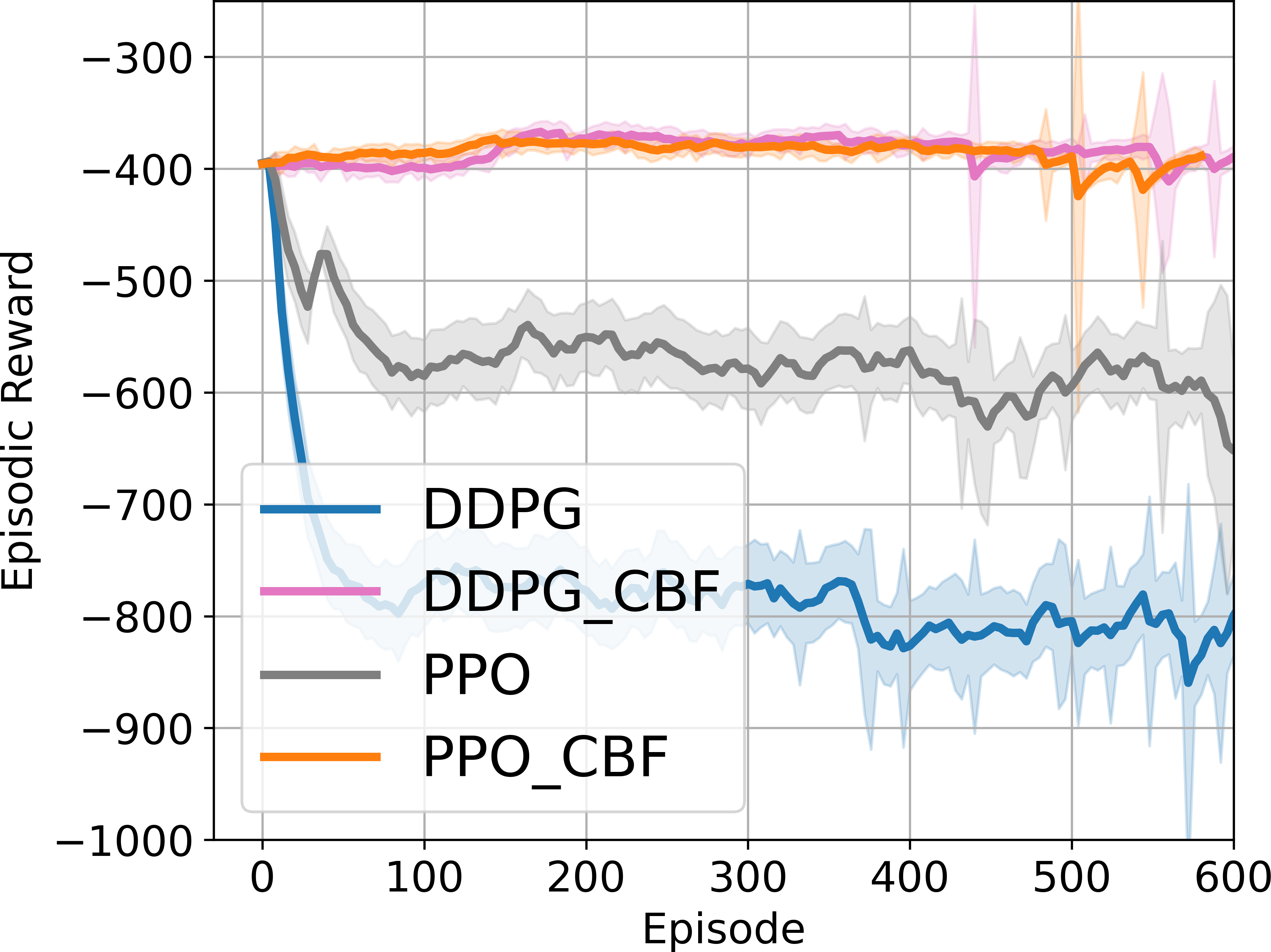}
    \caption{Episodic rewards on packet-level simulation using pre-trained models without fine-tuning.}
    \label{fig:ns3_epstrain_mt}
\end{figure}

\begin{figure}[ht!]
        \centering
        \begin{subfigure}[b]{0.43\textwidth}
            \centering
  \includegraphics[width=\textwidth]{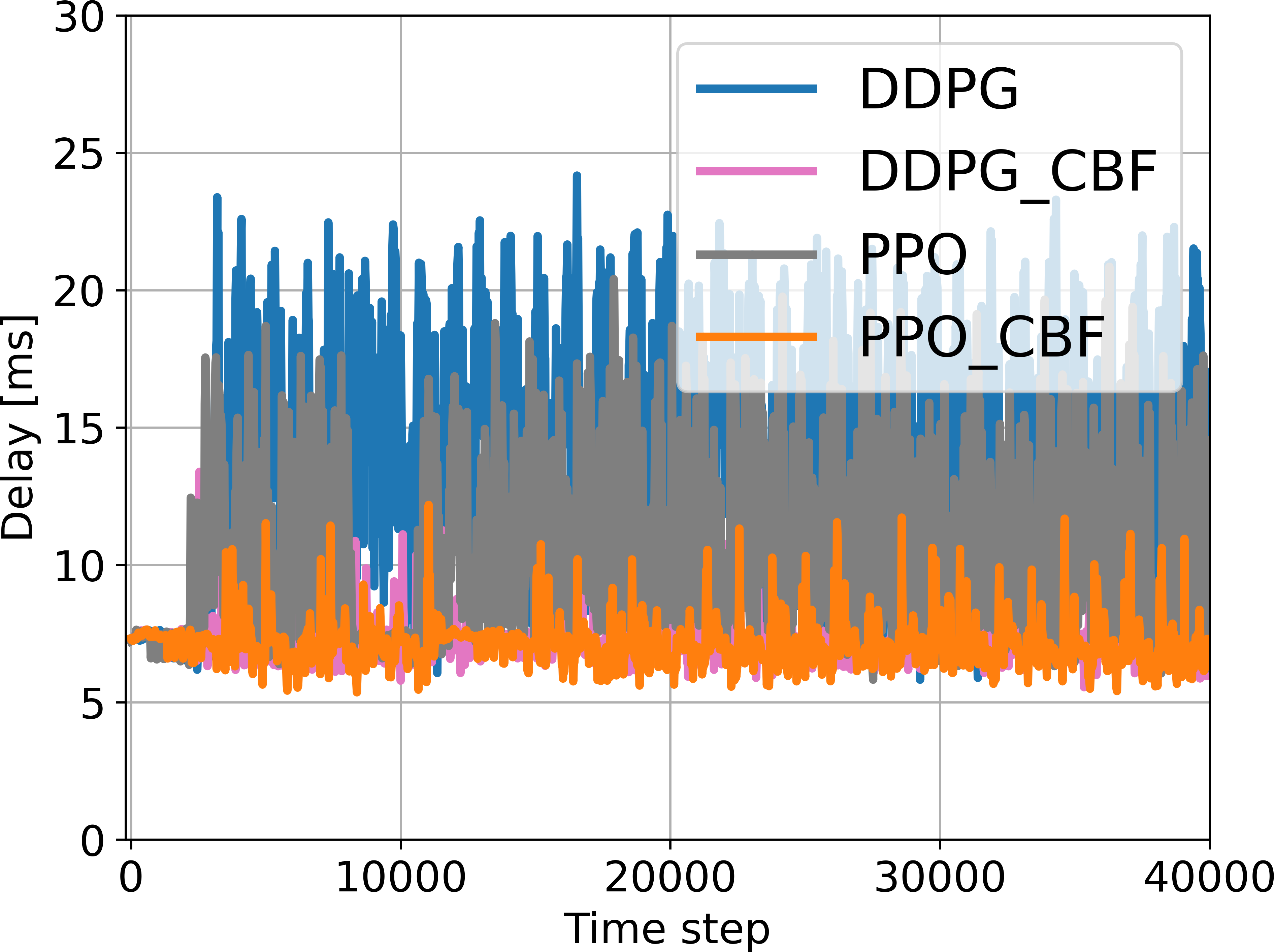}
  \caption{Delay evolution}
  \label{fig:ns3_deltrain_mt}
        \end{subfigure}
        \begin{subfigure}[b]{0.43\textwidth}  
            \centering
  \includegraphics[width=\textwidth]{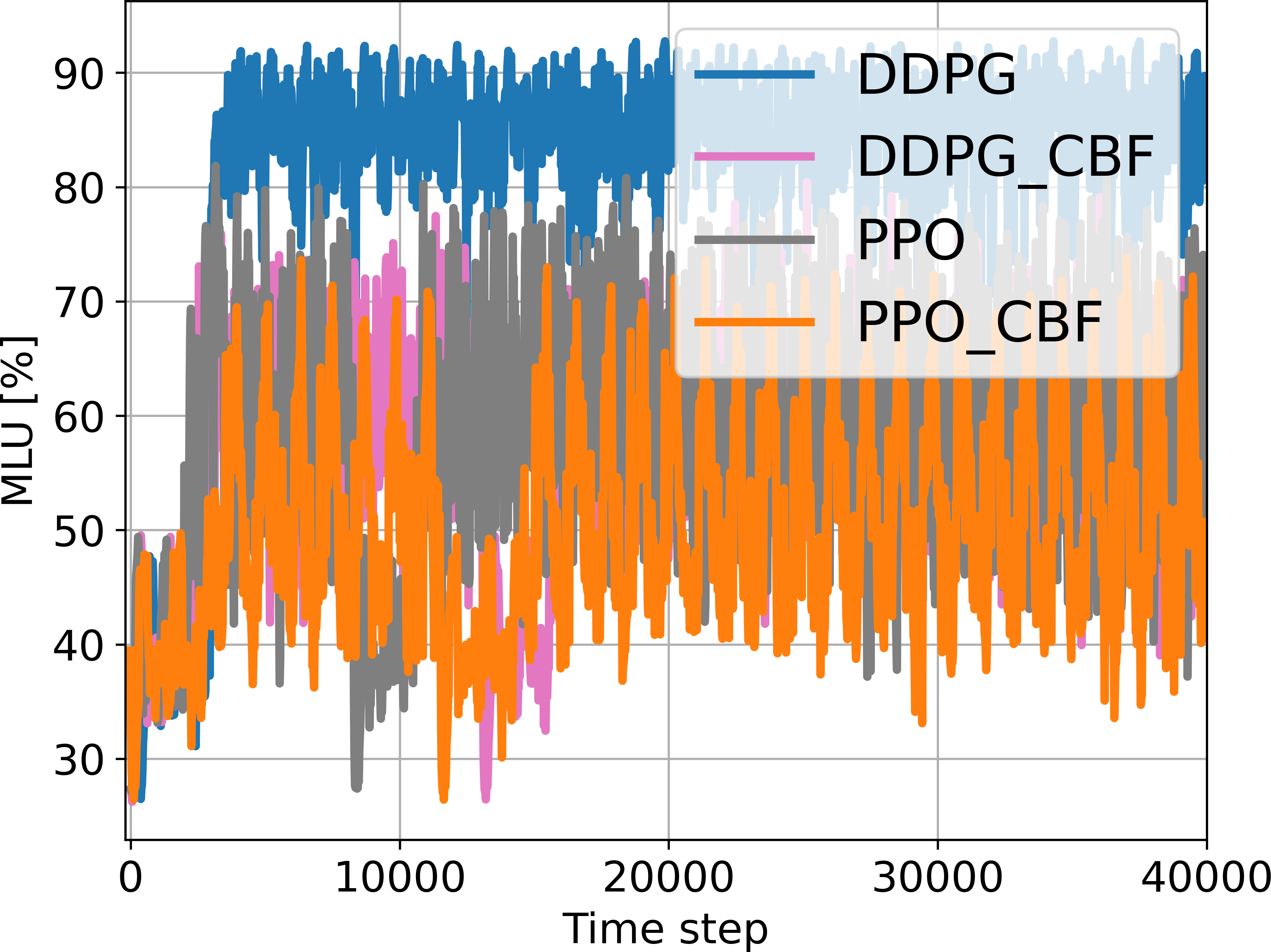}
  \caption{\acrshort{mlu} evolution}
  \label{fig:ns3_mlutrain_mt}
        \end{subfigure}
        \caption{Delay and \acrshort{mlu} measurements on packet-level simulations using pre-trained models without fine-tuning.} 
        \label{fig:ns3_sla_mt}
\end{figure}

Figure \ref{fig:ns3_epstrain_mt} displays the episodic rewards of pre-trained models on packet-level simulation without fine-tuning. As we can observe, algorithms with \acrshort{cbf} maintain stable rewards over episodes. It is worth noting that the models are not updated at different episodes because they are not fine-tuned in this case. Then, the \acrshort{cbf} function plays an instrumental role in keeping lower \acrshort{mlu} by favorably injecting traffic on the path with higher capacity and avoiding queuing delay upsurges. Although the obtained rewards are not optimal when they are compared to our regular training as shown in Figure \ref{fig:ns3_eps_rew}, it keeps stable rewards throughout the training process. When the \acrshort{cbf} function disappears, our learning algorithms experience lower rewards. Typically, \acrshort{ppo} rewards are better than \acrshort{ddpg}  rewards because the latter algorithm depends crucially on its replay buffer, whereas this is not the case for the former algorithm. It demonstrates that we can not barely use pre-trained models from our simple simulator in a more complex environment and that fine-tuning is a must to improve  performance.

Figure \ref{fig:ns3_sla_mt} confirms our observations on the rewards by showing the SLA performance (i.e., delay and \acrshort{mlu}) of those models in the packet-level simulation without fine-tuning. When the \acrshort{cbf} function is applied, a lower \acrshort{mlu} is obtained and delay measurements are lower than the models without CBF. These figures explain why our episodic rewards with \acrshort{cbf} function are better while delay and \acrshort{mlu} values are smaller. 
\begin{figure}[ht!]
    \centering
    \includegraphics[scale=0.5]{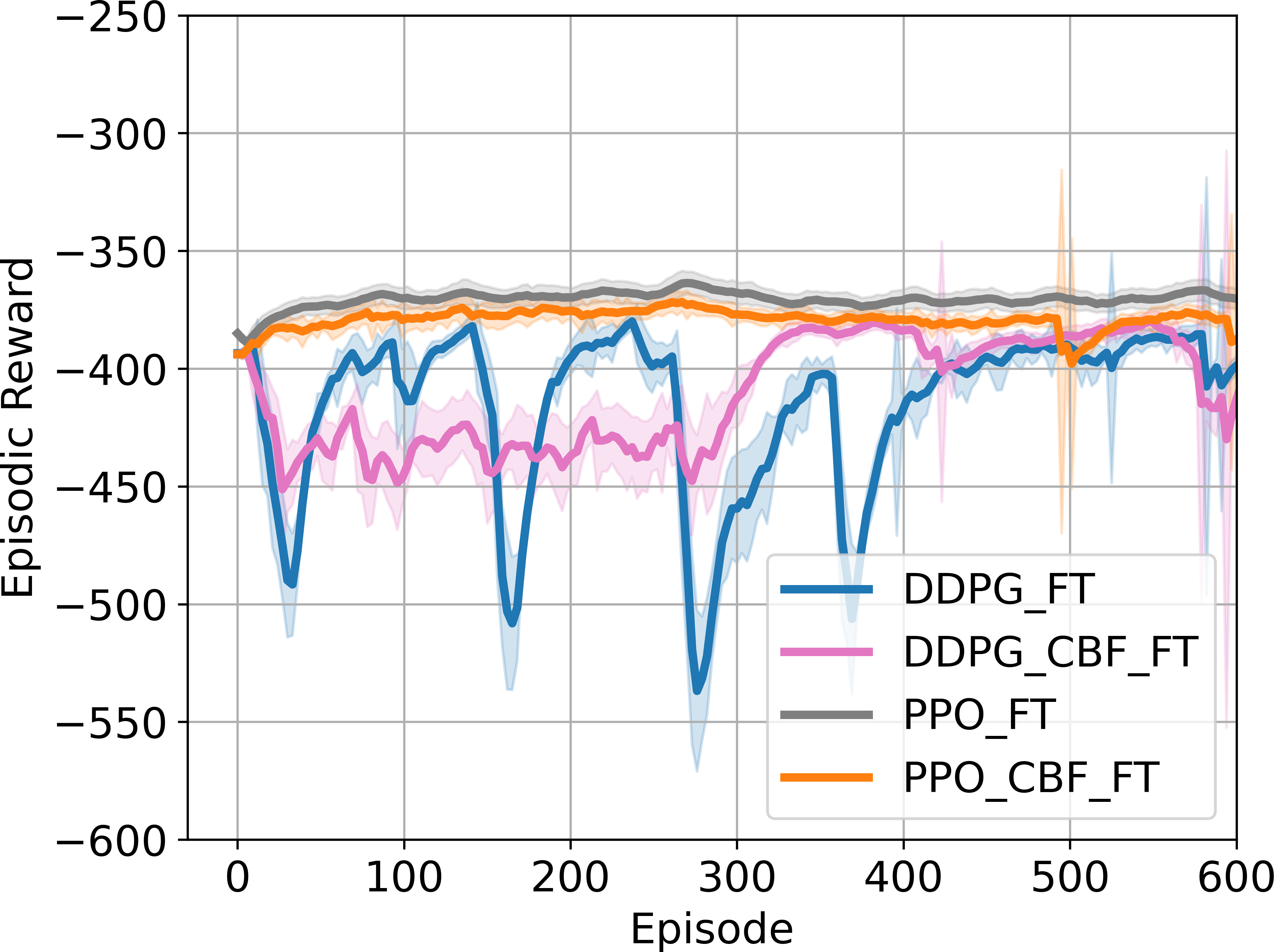}
    \caption{Episodic rewards in packet-level simulation by using models from flow-based simulation with fine-tuning.}
    \label{fig:ns3_epstrain_ft}
\end{figure}
\subsubsection{With fine-tuning}

When all the learning models are fine-tuned, Figure \ref{fig:ns3_epstrain_ft} shows the training rewards of the regular learning algorithms and with \acrshort{cbf} on top of them. In general, we can see that the training rewards of the models without the \acrshort{cbf} function (i.e., \acrshort{ddpg}  and \acrshort{ppo}) significantly improve compared to the Figure \ref{fig:ns3_epstrain_mt} when fine-tuning is not applied.
With respect to the off-policy algorithm (i.e., \acrshort{ddpg}  and \acrshort{ddpg}-CBF), it is important to mention that the replay buffer in the pre-trained models is copied to the packet-based environment for model's fine-tuning. When models are updated, the mixture of old training tuples and new training tuples causes unstable updates until episode 300 and episode 220 for \acrshort{ddpg}  and \acrshort{ddpg}-CBF, respectively.  However, the unstable updates are not seen in the \acrshort{ppo}  and \acrshort{ppo}-\acrshort{cbf} algorithms because they use fresh experiences from the \acrshort{ns3} environment to update the policy from the simplified environment.

\begin{figure}[ht!]
        \centering
        \begin{subfigure}[b]{0.45\textwidth}
            \centering
  \includegraphics[width=\textwidth]{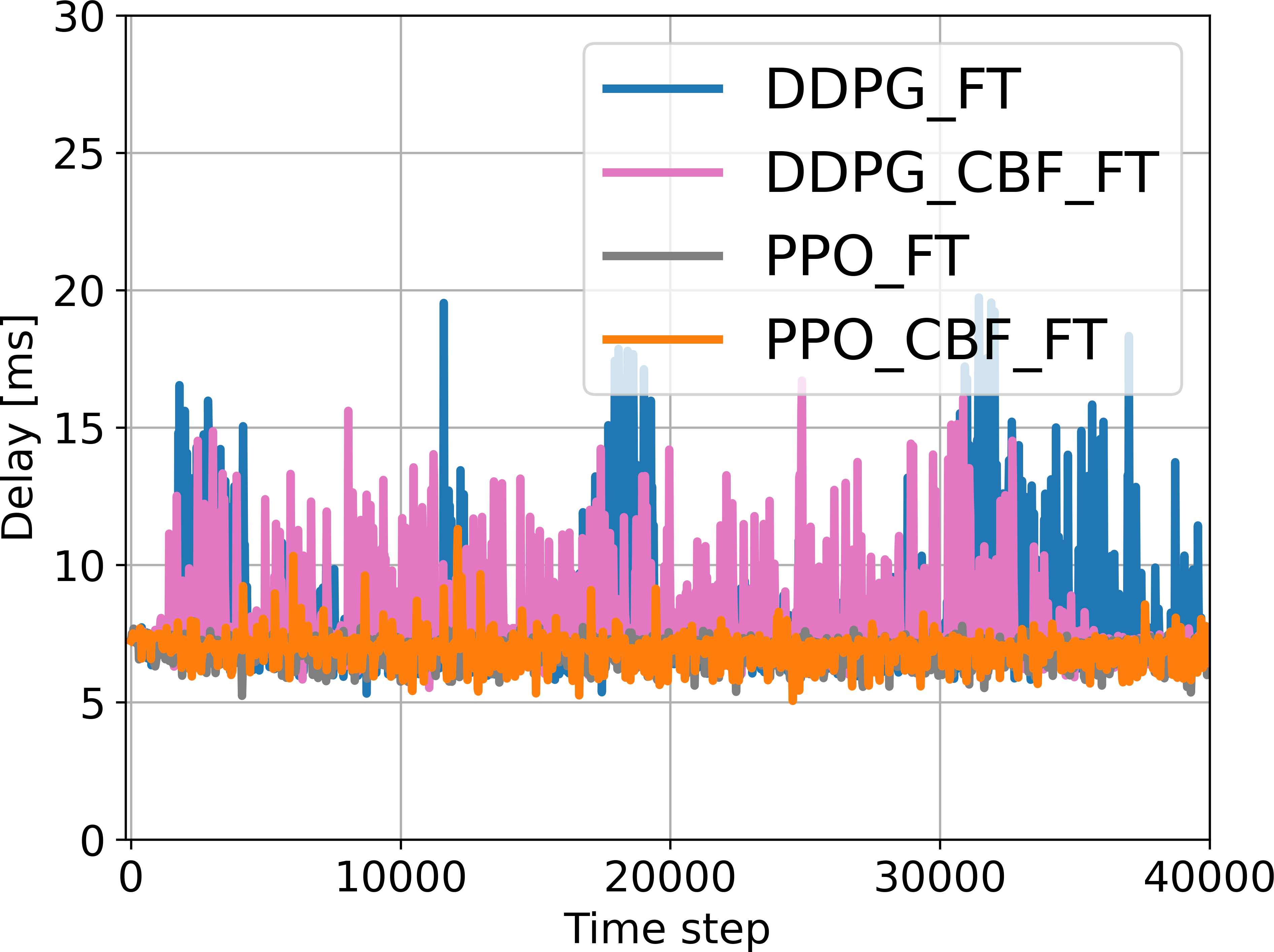}
  \caption{Delay evolution. }
  \label{fig:ns3_deltrain_ft}
        \end{subfigure}
        \begin{subfigure}[b]{0.45\textwidth}  
            \centering
  \includegraphics[width=\textwidth]{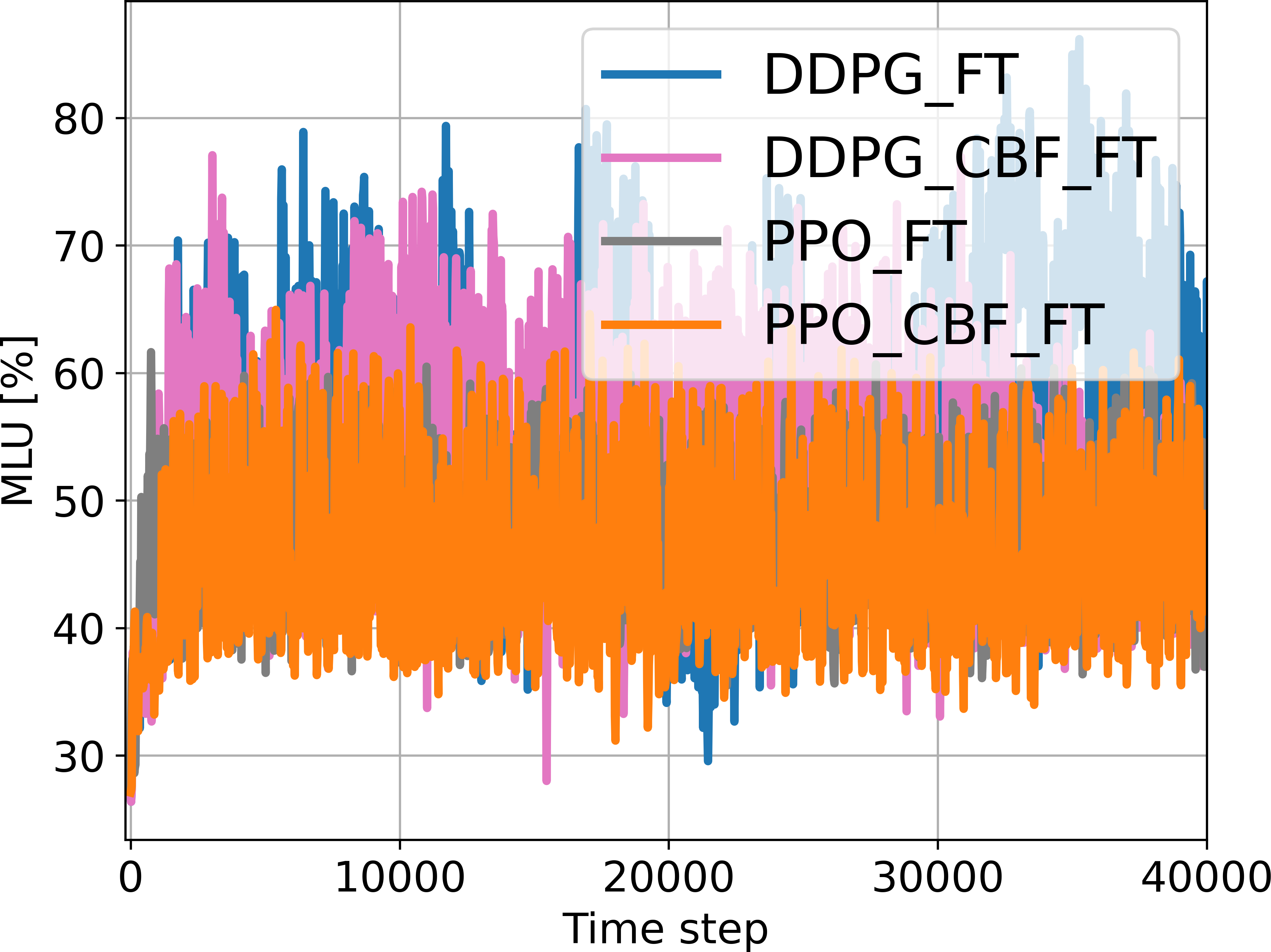}
  \caption{\acrshort{mlu} evolution}
  \label{fig:ns3_mlutrain_ft}
        \end{subfigure}
        \caption{Evolution of delay and \acrshort{mlu} with fine-tuning.} 
        \label{fig:ns3_sla_ft}
\end{figure}

\begin{figure}[ht!]
        \centering
        \begin{subfigure}[b]{0.45\textwidth}
            \centering
  \includegraphics[width=\textwidth]{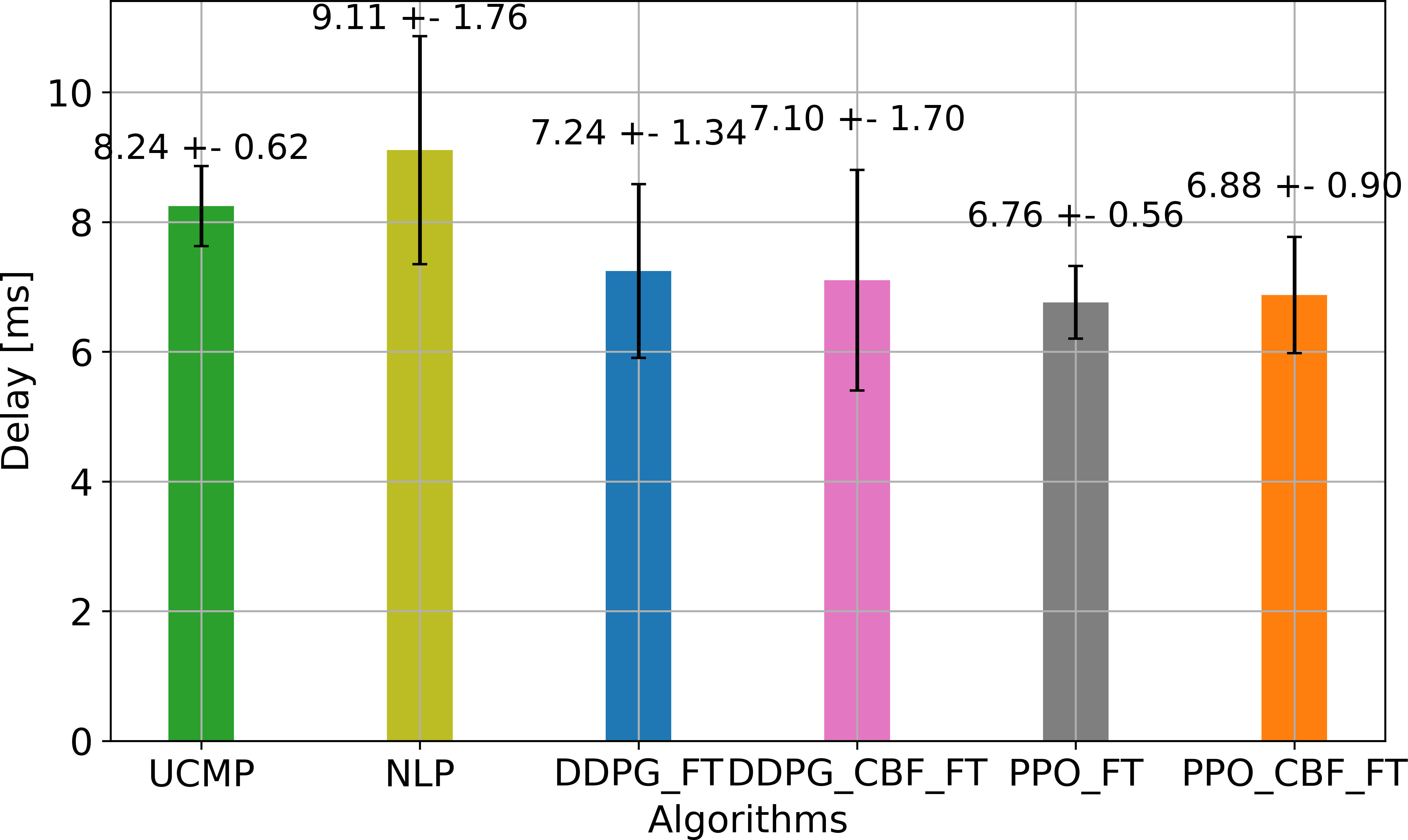}
  \caption{Average delay}
  \label{fig:ave_b_delay_c}
        \end{subfigure}
        \begin{subfigure}[b]{0.45\textwidth}  
            \centering
  \includegraphics[width=\textwidth]{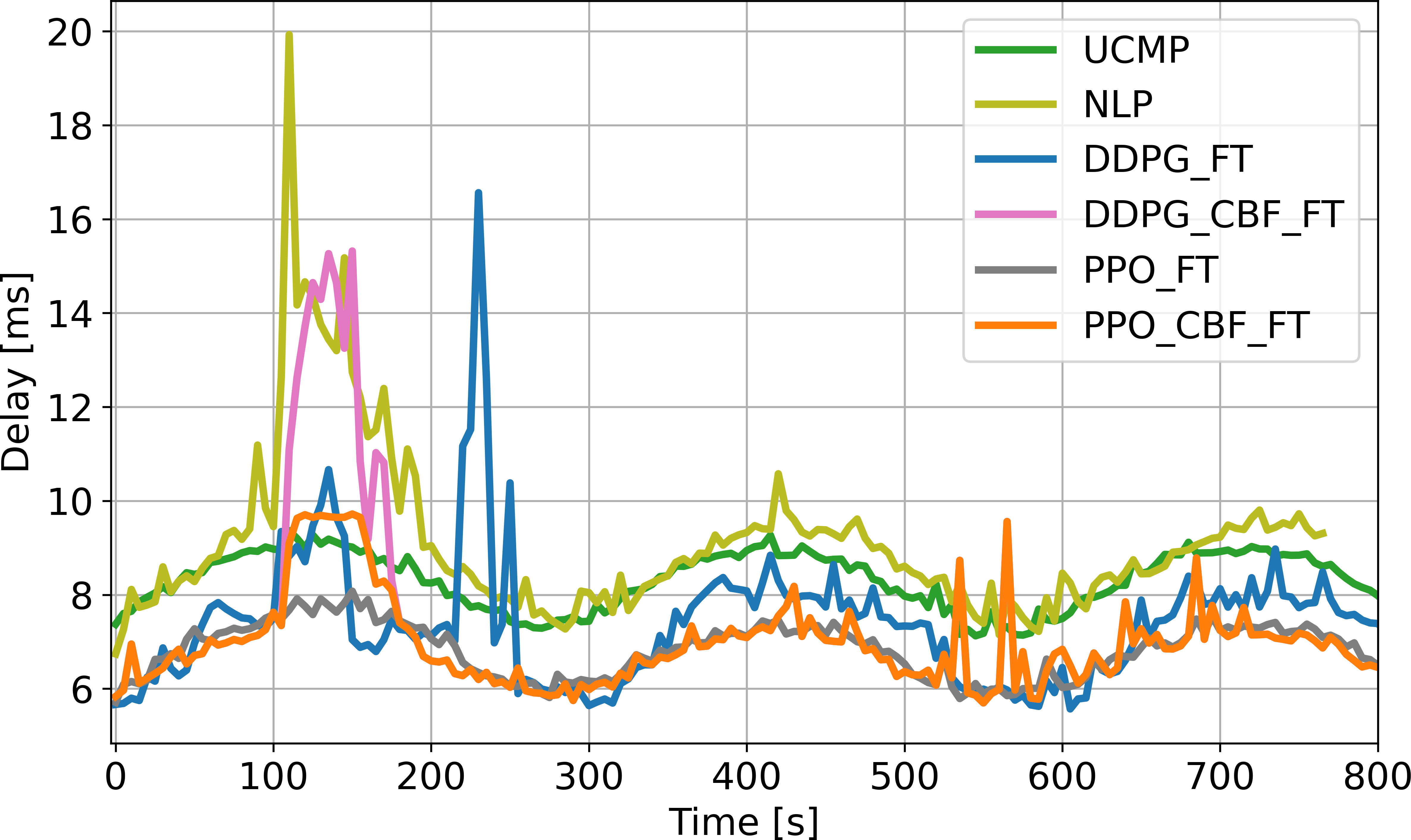}
  \caption{Delay evolution}
  \label{fig:temp_b_delay_cbf}
        \end{subfigure}
        \caption{Average (a) and delay evolution (b) with fine-tuning.} 
        \label{fig:delay_test_cbf}
\end{figure}

\begin{figure}[ht!]
        \centering
        \begin{subfigure}[b]{0.45\textwidth}
            \centering
  \includegraphics[width=\textwidth]{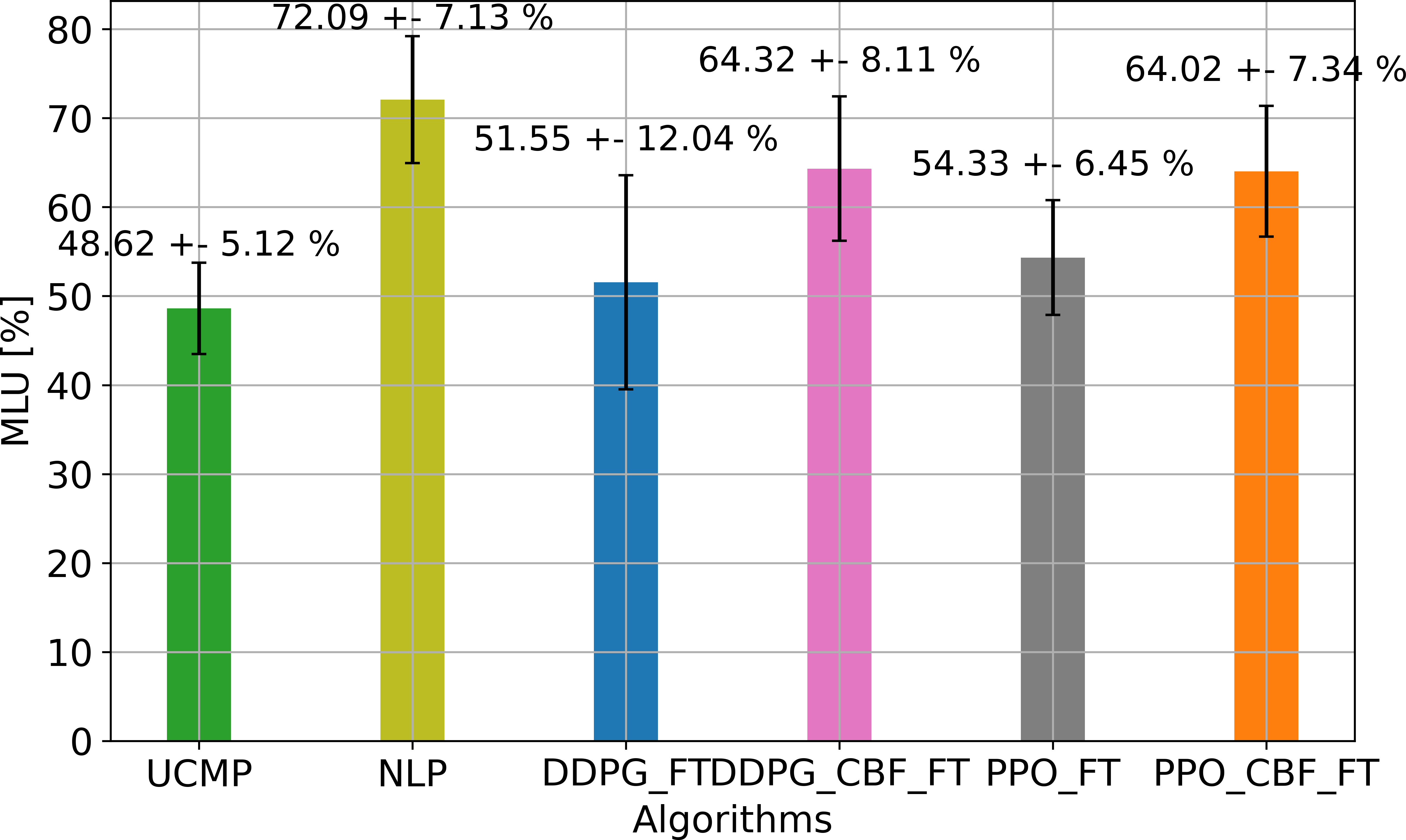}
  \caption{Average \acrshort{mlu}}
  \label{fig:ave_b_mlu_cbf}
        \end{subfigure}
        \begin{subfigure}[b]{0.45\textwidth}  
            \centering
  \includegraphics[width=\textwidth]{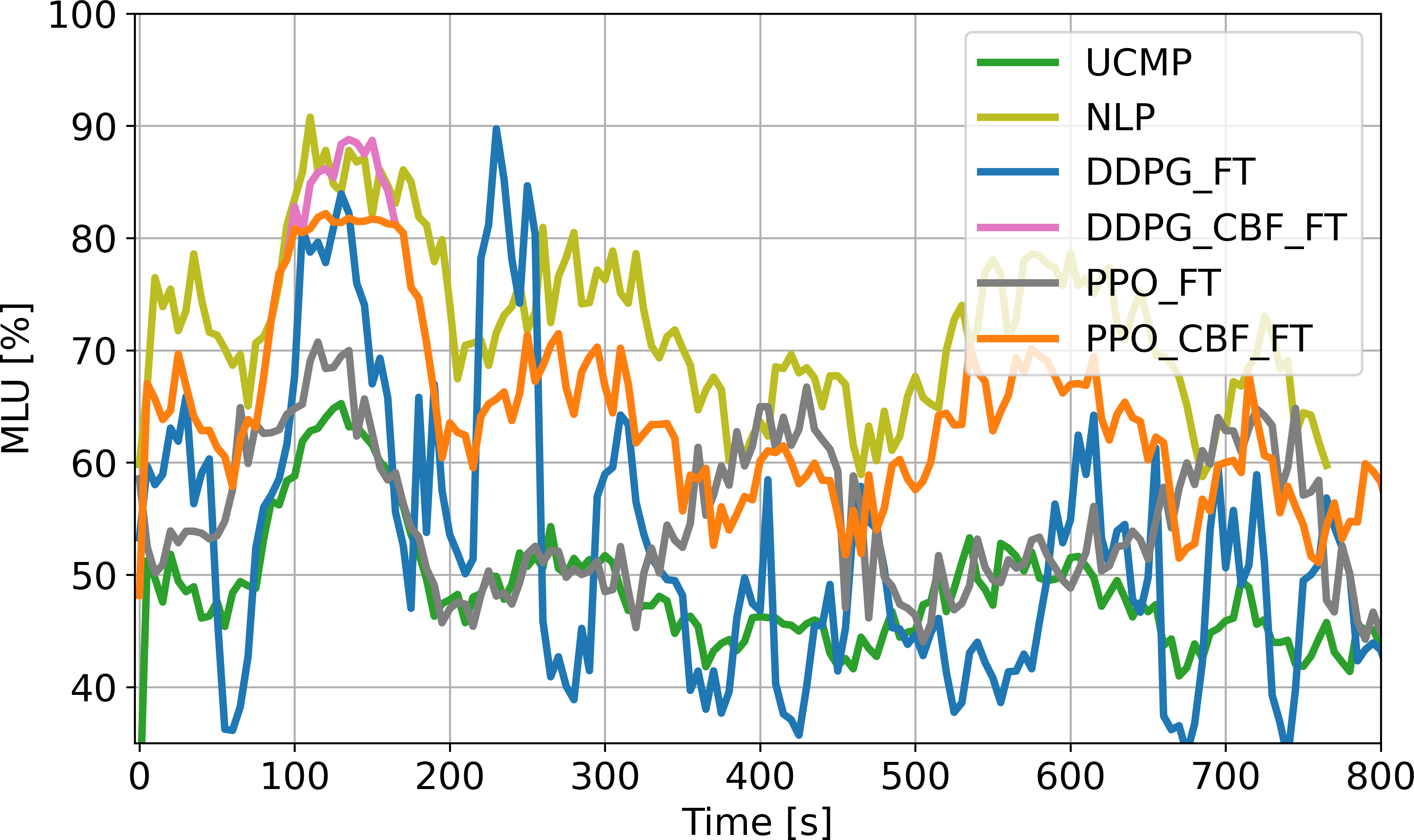}
  \caption{\acrshort{mlu} evolution}
  \label{fig:temp_b_mlu_cbf}
        \end{subfigure}
        \caption{Average (a) and  \acrshort{mlu} evolution (b)  with fine-tuning.} 
        \label{fig:mlu_test_cbf}
\end{figure}

\begin{figure}[ht!]
        \centering
        \begin{subfigure}[b]{0.45\textwidth}
            \centering
  \includegraphics[width=\textwidth]{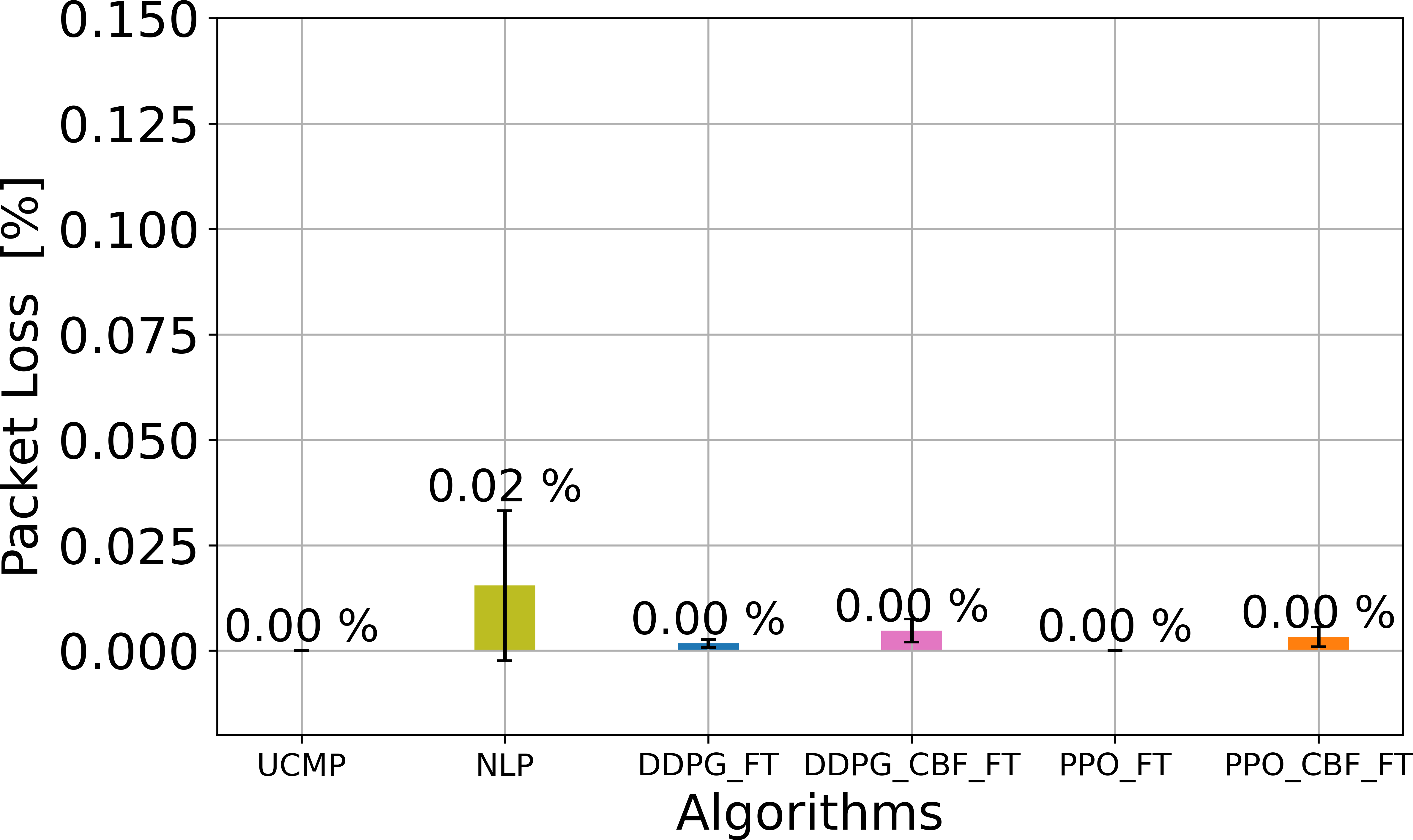}
  \caption{Average PKLoss}
  \label{fig:ave_b_pkl_cbf}
        \end{subfigure}
        \begin{subfigure}[b]{0.45\textwidth}  
            \centering
  \includegraphics[width=\textwidth]{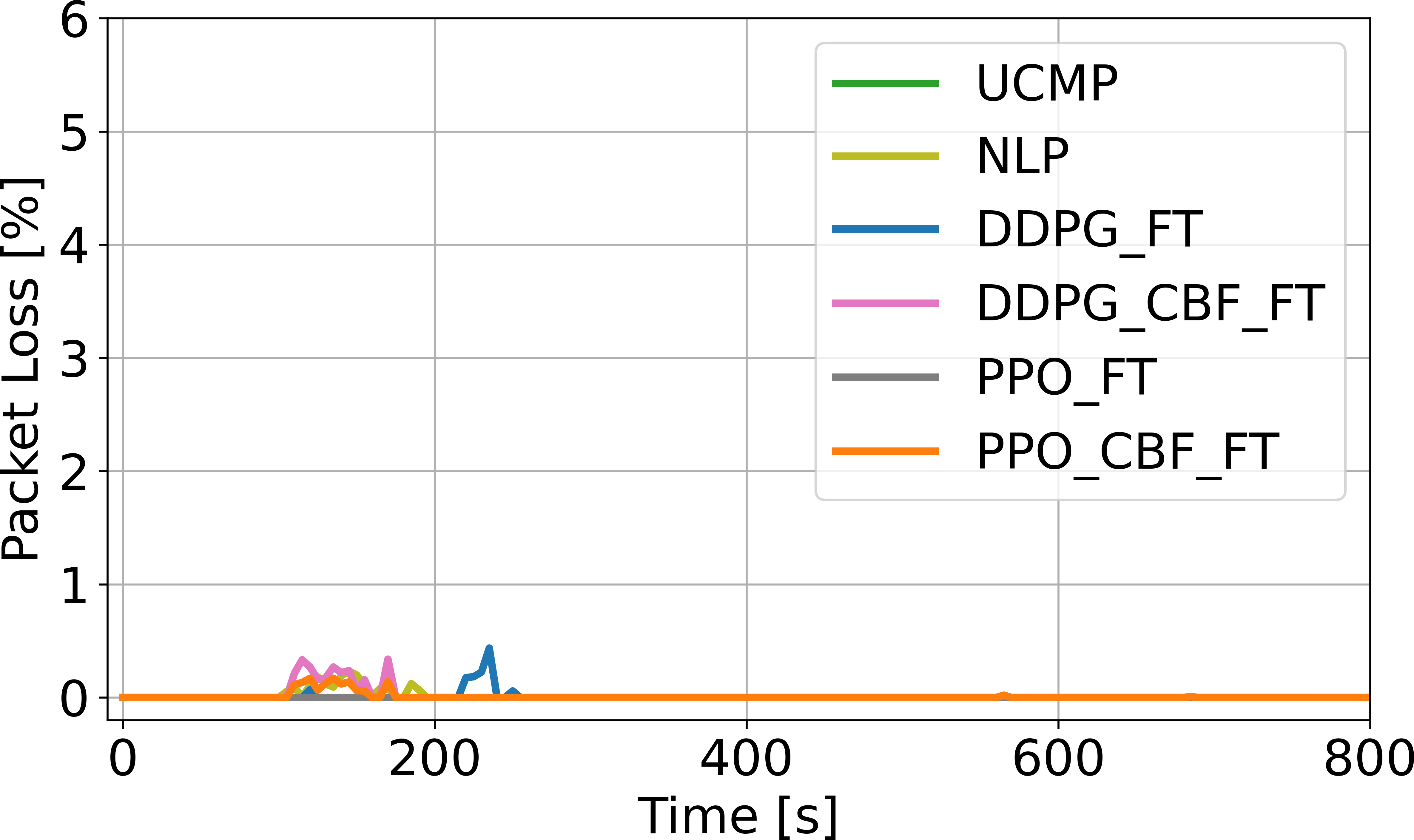}
  \caption{PKLoss evolution}
  \label{fig:temp_b_pkl_res}
        \end{subfigure}
        \caption{Average (a) and PKLoss evolution (b) with fine-tuning.} 
        \label{fig:pkl_test_cbf}
\end{figure}

Figures \ref{fig:ns3_deltrain_ft} and \ref{fig:ns3_mlutrain_ft} depict delay and \acrshort{mlu} of fine-tuned models. They can be compared to Figure \ref{fig:ns3_sla_mt}, in which fine-tuning is not applied and Figures \ref{fig:delay_training}/\ref{fig:mlu_training} in which the models are trained from scratch. We can see that smoother delay during training are obtained while \acrshort{mlu} is kept below 80 $\%$ (i.e., safety constraints are respected).

When the fine-tuned models have converged, they are applied in the testing environment for comparison with the baselines. Figures \ref{fig:delay_test_cbf}, \ref{fig:mlu_test_cbf} and \ref{fig:pkl_test_cbf}, display delay, \acrshort{mlu} and packet loss, for fine-tuned models, UCMP and NLP. They show that the achieved delays of our proposals are better than the benchmarks and no packet loss is seen during testing. Moreover, we can see that the outcomes of the fine-tuned models are similar to the models trained from scratch. It demonstrates the benefits of using fine-tuning approaches instead of regular training while the safety is ensured during the training and similar testing performance is obtained.

\section{Limitations and Future Perspectives}
\label{sec:limitations}
We believe that our study, which combines learning-based and safety-based methods, could still face a few limitations. 

\textbf{Implicit safety constraints:} In this work, we introduced an approach combining a \acrshort{cbf} function with \acrshort{drl} agents for an explicit safety constraint (i.e., \acrshort{mlu}). In addition to explicit safety constraints, implicit constraints can be more challenging to handle as they do not have closed-form expressions (e.g., probability of packet errors is upper bounded) \citep{dalalsafeexploration2018}. To deal with this, more complex \acrshort{cbf} functions should be considered to define a safe region where the learning algorithm can explore (e.g., which paths/links are statistically less congested). As safe regions can be difficult to be approximated, Gaussian Processes (GP) \citep{wachiSafeReinforcementLearning2020a},  can be effective in estimating unknown system dynamics and constructing a safety region.

\textbf{Towards Explainability} Our \acrshort{drl} algorithms heavily depend on fully connected neural networks (NN) for the approximation of policies with continuous network states. Although it is recognized as a powerful tool, it lacks insights into its internal structure to understand why decisions are taken \citep{zhangSurveyNeuralNetwork2021}. This poses not only a practical problem, but also an acceptability problem \citep{wangSurveyInterpretableMachine2022}. Explainable AI tools such as LIME \citep{ribeiroWhyShouldTrust2016}, SHAP\citep{lundbergUnifiedApproachInterpreting2017} can be useful to generate explanations for and help operators understand the logic of DRL agents.

\textbf{More comprehensive neural networks }
Besides explainability, more cutting-edge network architecture can also be considered as alternatives to our fully connected neural network. Recently, Kolmogorov-Arnold Network architecture (KAN) \citep{liuKANKolmogorovArnoldNetworks2024} is gaining more attention because of its capability of achieving higher training accuracy and interpretability. Relying on versatile learnable edges over nodes/neurons as Multi-Layer Perceptron (MLP), it removes non-linear activation function entirely, and replaces them with 1-D spline functions, which improve interpretability. Graph neural networks (GNN) \citep{scarselliGraphNeuralNetwork2009} are also a promising approach to better understand the complex relationship between network topology and input traffic for faster training and/or better quality solutions \citep{wuComprehensiveSurveyGraph2021}.



\section{Conclusion}
\label{sec:conclusion}
In this paper, we have proposed a safe load-balancing approach that uses a \acrfull{cbf} on top of \acrfull{drl} algorithms to optimize \acrshort{qos}.
We evaluated the solution in both flow- and  packet-based simulators.
In the flow-based simulator, in which  networking protocols are omitted for faster evaluation, we have shown that our safe algorithm is capable of achieving  near-optimal delay performance compared to \acrfull{nlp} solution, while the safety constraint with respect to \acrshort{mlu} is guaranteed. In the packet-based simulator, we  demonstrated that our proposal outperforms  non-learning baselines. Using transfer learning, we showed how pre-trained models in the flow-based simulator, which is faster for  learning, can be applied in the packet-based simulator with some fine-tuning.

\section*{Acknowledgment}
This work was financially supported by the French National Research Agency (ANR) under the grant ANR-21-CE25-0005.

\section*{Declaration of Competing Interest}
The authors declare that they have no known competing financial interests or personal relationships that could have appeared to influence the work reported in this paper.

\section*{CRediT authorship contribution statement}
\textbf{Lam Dinh}: Conceptualization, Methodology, Software, Validation, Formal analysis, Investigation, Data curation, Writing – original draft, Writing – review $\&$ editing, Visualization. \textbf{Pham Tran Anh Quang}: Conceptualization, Methodology, Formal analysis, Investigation, Writing – original draft. \textbf{Jérémie Leguay}: Supervision, Writing – original draft, Writing – review $\&$ editing, Methodology. 



 \bibliographystyle{elsarticle-harv} 
 \bibliography{biblio}







\end{document}